\documentclass[a4paper,11pt]{article}
\usepackage{amsmath}    
\usepackage{graphicx}   
\usepackage{amssymb,amscd}    
\usepackage{enumitem}
\usepackage{color}
\usepackage{mathrsfs}
\usepackage{bm}
\usepackage{bbm,wasysym}
\usepackage{multirow}
\usepackage{easyReview}
\usepackage{array}
\usepackage{empheq}
\usepackage{pifont}
\usepackage{soul}
\usepackage{svg}
\usepackage{float}

\usepackage{jheppub} 
\usepackage{lineno}

\arxivnumber{2311.18421}

\renewcommand{\(}{\left(}
\renewcommand{\)}{\right)}
\newcommand{\ec}{\mathtt{e}}
\newcommand{\bb}[1]{\mathbb{#1}}
\newcommand{\fc}[1]{\mathcal{#1}}
\newcommand{\scr}[1]{\mathscr{#1}}
\newcommand{\mc}[1]{\text{\fontencoding{U}\fontfamily{boondoxuprscr}\fontshape{n}\selectfont #1}}
\newcommand{\mf}[1]{\mathfrak{#1}}

\renewcommand{\d}{\partial}

\newcommand{\8}{\infty}
\newcommand{\assert}{\overset{!}{=}}

\newcommand{\p}{\hspace*{5ex}}
\newcommand{\ps}{\hspace*{2ex}}

\newcommand{\pvec}[1]{\vec{#1}\mkern2mu\vphantom{#1}'}

\newcommand{\OO}{\fc{O}}
\newcommand{\twid}[1]{\widetilde{#1}}
\newcommand{\eps}{\varepsilon}

\newcommand{\cmt}[1]{}
\newcommand{\ee}{\text{E}}
\newcommand{\stat}{\text{stat}}

\DeclareFontFamily{U}{boondoxuprscr}{\skewchar \font =45}
\DeclareFontShape{U}{boondoxuprscr}{m}{n}{
	<-> BOONDOXUprScr-Regular}{}
\DeclareFontShape{U}{boondoxuprscr}{b}{n}{
	<-> BOONDOXUprScr-Bold}{}

\newcommand{\rk}{$\sqrt{\text{Kerr}}$ }

\begin{document}

\title{Dynamical Implications of the Kerr Multipole Moments for Spinning Black Holes}

\author[1]{T. Scheopner\note{Corresponding author.}}
\author{and J. Vines}
\affiliation{Mani L. Bhaumik Institute for Theoretical Physics, University of California at Los Angeles,\\ Los Angeles, CA 90095, USA}

\emailAdd{trevor@physics.ucla.edu}

\abstract{
    Previously the linearized stress tensor of a stationary Kerr black hole has been used to determine some of the values of gravitational couplings for a spinning black hole to linear order in the Riemann tensor in the action (worldline or quantum field theory). In particular, the couplings on operators containing derivative structures of the form $(S\cdot\nabla)^n$ acting on the Riemann tensor were fixed, with $S^\mu$ the spin vector of the black hole. In this paper we find that the Kerr solution determines all of the multipole moments in the sense of Dixon of a stationary spinning black hole and that these multipole moments determine all linear in $R$ couplings. For example, additional couplings beyond the previously mentioned are fixed on operators containing derivative structures of the form $S^{2n}(p\cdot\nabla)^{2n}$ acting on the Riemann tensor with $p^\mu$ the momentum vector of the black hole. These additional operators do not contribute to the three-point amplitude, and so do not contribute to the linearized stress tensor for a stationary black hole. However, we find that they do contribute to the Compton amplitude. Additionally, we derive formal expressions for the electromagnetic and gravitational Compton amplitudes of generic spinning bodies to all orders in spin in the worldline formalism and evaluated expressions for these amplitudes to $\OO(S^3)$ in electromagnetism and $\OO(S^5)$ in gravity.
}

\maketitle
\flushbottom

\newpage

\section{Introduction} \label{sec:intro}

	\subsection{General Overview}
 
	The observation of gravitational waves by the LIGO/Virgo collaboration~\cite{LIGOScientific:2016aoc,  LIGOScientific:2017vwq} began a new era of gravitational physics, with implications for astronomy, cosmology, and possibly particle physics. Physical black holes and neutron stars generically carry significant spin angular momentum which affects their dynamics during mergers in binary systems and the gravitational wave signals they emit. These spin effects will play an increasingly important role in signal analysis as gravitational wave detectors become more sensitive~\cite{Punturo:2010zz,  LISA:2017pwj,  Reitze:2019iox} and also lead to rich theoretical structure for generic bodies, but especially so for black holes. 
 
	The study of the dynamics of generic spinning bodies in general relativity has a long history~\cite{Mathisson:1937zz,   Papapetrou:1951pa,   Pirani:1956tn,   Tulczyjew, Dixon:1970mpd, dixon2, dixon3}. Multiple successful field theoretic and worldline based approaches exist for the study of spinning bodies in both the post-Newtonian (PN) approximation~\cite{Barker:1970zr, Barker:1975ae, Kidder:1992fr, Kidder:1995zr, Blanchet:1998vx, Tagoshi:2000zg, Porto:2005ac, Faye:2006gx, Blanchet:2006gy, Damour:2007nc, Steinhoff:2007mb, Levi:2008nh, Steinhoff:2008zr, Steinhoff:2008ji, Marsat:2012fn,Hergt:2010pa, Porto:2010tr,  Levi:2010zu,  Porto:2010zg, Levi:2011eq, Porto:2012as,  Hergt:2012zx,  Bohe:2012mr,  Hartung:2013dza,  Marsat:2013wwa,  Levi:2014gsa, Vaidya:2014kza, Bohe:2015ana,  Bini:2017pee,  Siemonsen:2017yux, Porto:2006bt,  Porto:2007tt,  Porto:2008tb,  Porto:2008jj,  Levi:2014sba, Levi:2015msa, Levi:2015uxa,  Levi:2015ixa,  Levi:2016ofk,   Levi:2019kgk,  Levi:2020lfn, Levi:2020kvb,  Levi:2020uwu,  Kim:2021rfj,  Maia:2017gxn,  Maia:2017yok,  Cho:2021mqw,  Cho:2022syn, Kim:2022pou,  Mandal:2022nty, Kim:2022bwv, Mandal:2022ufb, Levi:2022dqm,  Levi:2022rrq}
    and the post-Minkowskian (PM) approximation~\cite{Bini:2017xzy, Bini:2018ywr, Maybee:2019jus,   Guevara:2019fsj, Chung:2020rrz,   Guevara:2017csg,   Vines:2018gqi,  Damgaard:2019lfh, Aoude:2020onz, Vines:2017hyw, Guevara:2018wpp, Chung:2018kqs, Chung:2019duq, Bern:2020buy, Kosmopoulos:2021zoq, Liu:2021zxr, Aoude:2021oqj, Jakobsen:2021lvp, Jakobsen:2021zvh, Chen:2021kxt, Chen:2022clh, Cristofoli:2021jas, Chiodaroli:2021eug, Cangemi:2022abk, Cangemi:2022bew, Haddad:2021znf, Aoude:2022trd, Menezes:2022tcs, Bern:2022kto, Chen:2022clh, Alessio:2022kwv, Alessio:2023kgf, Bjerrum-Bohr:2023jau, Damgaard:2022jem, Haddad:2023ylx, Aoude:2023vdk, Jakobsen:2023ndj, Jakobsen:2023hig, Heissenberg:2023uvo, Bianchi:2023lrg, Bern:2023ity, gmoocv:2021, Ben-Shahar:2023djm}. The electromagnetic~\cite{Westpfahl:1979gu, Damour:1990jh, Buonanno:2000qq, Kosower:2018adc, Saketh:2021sri, Bern:2021xze, Bern:2023ccb} and non-abelian gauge theory~\cite{delaCruz:2020bbn, delaCruz:2021gjp} cases are very similar in structure to gravity and can be used to develop helpful insights for the harder gravitational problem. 
    
    In both the field theoretic and worldline approaches when considering only the minimal Poincar\'e degrees of freedom, the interaction of the body with gravity is characterized by a tower of effective field theory operators in the action, each carrying a Wilson coefficient, some number of powers of the spin of the body, and some number of powers of the Riemann tensor and its derivatives. For generic bodies these Wilson coefficients take arbitrary values. We will specialize our interest in this paper exclusive to spinning black holes. This restriction in principle determines the values of all such Wilson coefficients. However, presently the values for these Wilson coefficients on linear and quadratic in Riemann tensor are only partly known. 
    
    The coefficients for operators of the form $(S\cdot\nabla)^n R_{\cdot\cdot\cdot\cdot}$ were fixed in Ref.~\cite{Levi:2015msa}. Such operators are precisely those which contribute to the three point amplitude. There are possible operators which are linear in the Riemann tensor but not of this form, such as those of the form $S^{2n} (p\cdot \nabla)^{2n} R_{\cdot\cdot\cdot\cdot}$, whose Wilson coefficients cannot be determined from the three point amplitude. Using the equations of motion in the action, one can see that such operators contribute at order $R^2$ for scattering processes. We find that for a black hole the coefficients for all operators which are linear in the Riemann curvature can be fixed by matching against the multipole moments of the Kerr solution. Several proposals~\cite{Aoude:2022trd, Bern:2022kto, Haddad:2023ylx,  Aoude:2023vdk} have appeared in the literature to fix the coefficients on quadratic in Riemann operators based on a shift-symmetry principle which is already true of the linear in Riemann results. Refs.~\cite{Bautista:2022wjf,Bautista:2023szu} find that the Compton amplitude derived by solving the Teukolsky equation agrees with the shift-symmetry principle through $\OO(S^4)$ but that tension with the shift-symmetry begins at $\OO(S^5)$ (though the results from the Teukolsky equation involve a subtle analytic continuation between the black-hole and naked-singularity regimes). The couplings we find based on consideration of multipole moments can be made consistent with spin-exponentiation, shift-symmetry, or the Teukolsky results through $\OO(S^5)$. As well, they can be made simultaneously consistent with spin-exponentiation and the Teukolsky results through $\OO(S^5)$. (Beginning at $\OO(S^5)$ one helicity combination develops a spurious pole in the the spin-exponentiated amplitude; when we say that we can match spin-exponentiation at $\OO(S^5)$ or beyond we only mean that we match to the helicity combination without a spurious pole after $\OO(S^4)$.)
	
	\subsection{Summary of Method and Results}

    In Dixon's landmark papers Ref.~\cite{dixon2, dixon3} on the worldline formalism, among other results, he proves that there is a unique way to define the multipole moments of the current density or stress tensor for an extended body in general relativity so that those multipole moments can be made into a generating function for the current density/stress tensor in the usual way and so that those multipole moments are fully reduced (there are no interdependencies between moments of different orders required by either the continuity of charge or local conservation of energy-momentum). The definitions of these moments are highly nontrivial and generally do not coincide with the ``naive'' moments (from integrating powers of displacement against the current density/stress tensor over a spatial slice). 
    
    The higher multipole moments of Dixon's formalism (quadrupole and beyond) have time evolution completely unconstrained by the continuity of charge and local conservation of energy-momentum while the low order multipole moments (essential the total linear momentum and spin) have time evolution completely determined by the same two principles. Their equations of motion are the Mathisson-Papapetrou-Dixon (MPD) equations. In order to determine the time evolution of the higher multipole moments, each higher moment requires either a supplementary equation of motion or constitutive relation in terms of lower moments. It is possible to provide such constitutive relations in terms of the minimal Poincare degrees of freedom $(z,p,S)$ by deriving them from a variational principle in terms of these variables. This minimal set of degrees of freedom is especially natural to consider from an effective theory perspective as at large distance from the body it is effectively point-like and these degrees of freedom form the generators Poincare group for which symmetry is broken by the existence of such a point-like source. The resulting constitutive relations will give spin-induced higher multipole moments.
    
    From the Kerr solution, following Israel's analysis~\cite{israel:1970}, we compute the stress tensor which acts as its source (in the maximally causally extended spacetime) and from that source we compute the multipole moments of a spinning black hole using Dixon's definitions of the multipole moments. Those multipole moments can then be used to determine a stress tensor, which in turn can be used to determine an action for the spinning black hole, up to couplings to operators which are quadratic in the Riemann tensor. The action we find is put in dynamical mass function form in \eqref{eq:grmpdact} and the specific dynamical mass function we find for spinning black holes is given in \eqref{eq:grdmf}.

    The dynamical mass function we find contains all of the equivalent black hole couplings identified in Ref.~\cite{Levi:2015msa}, which can be found by comparison to the three-point amplitude (the stationary stress tensor), as well as many new terms. Dixon's formalism specifies the unique way to lift those naive moments to proper multipole moments, and that lifting fixes the additional couplings in \eqref{eq:grdmf} relative to Ref.~\cite{Levi:2015msa}. This lifting is only able to fix linear in Riemann couplings in the action because no information about higher order in Riemann operators is contained in the stationary moments. We find that the dynamical mass function in \eqref{eq:grdmf} can be made consistent with the spin-exponentiation proposed by Ref.~\cite{Guevara:2018wpp} and shift-symmetry proposed by Refs.~\cite{Aoude:2022trd, Bern:2022kto, Haddad:2023ylx,  Aoude:2023vdk} through $\OO(S^4)$ for the appropriate choice of quadratic in Riemann couplings. As well, it can be made consistent with any one of the three principles of spin-exponentiation, shift-symmetry, or the Teukolsky equation results found in Ref~\cite{Bautista:2022wjf} through $\OO(S^5)$ and made consistent with any pair of them except the combination of shift-symmetry and the Teukolsky equation, which are incompatible. To facilitate this analysis, we find a formal expression for the gravitational Compton amplitude for a generic spinning body to all orders in spin and explicitly compute that amplitude in terms of all possible Wilson coefficients in the action through $\OO(S^5)$. We find that there is one linearly independent structure in the amplitude at $\OO(S^2)$, with one more appearing at $\OO(S^3)$, seven more at $\OO(S^4)$, and eleven more at $\OO(S^5)$. 

    In order to understand these gravitational results, it is instructive to first follow all of the same steps of analysis for a \rk particle in electromagnetism. In section \ref{sec:emmpd} we review the worldline formalism with a dynamical mass function for electromagnetism. In section \ref{sec:dixon} we review the basics of Dixon's theory of multipole moments and specialize his results to the current-density in flat spacetime. In section \ref{sec:rkmoments} we use Dixon's formalism to compute the multipole moments of a \rk particle and from those moments we compute the necessary dynamical mass function for such a particle, up to corrections which are quadratic in the field strength. Our electromagnetic analysis culminates in section \ref{sec:emcompton} in which we compute the electromagnetic Compton amplitude for a generic spinning body to all orders in spin in terms of its dynamical mass function. We then specialize that all orders result to cubic order in spin by enumerating all possible operators in the action and study the spin-exponentiation and shift-symmetry properties of the resultant amplitude. We find that for electromagnetism, it is possible to simultaneously demand spin-exponentiation, shift-symmetry, and consistency with the \rk multipole moment based dynamical mass function through $\OO(S^3)$.

    In the second half of the paper, we perform the same analysis for gravity. We begin in \ref{sec:grmpd} by reviewing the worldline formalism with a dynamical mass function in general relativity. In section \ref{sec:kerrmoments} we use Dixon's formalism to compute the multipole moments of a Kerr particle and from those moments we compute the necessary dynamical mass function for such a particle, up to corrections which are quadratic in the Riemann tensor. Then in \ref{sec:grcompton} we derive a formal expression for the gravitational Compton amplitude for a generic spinning body to all orders in spin, which is unfortunately much more complex than the corresponding electromagnetic formula. We then specialize that all orders result to quintic order in spin by enumerating all possible operators in the action and study the requirements imposed on that amplitude by spin-exponentiation, shift-symmetry, and matches to the Teukolsky equation. We find that for gravity, the multipole moment based dynamical mass function is consistent with the combination of spin-exponentiation and the Teukolsky equation at $\OO(S^5)$ and that requiring these fixes all available coefficients in the dynamical mass function at this order.
    
	\subsection{Notation}
	
	We call the spacetime manifold $\bb{T}$. Beginning alphabet Greek letter indices $\alpha,\beta,\gamma,\delta,...$ are used for spacetime indices at a generic event $X \in \bb{T}$. The tangent space to $\bb{T}$ at $X$ is written $T_X(\bb{T})$. Late alphabet Greek letter indices $\mu, \nu, \rho, \kappa, ...$ are used for spacetime indices at a particular event of interest $Z \in \bb{T}$. Indices are symmetrized using parentheses and antisymmetrized with brackets, both with the typical symmetry factors:
	\begin{equation}
		M^{(\alpha\beta)} = \frac{M^{\alpha\beta} + M^{\beta\alpha}}{2}, \p M^{[\alpha\beta]} = \frac{M^{\alpha\beta} - M^{\beta\alpha}}{2}.
	\end{equation}
	For a generic vector $v^\alpha$, we define:
	\begin{equation}
		|v| = \sqrt{|g_{\alpha\beta} v^\alpha v^\beta|}, \p \hat v^\alpha = \frac{v^\alpha}{|v|}.
	\end{equation}
	
    Let $\zeta(s, Z, v)$ be a geodesic with affine parameter $s$ so that $\zeta^\mu(0, Z, v) = z^\mu$ and $\frac{d\zeta^\mu}{ds}(0, Z, v) = v^\mu$. Then, the exponential map is defined by the event:
	\begin{equation}
		\exp_Z(v) = \zeta(1, Z, v).
	\end{equation}
	Consider the geodesic $\zeta(s)$ with affine parameter $s$ so that $\zeta^\mu(0) = z^\mu$ and $\zeta^\alpha(1) = x^\alpha$ (for $Z$ and $X$ sufficiently close for one only such geodesic to exist). Then, Synge's worldfunction $\sigma(Z, X)$ is defined by:
	\begin{equation}
		\sigma(Z,X) = \frac{1}{2}\int_0^1 g_{\alpha\beta}(\zeta) \frac{d\zeta^\alpha}{ds} \frac{d\zeta^\beta}{ds} ds.
	\end{equation}
	Instead viewing $\sigma$ as a functional or the path, under variation with respect to the path $\zeta$, we find:
	\begin{equation}
		\delta \sigma = \frac{d\zeta_\alpha}{ds}(1) \delta x^\alpha - \frac{d\zeta_\mu}{d s}(0) \delta z^\mu.
	\end{equation}
	We place indices on $\sigma$ to indicate covariant derivatives with $\alpha,\beta,...$ indices for $x$ and $\mu,\nu,...$ indices for $z$:
	\begin{equation}
		\sigma_\alpha = \nabla_\alpha \sigma = \frac{\d\sigma}{\d x^\alpha} = \frac{d \zeta_\alpha}{ds}(1) , \p \sigma_\mu = \nabla_\mu \sigma = \frac{\d\sigma}{\d z^\mu} = -\frac{d\zeta_\mu}{ds}(0). \label{eq:geotan}
	\end{equation}
	If more that two indices of the same type were to be placed on $\sigma$ the order would be important due to the noncommutativity of covariant derivatives, however we will have no need to do this. As well, we introduce the inverse matrix $\sigma^{-1}$:
	\begin{equation}
		\sigma^{-\alpha}{}_\mu\sigma^\mu{}_\beta = \delta^\alpha_\beta.
	\end{equation}
    Consider the set of events $X$ close enough to $Z$ to only be reachable by a single geodesic beginning at $Z$. Such $X$ are then determined by $Z$ and the initial tangent vector $v^\mu$ at $Z$ of the geodesic by some function $x^\alpha(z,v)$. So, $\frac{\d x^\alpha}{\d v^\mu}$ exists. Equation \eqref{eq:geotan} guarantees that $v^\mu = -\sigma^\mu(x,z)$ and so $\frac{\d v^\mu}{\d x^\alpha} = -\sigma^\mu{}_\alpha$. The existence of $\frac{\d x^\alpha}{\d v^\mu}$ then demands $\frac{\d x^\alpha}{\d v^\mu} = -\sigma^{-\alpha}{}_\mu$ and so the inverse matrix $\sigma^{-1}$ is guaranteed to exist for such $X$. 
	
	We will describe the motion of the spinning body so that the worldline $z^\mu(\lambda)$ tracks the center of momentum of the body with $\lambda$ the worldline time parameter. $u^\mu(\lambda)$ will be a smooth one parameter family of future oriented timelike unit vectors which later will be specialized to be $\hat p^\mu(\lambda)$, the unit vector in the direction of the linear momentum $p^\mu(\lambda)$ of the body. We let $\Sigma(\lambda)$ be the Cauchy slice formed by shooting out geodesics based at $z^\mu(\lambda)$ which are orthogonal to $u^\mu(\lambda)$. Explicitly:
	\begin{equation}
		\Sigma(\lambda) = \{ X \in \bb{T} : u^\mu(\lambda)\sigma_\mu(Z(\lambda), X) = 0\}. \label{eq:SigmaDef}
	\end{equation}
	Let $\tau(X)$ be the value of $\lambda$ so that $X \in \Sigma(\lambda)$. The existence and uniqueness of the function $\tau(X)$ is a condition on the well-definedness of our time slicing. We require each point in spacetime $X$ which is sufficiently close to the worldline $Z$ to enter any requisite integrals which appear later to fail into one and only one equal-time slice $\Sigma(\lambda)$. Let $w_1^\alpha(X)$ be any vector field satisfying:
	\begin{equation}
		\lambda = \tau(X) \implies \lambda+\delta\lambda = \tau(\exp_X(w_1\delta\lambda)) + \OO(\delta\lambda^2).
	\end{equation}
	That is, if each point of $\Sigma(\lambda)$ is displaced by $w_1^\alpha\delta\lambda$ then it produces a point in $\Sigma(\lambda+\delta\lambda)$ for sufficiently small $\delta\lambda$. Automatically:
	\begin{equation}
		w_1^\alpha \nabla_\alpha \tau = 1.
	\end{equation}
	Let $D^4 x$ be the invariant spacetime volume measure:
	\begin{equation}
		D^4 x = \sqrt{-\det g} d^4 x.
	\end{equation}
	Let $d\Sigma_\alpha$ be the future oriented invariant volume measure on $\Sigma$. For any scalar function $f(X)$ then:
	\begin{equation}
		\int_{\bb{T}} f(X) D^4x = \int_{-\8}^\8 \int_{\Sigma(\lambda)} f(X) w_1^\alpha d\Sigma_\alpha d\lambda.
	\end{equation}
	
    Because $\lambda$ is an arbitrary parameter for the worldline, it is useful to introduce the einbein $\ec(\lambda)$ to help manage reparameterization invariance manifestly. The einbein is an arbitrary function defined so that under a smooth monotone increasing reparameterization $\lambda' = \lambda'(\lambda)$:
	\begin{equation}
		\ec'(\lambda') = \frac{d\lambda}{d\lambda'} \ec(\lambda).
	\end{equation}
	The reparameterization invariant worldline measure $D\lambda$ is then defined by:
	\begin{equation}
		D\lambda = \ec(\lambda) d\lambda.
	\end{equation}
	We also define the reparameterization invariant version of the vector field $w_1^\alpha(X)$:
	\begin{equation}
		w^\alpha(X) = \frac{w_1^\alpha(X)}{\ec(\tau(X))}.
	\end{equation}
	Then, we have:
	\begin{equation}
		\int_{\bb{T}} f(X) d\bb{T} = \int_{-\8}^\8 \int_{\Sigma(\lambda)} f(X) w^\alpha d\Sigma_\alpha D\lambda.
	\end{equation}
	Selection of a worldline parameter then amounts to choosing $\ec(\lambda)$ as an arbitrary function.

\section{Electromagnetic MPD Equations}	\label{sec:emmpd}
	
	We begin with an analysis of the \rk electromagnetic particle in flat spacetime to develop a road-map for the gravitational analysis which follows. The motion of a generic spinning body under the influence of electromagnetism in Minkowski space is described by the electromagnetic flat space Mathisson Papapetrou Dixon (MPD) equations~\cite{dixon3}. It is well established~\cite{dixon3, Ehlers:1977rud, bailey1975, Vines:2017hyw, JanSteinhoff:2015ist, gmoocv:2021} that the electromagnetic MPD equations can be derived from a variational principle through an action $\fc{S}$ of the form:
	\begin{equation}
		\fc{S}[z, p, \Lambda, S, \alpha, \beta] = \int_{-\8}^\8\(p_\mu\dot z^\mu + q A_\mu \dot z^\mu + \frac{1}{2}\epsilon_{\mu\nu\rho\sigma} u^\mu S^\nu \Omega^{\rho\sigma} - \frac{\alpha}{2}(p^2 + \fc{M}^2) + \beta p\cdot S\)d\lambda	\label{eq:emmpdact}.
	\end{equation}
	In this action, $z^\mu(\lambda)$ is the center of momentum worldline of the body, its conjugate is the linear momentum carried by the body $p_\mu(\lambda)$, $q$ is the charge of the body, $A_\mu$ is the vector potential, $u^\mu = \hat p^\mu$, $\alpha = \frac{\ec}{\fc{M}}$ is a Lagrange multiplier which reinforces reparameterization invariance, and $S^\mu$ is the spin vector of the body. The spin vector is defined so that it becomes the angular momentum vector of the body in its center of momentum frame. Automatically, then $S\cdot p = 0$ and so $\beta$ is a Lagrange multiplier included to enforce this constraint. Also appearing is ${\Lambda^\mu}_A(\lambda)$, a tetrad tracking the orientation of the body, which satisfies:
	\begin{equation}
		\eta^{\mu\nu} = \Lambda^\mu{}_A\Lambda^\nu{}_B \eta^{AB}, \p \eta_{AB} = \eta_{\mu\nu}\Lambda^\mu{}_A\Lambda^\nu{}_B.
	\end{equation}
	Without loss of generality, we take:
	\begin{equation}
		 {\Lambda^\mu}_0(\lambda) = u^\mu(\lambda), \p {\Lambda^\mu}_3(\lambda) = \hat S^\mu(\lambda).
	\end{equation}
	(If these are set as initial conditions, they are maintained dynamically automatically.) Capital Latin indices $A, B, C, D, ...$ are always used for Lorentz indices in the default frame of the body so that ${\Lambda^\mu}_A$ represents the Lorentz transformation the body has undergone in its motion relative to an arbitrary default orientation. The angular velocity tensor of the body is defined by:
	\begin{equation}
		\Omega^{\mu\nu} = \eta^{AB}{\Lambda^\mu}_A\frac{d{\Lambda^\nu}_B}{d\lambda}	\label{eq:angveldef}.
	\end{equation}
	Finally, $\fc{M}(z, u, S)$, called the dynamical mass function of the body, encodes the free mass of the body and all of its nonminimal couplings to electromagnetism. In particular, it takes the form:
	\begin{equation}
		\fc{M}^2(z, u, S) = m^2 + \OO(q\fc{F})
	\end{equation}
	where $m$ is the mass of the body in vacuum. 
	
	For variations of the action it is useful to define the antisymmetric tensor:
	\begin{equation}
		\delta\theta^{\mu\nu} = \eta^{AB}{\Lambda^\mu}_A \delta\Lambda^\nu{}_B.
	\end{equation}
	Then, the variation of the above action gives:
	\begin{align}
		\delta \fc{S} = &\int_{-\8}^\8\(\delta z^\mu\( -\dot p_\mu + q\fc{F}_{\mu\nu}\dot z^\nu - \ec\frac{\d\fc{M}}{\d z^\mu}\)\right.\nonumber \\
		&\left.+\delta p_\mu\(\dot z^\mu - \ec u^\mu - \frac{\ec}{|p|}\frac{\d\fc{M}}{\d u^\nu}(\eta^{\mu\nu}+u^\mu u^\nu) + \beta S^\mu + \frac{1}{2|p|}(\delta^\mu_\alpha+u^\mu u_\alpha)\epsilon^{\alpha\nu\rho\sigma}S_\nu\Omega_{\rho\sigma}\)\right. \nonumber \\
		& \left.+\frac{1}{2}\delta\theta^{\rho\sigma}\(-\frac{d}{d\lambda}\(\epsilon_{\mu\nu\rho\sigma} u^\mu S^\nu\) + \epsilon_{\mu\nu\rho\alpha} u^\mu S^\nu \Omega^\alpha{}_\sigma - \epsilon_{\mu\nu\sigma\alpha} u^\mu S^\nu \Omega^\alpha{}_\rho\)\right. \nonumber \\
		& \left.+\delta S^\mu\(-\frac{1}{2}\epsilon_{\mu\nu\rho\sigma} u^\nu \Omega^{\rho\sigma} - \ec\frac{\d\fc{M}}{\d S^\mu} + \beta p_\mu\) - \frac{\delta\alpha}{2}\(p^2+\fc{M}^2\) + \delta\beta p\cdot S\) d\lambda.
	\end{align}
	Using the $\delta S^\mu$ variation to solve for the angular velocity tensor, one can then determine the value of $\beta$. That value of $\beta$ can then be used to simplify the spin and trajectory equations of motion. Explicitly, these give:
	\begin{align}
		\Omega^{\mu\nu} &= \dot u^\mu u^\nu - u^\mu \dot u^\nu + \ec \epsilon^{\mu\nu\rho\sigma} u_\rho \frac{\d\fc{M}}{\d S^\sigma} \label{eq:angvelsol} \\
		\beta &= -\frac{\ec}{\fc{M}}u^\mu \frac{\d\fc{M}}{\d S^\mu}\\
		\dot S^\mu &= u^\mu \dot u\cdot S + \ec \epsilon^{\mu\nu\rho\sigma} u_\nu S_\rho \frac{\d\fc{M}}{\d S^\sigma} \\
		\dot z^\mu &= \ec u^\mu + \frac{\ec}{\fc{M}}\frac{\d\fc{M}}{\d u_\mu} + \frac{\ec}{\fc{M}} u^\mu u^\nu \frac{\d\fc{M}}{\d u^\nu} + \frac{\ec}{\fc{M}} S^\mu u^\nu \frac{\d\fc{M}}{\d S^\nu} + \frac{1}{\fc{M}^2} \epsilon^{\mu\nu\rho\sigma} S_\nu u_\rho \dot p_\sigma.	\label{eq:velsol1}
	\end{align}
    If the dynamical mass function is kept only through linear order in spin, the resulting equations of motion are equivalent to the Bargmann-Michel-Telegdi (BMT) equation. In order to determine the trajectory evolution explicitly we must insert the momentum equation of motion into \eqref{eq:velsol1}. To simplify, it is useful to introduce the dual field strength:
	\begin{equation}
		{}^\star\fc{F}_{\mu\nu} = \frac{1}{2}\epsilon^{\mu\nu\rho\sigma} \fc{F}_{\rho\sigma} \implies \fc{F}_{\mu\nu} = -\frac{1}{2}\epsilon_{\mu\nu\rho\sigma} {}^\star\fc{F}^{\rho\sigma}.
	\end{equation}
	Simplifying finally gives the electromagnetic MPD equations of motion for the spinning body:
	\begin{align}
		\(1 - \frac{q}{\fc{M}^2}{}^\star{\fc{F}}^{\alpha\beta}u_\alpha S_\beta\)\frac{\dot z^\mu}{\ec} =&\ u^\mu + \frac{1}{\fc{M}}\frac{\d\fc{M}}{\d u_\mu} + u^\mu\frac{u^\nu}{\fc{M}}\frac{\d\fc{M}}{\d u^\nu} + S^\mu \frac{u^\nu}{\fc{M}}\frac{\d\fc{M}}{\d S^\nu} + \frac{1}{\fc{M}^2} \epsilon^{\mu\nu\rho\sigma} u_\nu S_\rho \frac{\d \fc{M}}{\d z^\sigma} \nonumber \\
		&\ \phantom{u^\mu}+ \frac{q}{\fc{M}^2}{}^\star{\fc{F}}^{\mu\nu}S_\nu + \frac{q}{\fc{M}^3}\(S^\rho \frac{\d\fc{M}}{\d u^\rho} + S^2 u^\rho \frac{\d\fc{M}}{\d S^\rho}\){}^\star{\fc{F}}^{\mu\nu}u_\nu \label{eq:velsol2}\\
		\dot p_\mu =&\ q\fc{F}_{\mu\nu}\dot z^\nu - \ec\frac{\d\fc{M}}{\d z^\mu} \\
		\dot S^\mu =&\ u^\mu \dot u\cdot S + \ec \epsilon^{\mu\nu\rho\sigma} u_\nu S_\rho \frac{\d\fc{M}}{\d S^\sigma}.
	\end{align}
	For solving these equations of motion we will always choose $\lambda$ so that $\ec = 1$. 
	
	To understand how the dynamical mass function relates to the multipole moments of the body, we will need to study the current produced by our action. Define the $\fc{Q}_n$ moments:
	\begin{equation}
		\fc{Q}_n^{\rho_1...\rho_n\mu\nu} = \fc{Q}_n^{(\rho_1...\rho_n)[\mu\nu]} = \frac{\d\fc{M}}{\d \d^n_{\rho_1...\rho_n}\fc{F}_{\mu\nu}}	\label{eq:Qmoment}
	\end{equation}
	Then, our action produces a formal distributional expression for $J^\mu$:
	\begin{equation}
		J^\mu(X) = \frac{\delta\fc{S}}{\d A_\mu} = \int_{-\8}^\8\(q\dot z^\mu \delta(X-Z) - 2\ec\sum_{n=0}^\8(-1)^n\fc{Q}_n^{\rho_1...\rho_n\mu\nu} \d^{n+1}_{\rho_1...\rho_n\nu}\delta(X-Z)\) d\lambda	\label{eq:current}
	\end{equation}

\section{Dixon's Multipole Moments} \label{sec:dixon}
	
	In this section we summarize some of the ingredients and results of Dixon's definition of multipole moments~\cite{dixon2}. The multipole moments of the current density (in the case of electromagnetism) and of the energy-momentum tensor (in the case of gravity) directly enter the equations of motion of the body and can be computed from the stationary fields produced by the body when isolated. We find that these multipole moments determine all linear in $\fc{F}$ (for electromagnetism) or linear in $R$ (for gravity) operators in the dynamical mass function. The necessary matching is similar to the three-point amplitude matching performed in Ref.~\cite{Vines:2017hyw} in the worldline or in Ref.~\cite{Bern:2020buy} in the field theory. However, using Dixon's multipole construction we are able to extract more physical information from the stationary \rk or Kerr solutions than is contained in the three-point amplitude, which allows the determination of an increased number of Wilson coefficients.
 
    Following Dixon's discussion, we first consider how multipole moments are defined for a generic scalar field, then a generic tensor field. Then, we see how the general multipole moments of a vector field are constrained in complicated ways if that vector field satisfies the continuity equation. This leads to Dixon's definition of the reduced multipole moments of a conserved vector field which we apply to the current density. In this section, in all but the final subsection, we keep the spacetime generic in anticipation of applying our analysis to gravity. In the final subsection, we specialize to Minkowski space in preparation for electromagnetic calculations.
	
	\subsection{Moments of a scalar field}
	
	Using the exponential map we can take functions on $\bb{T}$ and turn them into functions on $T_z(\bb{T})$. For a scalar field $\phi(X)$ ($X \in \bb{T}$) we may define the function $\phi'(Z, v)$ ($v \in T_Z(\bb{T})$) by:
	\begin{equation}
		\phi'(Z, v) = \phi(\exp_Z(v)).
	\end{equation}
	Because $\phi'(Z,v)$ is a function on the flat tangent space $T_Z(\bb{T})$, it is simple to define the Fourier transform $\twid\phi(Z, k)$ and inverse Fourier transform:
	\begin{equation}
		\twid{\phi}(Z,k) = \int e^{-ik\cdot v} \phi'(Z, v) \frac{D^4 v}{(2\pi)^2}, \p \phi'(Z, v) = \int e^{ik\cdot v} \twid{\phi}(Z, k) \frac{D^4k}{(2\pi)^2}
	\end{equation}
	where the invariant tangent space measure and Fourier space measure are defined as:
	\begin{equation}
		D^4v = \sqrt{-\det g(Z)} d^4 v, \p D^4k = \frac{d^4 k}{\sqrt{-\det g(Z)}}.
	\end{equation}
	Using that $v^\mu = -\sigma^\mu(Z, X)$ we can express the inverse Fourier transform for our original scalar function as:
	\begin{equation}
		\phi(X) = \int e^{-ik_\mu\sigma^\mu(Z, X)} \twid{\phi}(Z, k) \frac{D^4k}{(2\pi)^2}	\label{eq:invfourierscalar}
	\end{equation}
	It immediately follows that for scalar functions $\phi(X)$ and $\psi(X)$ and their associated $\phi'(Z, v), \psi'(Z, v)$:
	\begin{equation}
		\int_{\Sigma(\lambda)} \psi^*(X) \phi(X) w^\alpha d\Sigma_\alpha = \int \sum_{n=0}^\8 \frac{i^n}{n!} k_{\mu_1}...k_{\mu_n} \twid{\psi}^*(Z, k)\int_{\Sigma} \sigma^{\mu_1}...\sigma^{\mu_n} \phi(X) w^\alpha d\Sigma_\alpha \frac{D^4 k}{(2\pi)^2}.
	\end{equation}
	Define then the moments $F^{\mu_1...\mu_n}_n$ and the moment generating function $F$ associated to the scalar function $\phi$:
	\begin{gather}
		F_n^{\mu_1...\mu_n}[\phi](\lambda) = \int_\Sigma(-\sigma^{\mu_1})...(-\sigma^{\mu_n})\phi(X) w^\alpha d\Sigma_\alpha \\
		F[\phi](\lambda, k) = \sum_{n=0}^\8 \frac{(-i)^n}{n!} F_n^{\mu_1 ... \mu_n} k_{\mu_1} ... k_{\mu_n}.
	\end{gather}
	Automatically the moments of $\phi$ satisfy the conditions:
	\begin{equation}
		F^{\mu_1...\mu_n}_n = F^{(\mu_1...\mu_n)}_n, \p u_{\mu_1} F_n^{\mu_1...\mu_n} = 0
	\end{equation}
	where for this last condition it is useful to remember equation \eqref{eq:SigmaDef}. The moment generating function determines how $\phi(X)$ behaves against test functions and thus determines $\phi(X)$ completely according to:
	\begin{equation}
		\int_{\Sigma(\lambda)} \psi^*(X) \phi(X) w^\alpha d\Sigma_\alpha = \int \twid{\psi}^*(Z,k) F(\lambda, k) \frac{D^4 k}{(2\pi)^2}.
	\end{equation}
	
	\subsection{Moments of a general tensor field}
	
	Suppose we now consider a tensor field $\phi^{\alpha_1...\alpha_m}{}_{\beta_1...\beta_n}(X)$. Using Synge's world function and the exponential map we can translate this to a tensor function on the tangent space at $z$ by:
	\begin{equation}
		\phi'^{\mu_1...\mu_m}{}_{\nu_1...\nu_n}(Z, v) = (-\sigma^{\mu_1}{}_{\alpha_1})...(-\sigma^{\mu_m}{}_{\alpha_m})(-\sigma^{-\beta_1}{}_{\nu_1})...(-\sigma^{-\beta_n}{}_{\nu_n})\phi^{\alpha_1...\alpha_m}{}_{\beta_1...\beta_n}(\exp_Z(v)).	\label{eq:bitensortransport}
	\end{equation}
	The Fourier transform and its inverse may now be defined as:
	\begin{align}
		\twid{\phi}^{\mu_1...\mu_m}{}_{\nu_1...\nu_n}(Z, k) &= \int \frac{e^{-ik\cdot v}}{(2\pi)^2} \phi'^{\mu_1...\mu_m}{}_{\nu_1...\nu_n}(Z, v) D^4 v \\
		\phi'^{\mu_1...\mu_m}{}_{\nu_1...\nu_n}(Z, v) &= \int \frac{e^{ik\cdot v}}{(2\pi)^2}\twid{\phi}^{\mu_1...\mu_m}{}_{\nu_1...\nu_n}(Z, k) D^4 k
	\end{align}
	For a given $\phi$, suppose we define the lower rank tensor $\varphi$ by contracting two indices:
	\begin{gather}
		\varphi^{\alpha_1...\alpha_{m-1}}{}_{\beta_1...\beta_{n-1}} = \phi^{\alpha_1...\alpha_{m-1} \alpha_m}{}_{\beta_1...\beta_{n-1}\alpha_m} \nonumber \\
  \implies \varphi'^{\mu_1...\mu_{m-1}}{}_{\nu_1...\nu_{n-1}} = \phi'^{\mu_1...\mu_{m-1}\mu_m}{}_{\nu_1...\nu_{n-1}\mu_m}.	\label{eq:tenscontract}
	\end{gather}
	This property only holds because the upper index bitensor propagator in \eqref{eq:bitensortransport} $(-\sigma^\mu{}_\alpha)$ is the matrix inverse of the lower index bitensor propagator $(-\sigma^{-\beta}{}_\nu)$. Alternatively, suppose we consider the moments of a vector field $\phi_\beta$ which is itself the gradient of a scalar field $\varphi$:
	\begin{equation}
		\phi_\beta =\nabla_\beta\varphi \implies \twid{\phi}_\nu = ik_\nu \twid{\varphi}. \label{eq:fourierdiff}
	\end{equation} 
	This property only holds if the lower index bitensor propagator is $-\sigma^{-\beta}{}_\nu$. Therefore while one could have imagined other bitensor propagators to use for transporting the components of $\phi$, such as parallel transport, the choice in \eqref{eq:bitensortransport} is unique in satisfying both \eqref{eq:tenscontract} and \eqref{eq:fourierdiff}. Unfortunately, the expression given in \eqref{eq:bitensortransport} has the property that translation to a tensor function on tangent space does not commute with raising/lowering indices of the original tensor field. This causes no real inconvenience for our calculations but one should be aware that $\phi'^\mu \neq g^{\mu\nu}(Z)\phi'_\nu$ if $\phi'^\mu$ is defined as above from $\phi^\mu$ and $\phi'_\mu$ is defined as above from $\phi_\mu$.
	
	For a generic $\phi$, the moments and moment generating function of $\phi$ are defined by:
	\begin{align}
		F_N^{\mu_1...\mu_N \rho_1...\rho_m}{}_{\nu_1...\nu_n}[\phi](\lambda) &= \int_{\Sigma}(-\sigma^{\mu_1})....(-\sigma^{\mu_N})\phi'^{\rho_1...\rho_m}{}_{\nu_1...\nu_n} w^\alpha d\Sigma_\alpha \\
		F^{\rho_1...\rho_m}{}_{\nu_1...\nu_n}[\phi](\lambda, k) &= \sum_{N=0}^\8 \frac{(-i)^N}{N!} F_N^{\mu_1 ... \mu_N\rho_1...\rho_m}{}_{\nu_1...\nu_n} k_{\mu_1} ... k_{\mu_N}.
	\end{align}
	Automatically the moments of $\phi$ satisfy the conditions:
	\begin{equation}
		F_N^{\mu_1...\mu_N\rho_1...\rho_m}{}_{\nu_1...\nu_n} = F_N^{(\mu_1...\mu_N)\rho_1...\rho_m}{}_{\nu_1...\nu_n}, \p u_{\mu_1} F_N^{\mu_1...\mu_N\rho_1...\rho_m}{}_{\nu_1...\nu_n} = 0.
	\end{equation}
	As well, $\phi$'s behavior against test functions is determined by:
	\begin{equation}
		\int_{\Sigma}\psi^{*\beta_1...\beta_n}{}_{\alpha_1...\alpha_m} \phi^{\alpha_1...\alpha_m}{}_{\beta_1...\beta_n} w^\gamma d\Sigma_\gamma = \int \twid{\psi}^{*\nu_1...\nu_n}{}_{\mu_1...\mu_m} F^{\mu_1...\mu_m}{}_{\nu_1...\nu_n}(\lambda, k) \frac{D^4k}{(2\pi)^2}.
    \end{equation}
    Thus, just as in the scalar case the moment generating function determines the original tensor field.
	
	\subsection{Moments of a conserved vector field}
	
	If we consider a vector field $\phi^\alpha(X)$ with moments as defined above then a brief calculation reveals:
	\begin{equation}
		\int_\bb{T} \psi^*\nabla_\alpha\phi^\alpha d\bb{T} = -\int_{-\8}^\8\int\twid{\psi}^*(z, k)\sum_{n=1}^\8 \frac{(-i)^n}{(n-1)!} k_{\mu_1}...k_{\mu_n} F^{(\mu_1...\mu_n)}_{n-1} \frac{D^4 k}{(2\pi)^2} D\lambda 
	\end{equation}
	If $\phi^\alpha$ is a conserved vector field so that $\nabla_\alpha\phi^\alpha = 0$, then the left hand side is 0 for all $\psi$. One may then wish based on this to conclude that $F^{(\mu_1...\mu_n)}_{n-1}$ is 0 for each $n\ge 1$. However, this is not a valid deduction. $\twid{\psi}^*(Z, k)$ at each fixed $\lambda$ determines the function $\psi(X)$ through the inverse Fourier transform and so $\twid{\psi}^*(Z(\lambda_1), k)$ and $\twid{\psi}^*(Z(\lambda_2), k)$ for $\lambda_1\neq \lambda_2$ are not independent of each other. Because of this, one cannot conclude that the integrand above at each $\lambda$ must be individually 0 as they may conspire to cancel at different $\lambda$ for arbitrary $\psi^*(X)$. Consequently, it is difficult to conclude anything explicit about the moments of $\phi^\alpha$ from the condition $\nabla_\alpha \phi^\alpha = 0$.
	
	Because of the mentioned difficulty, Dixon helpfully introduced an alternate set of reduced moments for a conserved vector field. For us the vector field of interest will always be $J^\alpha$. The goal of this reduced set of moments is precisely to produce a set which does not have the entanglements of the naive moments for a conserved vector field. Due to the absence of entanglements between the moments, when using Dixon's moments it is valid to equate integrands moment-by-moment. To arrive at Dixon's reduced multipole moments, which for a conserved vector field are written $m^{\lambda_1...\lambda_n \mu}_n$, we first need to introduce a few building blocks. Define:
	\begin{align}
		\Theta^{\kappa\lambda}_n(Z, X) &= (n+1)\int_0^1 {\sigma^\kappa}_\alpha(Z, \zeta(t)) \sigma^{\alpha\lambda}(Z, \zeta(t)) t^n dt \label{eq:Thetadef}\\
		\mf{q}_n^{\lambda_1...\lambda_n\mu \nu} &= (-1)^n \int_\Sigma \sigma^{\lambda_1}...\sigma^{\lambda_n} \Theta^{\mu\nu}_{n-1} J^\alpha d\Sigma_\alpha \p (n \ge 1) \\
		\mf{j}_n^{\lambda_1...\lambda_n\mu} &= (-1)^n \int_\Sigma \sigma^{\lambda_1}...\sigma^{\lambda_n}{\sigma^\mu}_\alpha J^\alpha w^\beta d\Sigma_\beta \\
		Q_n^{\lambda_1...\lambda_n\mu\nu} &= \mf{j}_{n+1}^{\lambda_1...\lambda_n[\mu\nu]} + \frac{1}{n+1} \mf{q}_{n+1}^{\lambda_1...\lambda_n[\mu \nu] \kappa} \frac{\dot z^\kappa}{\ec} \\
		m_n^{\lambda_1...\lambda_n\mu} &= \frac{2n}{n+1} Q^{(\lambda_1...\lambda_n)\mu}_{n-1}	\p (n\ge 1).	\label{eq:reducedef}
	\end{align}
	The $m^{\lambda_1...\lambda_n\mu}_n$ moments (called the reduced moments) will be the actual moments of interest. The other quantities defined are useful intermediate pieces for calculation. The reduced moments  automatically satisfy:
	\begin{gather}
		m_n^{\lambda_1...\lambda_n\mu} = m_n^{(\lambda_1...\lambda_n)\mu}, \p m^{(\lambda_1...\lambda_n\mu)}_n = 0, \label{eq:mnindex1}\\
		u_{\lambda_1} m^{\lambda_1...\lambda_{n-1}[\lambda_n\mu]}_n = 0 \p (n\ge 2).	\label{eq:mnindex2}
	\end{gather}
	Dixon finds that beyond these conditions, the reduced moments are not restricted by the conservation of $J^\alpha$ and that they are independent of each other for different values of $n$. It is useful to define the $0^\text{th}$ moment:
	\begin{equation}
		m^\mu_0 = q\frac{\dot z^\mu}{\ec}
	\end{equation}
	where $q$ is the total charge of the body defined by:
	\begin{equation}
		q = \int_\Sigma J^\alpha d\Sigma_\alpha.
	\end{equation}
	Due to the conservation of $J^\alpha$, $q$ is independent of $\lambda$. Dixon finds also that the reduced moments are independent of this $0^\text{th}$ moment. Then, define the reduced moment generating function:
	\begin{equation}
		M^\mu(\lambda, k) = \sum_{n=0}^\8 \frac{(-i)^n}{n!} m^{\lambda_1...\lambda_n \mu}_n k_{\lambda_1}...k_{\lambda_n}.
	\end{equation}
	Like the naive moments, the reduced moment generating functions determine the behavior of $J^\alpha$ against test functions. In particular, for an arbitrary vector field $A_\alpha(X)$:
	\begin{equation}
		\int_\Sigma A^*_\alpha(X) J^\alpha(X) w^\beta d\Sigma_\beta = \int \twid{A}^*_\mu(Z, k) M^\mu(\lambda, k) \frac{D^4k}{(2\pi)^2}.	\label{eq:intcurrent}
	\end{equation}
	The moment generating function automatically satisfies:
	\begin{equation}
		M^\mu k_\mu = \frac{q}{\ec}\dot z\cdot k.
	\end{equation}
	This implies that for the gradient of a scalar function, using the definition of the reduced moment generating function:
	\begin{equation}
		\int_\Sigma J^\alpha \nabla_\alpha f w^\beta d\Sigma_\beta = \frac{q}{\ec} \frac{d}{d\lambda}f(Z).
	\end{equation}
	For $f$ which decay sufficiently quickly for no boundary term to be necessary under integration by parts we immediately have:
	\begin{equation}
		0 = \int_{\bb{T}} f \nabla_\alpha J^\alpha d\bb{T} = -q \left.f(Z)\right|^{\lambda\to\8}_{\lambda\to-\8}
	\end{equation}
	which is true without placing any restrictions on the reduced moments of $J^\alpha$ beyond $m_0^\mu$. Dixon proved~\cite{dixon2} that these reduced moments are the unique set of moments which are independent of each other for different $n$, have only $m_0^\mu$ restricted by the conservation law, and satisfy the index symmetry conditions in equations \eqref{eq:mnindex1} and \eqref{eq:mnindex2}.

    Through \eqref{eq:intcurrent}, the current density is determined in terms of the reduced multipole moments. Explicitly comparing that behavior against test functions to \eqref{eq:current} and using crucially that the reduced multipole moments are unique and contain no interdependencies, we can identify:
    \begin{equation}
        m_n^{\rho_1...\rho_n\mu} = -2\, n! \fc{Q}_{n-1}^{(\rho_1...\rho_n)\mu}
    \end{equation}
    which gives the reduced multipole moments from the couplings in the action. Alternatively, this can be nicely inverted using the index symmetry conditions of both quantities to find:
    \begin{equation}
        \fc{Q}_n^{\rho_1...\rho_n\mu\nu} = -\frac{1}{n!}\frac{1}{n+2} m_{n+1}^{\rho_1...\rho_n[\mu\nu]}.    \label{eq:emmomrecovery}
    \end{equation}
    This allows the direct determination of the coupling of the body to the field strength in the action from its exact reduced multipole moments. 
	
	\subsection{Moments of the current density in Minkowski space}

    We now specialize our calculations to Minkowski space. We can represent an arbitrary element $v^\mu$ of $T_z(\bb{T})$ by a vector $Y = y^A\vec\iota_A = y^A \Lambda^\mu{}_A\vec e_\mu$ where $\vec\iota_A$ are local Minkowskian basis vectors ($\vec\iota_A\cdot\vec\iota_B = \eta_{AB}$) and $\vec e_\mu$ are coordinate basis vectors. Thus:
	\begin{equation}
		v^\mu = {\Lambda^\mu}_A y^A.
	\end{equation}
	Just as in section \ref{sec:emmpd} we continue to always choose the tetrad so that $\Lambda^\mu{}_0 = u^\mu$. In flat space:
	\begin{equation}
		\sigma(Z, X) = \frac{1}{2}(x-z)^2, \p \sigma^\mu = -(x^\mu -z^\mu), \p {\sigma^\mu}_\alpha = -\delta^\mu_\alpha \implies \Theta^{\mu\nu}_n = \eta^{\mu\nu}
	\end{equation}
	We always use lowercase beginning Latin alphabet indices $a, b, c,...$ for values $1,2,3$ on the tangent space. Then, we have:
	\begin{equation}
		X \in \Sigma(\lambda) \implies \exists\ y^a \in \bb{R}^3\ :\ x^\mu = z^\mu + {\Lambda^\mu}_a y^a.
	\end{equation}
	Using the $y^a$ coordinates and the definition of $\tau(X)$ we may identify explicit flat space expressions for $w^\alpha$ and $d\Sigma_\alpha$:
	\begin{gather}
		\lambda = \tau(z + \lambda u) \implies 1 = (\dot z^\alpha + {\dot \Lambda^\alpha}{}_a y^a) \nabla_\alpha \tau \\
		w^\alpha = \frac{\dot z^\alpha + {\dot\Lambda^\alpha}{}_a y^a}{\ec}, \p \p d\Sigma_\alpha = -u_\alpha d^3 y.
	\end{gather}
	With these the $\mc{q}_n$ and $\mc{j}_n$ moments become:
	\begin{align}
		\mf{q}^{\lambda_1...\lambda_n \mu\nu}_n &= \eta^{\mu\nu} {\Lambda^{\lambda_1}}_{A_1}... {\Lambda^{\lambda_n}}_{A_n}\int_\Sigma y^{A_1}...y^{A_n} (-J\cdot u) d^3 y \\
		\mf{j}^{\lambda_1...\lambda_n \mu}_n &= {\Lambda^{\lambda_1}}_{A_1} ... {\Lambda^{\lambda_n}}_{A_n}{\Lambda^\mu}_B \int_\Sigma y^{A_1}...y^{A_n} {\Lambda_\nu}^B J^\nu \frac{-u\cdot \dot z - u\cdot \dot \Lambda \cdot y}{\ec} d^3 y.
	\end{align}
	Using the explicit formula for the worldline velocity $\dot z^\mu$ in \eqref{eq:velsol2}, $u\cdot \dot z = -\ec$. As well, due to $\eqref{eq:angvelsol}$, $\dot\Lambda = \OO(\fc{F})$. Therefore:
	\begin{equation}
		\mf{j}^{\lambda_1...\lambda_n \mu}_n = {\Lambda^{\lambda_1}}_{A_1} ... {\Lambda^{\lambda_n}}_{A_n}{\Lambda^\mu}_B \int_\Sigma y^{A_1}...y^{A_n} {\Lambda_\nu}^B J^\nu d^3 y + \OO(\fc{F}).
	\end{equation}
	It is useful to define one last class of moments:
	\begin{align}
		K_n^{a_1...a_n B} &= \int_\Sigma y^{a_1}...y^{a_n} {\Lambda_\nu}^B J^\nu d^3 y	\label{eq:kndef} \\
		\implies \mf{q}^{\lambda_1...\lambda_n \mu\nu}_n &= \eta^{\mu\nu} {\Lambda^{\lambda_1}}_{a_1}... {\Lambda^{\lambda_n}}_{a_n} K^{a_1...a_n 0}_n\\
		\implies\mf{j}^{\lambda_1...\lambda_n \mu}_n &= {\Lambda^{\lambda_1}}_{a_1} ... {\Lambda^{\lambda_n}}_{a_n}{\Lambda^\mu}_B K^{a_1...a_n B}_n + \OO(\fc{F}).
	\end{align}
	For a stationary body in its center of momentum frame, the $K_n$ moments coincide with the naive moments of the stationary current density. By following the chain of definitions from \eqref{eq:Thetadef} to \eqref{eq:reducedef} these $K_n$ moments determine the full reduced moments $m_n$. Therefore, computing the naive moments of a stationary body allows the determination of the (coupling independent part of the) full set of reduced moments.

\section{Root-Kerr Multipole Moments} \label{sec:rkmoments}
		
	In this section we determine the linear in $\fc{F}$ couplings in the dynamical mass function to all orders in spin for a \rk particle. We begin by reviewing how the vector potential which defines the \rk particle arises from the Kerr-Newman solution and computing some basic mechanical properties of the \rk fields. Then, we use the \rk fields to determine the charge and current densities which are the source of the \rk solution using analysis which parallels the calculation of the source of the Kerr metric performed in Ref.~\cite{israel:1970}. From these charge and current densities we are then able to determine the $K_n$ moments for \rk and consequently determine the $m_n$ reduced multipole moments up to corrections of $\OO(\fc{F})$. Comparing these reduced moments to those produced by the current density produced by \eqref{eq:current} fully determines all linear in $\fc{F}$ terms in the dynamical mass function for a \rk particle.
		
	\subsection{From Kerr-Newman to Root-Kerr}
	
	The Kerr-Newman solution for a stationary charged spinning black hole is defined by the line element and vector potential:
	\begin{align}
		-d\tau^2 =& \ -\frac{r^2 - 2Gm r + a^2 + \frac{q^2 G}{4\pi}}{r^2 + a^2\cos^2\theta}(dt-a\sin^2\theta d\varphi)^2 + \frac{\sin^2\theta}{r^2+a^2\cos^2\theta}((r^2+a^2)d\varphi - a dt)^2 \nonumber \\
		&\ +(r^2 +a^2\cos^2\theta)\(\frac{dr^2}{r^2 - 2Gm r + a^2 + \frac{q^2 G}{4\pi}} + d\theta^2\)\\
		A_\mu dx^\mu =& \ \ \frac{q}{4\pi}\frac{-r dt + a r \sin^2\theta d\varphi}{r^2 + a^2\cos^2\theta}.
	\end{align}
	The $G\to 0$ limit of the Kerr-Newman metric produces the line element:
	\begin{equation}
		-d\tau^2 = -dt^2 + \frac{r^2 + a^2\cos^2\theta}{r^2 + a^2} dr^2 + (r^2 + a^2\cos^2\theta)d\theta^2 + (r^2+a^2)\sin^2\theta d\varphi^2.
	\end{equation}
	This is just the Minkowski line element in a particular choice of coordinates. The associated spatial metric $g^{(3)}_{ab}$ implied by this line element is the natural metric of oblate-spheroidal coordinates, related to Cartesian coordinates by:
	\begin{align}
		x = \sqrt{r^2 + a^2}\sin\theta\cos\varphi, \p y = \sqrt{r^2 + a^2}\sin\theta\sin\varphi, \p z = r\cos\theta
	\end{align}
    and with corresponding coordinate basis vectors:
    \begin{gather}
        \vec e_r = \frac{\d\vec x}{\d r}, \p \vec e_\theta = \frac{\d\vec x}{\d \theta}, \p \vec e_\varphi = \frac{\d\vec x}{\d\varphi}, \p g^{(3)}_{ab} = \vec e_a\cdot\vec e_b.
    \end{gather}
	The vector potential of this system defines the vector potential of a stationary \rk particle:
	\begin{align}
		\phi &= -A_0 = \frac{q}{4\pi}\frac{r}{r^2 + a^2\cos^2\theta} \\
		\vec A &= g^{(3)ab} A_a \vec e_b =  \frac{q}{4\pi}\frac{ar}{(r^2+a^2)(r^2+a^2\cos^2\theta)} \vec e_\varphi.
	\end{align}
    These coincide with the vector potential in equation 2.1 of~\cite{Chung:2019yfs}. From these potentials, we find the electric and magnetic fields:
	\begin{align}
		\vec E &= \frac{q}{4\pi} \frac{(r^2+a^2)(r^2-a^2\cos^2\theta)}{(r^2+a^2\cos^2\theta)^3}\vec e_r - \frac{q}{4\pi} \frac{2ra^2\sin\theta\cos\theta}{(r^2+a^2\cos^2\theta)^3}\vec e_\theta \label{eq:efield} \\ 
		\vec B &= \frac{q}{4\pi}\frac{2ar(r^2+a^2)\cos\theta}{(r^2+a^2\cos^2\theta)^3} \vec e_r + \frac{q}{4\pi} \frac{a (r^2-a^2\cos^2\theta)\sin\theta}{(r^2+a^2\cos^2\theta)^3}\vec e_\theta. \label{eq:bfield}
	\end{align}
	
	Away from any singularities, this magnetic field satisfies $\nabla\times\vec B = 0$ and so can be expressed in terms of a magnetic scalar potential $\vec B = -\nabla\psi$. In general coordinates in Minkowski space, the condition that the magnetic field (away from any current sources) be given by both a magnetic scalar potential and vector potential is:
	\begin{equation}
		-\sqrt{\det g^{(3)}} g^{(3)ab}\frac{\d\psi}{\d x^b} = \epsilon^{abc}\frac{\d A_c}{\d x^b}.
	\end{equation}
	Using the stationary \rk vector potential, one may use this to explicitly compute the associated magnetic scalar potential. Doing so gives:
	\begin{equation}
		\psi = \frac{q}{4\pi}\frac{a\cos\theta}{r^2+a^2\cos^2\theta}.
	\end{equation}
	Defining $\vec a = a\vec e_z$, in oblate-spheroidal coordinates we have the identity:
	\begin{equation}
		|\vec x-i\vec a| = r-ia\cos\theta.
	\end{equation}
	Consequently, the stationary $\sqrt{\text{Kerr}}$ particle is equivalently defined~\cite{Lynden-Bell:2002dvr} by producing the electric and magnetic scalar potentials or electric and magnetic fields given by:
	\begin{equation}
		\phi + i\psi = \frac{q}{4\pi|\vec x- i\vec a|} \implies \vec E + i \vec B = -\nabla\(\frac{q}{4\pi|\vec x-i\vec a|}\).
	\end{equation}
	
	\subsection{Mechanical Properties of the Stationary Root-Kerr Solution}
	
	Consider a surface of constant $r = R$ for integration. If we take $R\to\8$ this becomes the spatial boundary. The surface area element is:
	\begin{equation}
		d\vec{\bb{A}} = (r^2+a^2)\sin\theta \vec e_r d\theta d\varphi.
	\end{equation}
	Then, by Gauss' law the total charge of the $\sqrt{\text{Kerr}}$ particle is given by:
	\begin{equation}
		\int_{\bb{S}} \rho d^3\vec x = \int_{\d\bb{S}} \vec E\cdot d\vec{\bb{A}} = \lim_{R\to\8} \int_0^\pi \int_0^{2\pi} \frac{q}{4\pi} \frac{R^2-a^2\cos^2\theta}{(R^2+a^2\cos^2\theta)^2}(R^2+a^2)\sin\theta d\varphi d\theta = q
	\end{equation}
	confirming our consistent use of the symbol $q$ for the total charge of the distribution. 
	
	The magnetic dipole moment of a charge distribution is by definition:
	\begin{equation}
		\vec\mu = \frac{1}{2}\int_{\bb{S}}\vec x\times\vec J d^3\vec x.
	\end{equation}
	Because the distribution is stationary, rewriting $\vec J = \nabla\times\vec B$ and integrating by parts sufficiently many times this is equivalently:
	\begin{equation}
		\vec\mu = -\frac{3}{2}\int_{\d\bb{S}} \vec A\times d\vec{\bb{A}}.
	\end{equation}
	With the same surface of constant $r = R$:
	\begin{equation}
		\vec\mu = -\frac{3}{2}\lim_{R\to\8}\int_0^\pi\int_0^{2\pi} \frac{qa}{4\pi} \frac{R}{R^2+a^2\cos^2\theta^2} \vec e_\varphi\times\vec e_r \sin\theta d\varphi d\theta = q \vec a.
	\end{equation}
	So, $\vec a$ is the magnetic dipole moment per unit charge. 
	
	Like for a point charge, the total energy in the electromagnetic field for this system is divergent. To regulate this instead of integrating all the way down to $r = 0$ for the total energy, we integrate down to a cutoff $r = \eps$. Then, the total energy is:
	\begin{align}
		\text{E} &= \int_{\bb{S}}\frac{|\vec E|^2+|\vec B|^2}{2} d^3\vec x = \frac{q^2}{16\pi\eps}\(1 + \frac{a}{\eps}\arctan\(\frac{a}{\eps}\) + \frac{\eps}{a}\arctan\(\frac{a}{\eps}\)\).
	\end{align}
	The Poynting vector for the particle is:
	\begin{equation}
		\vec \wp = \vec E\times\vec B = \frac{q^2}{16\pi^2} \frac{a\vec e_\varphi}{(r^2+a^2\cos^2\theta)^3}.
	\end{equation}
	This leads to the total linear momentum:
	\begin{equation}
		\vec p = \int_{\bb{S}}\vec\wp d^3\vec x = 0.
	\end{equation}
	So, the stationary $\sqrt{\text{Kerr}}$ fields given are in fact in the center of momentum frame. Consequently, the energy found before is the invariant mass of the distribution:
	\begin{equation}
		m = \frac{q^2}{16\pi\eps}\(1 + \frac{a}{\eps}\arctan\(\frac{a}{\eps}\) + \frac{\eps}{a}\arctan\(\frac{a}{\eps}\)\).
	\end{equation}
	
	The Poynting vector we found leads to the total angular momentum (which because we are in the center of momentum frame is the spin angular momentum):
	\begin{align}
		\vec S &= \int_{\bb{S}} \vec x\times\vec \wp d^3\vec x = \frac{q^2}{16\pi\eps}\(1 + \frac{a}{\eps} \arctan\(\frac{a}{\eps}\) + \frac{2\eps}{a}\arctan\(\frac{a}{\eps}\) - \frac{\eps^2}{a^2} + \frac{\eps^3}{a^3} \arctan\(\frac{a}{\eps}\)\) \vec a.
	\end{align}

    While both the spin and the mass of the solution are divergent in $\eps$, their ratio has a finite $\eps$ to 0 limit. Therefore, the physical value of the ratio of the spin to the mass is:
	\begin{equation}
		\frac{\vec S}{m} = \vec a.
	\end{equation}
	Immediately it follows that for this distribution:
	\begin{equation}
		\vec \mu = \frac{q}{m}\vec S.
	\end{equation}
	Thus, the $\sqrt{\text{Kerr}}$ particle has a gyromagnetic ratio of exactly $2$.
	
	\subsection{Charge and Current density}
	
	The electric and magnetic fields as well as scalar and vector potential given diverge whenever $r^2 + a^2\cos^2\theta = 0$ and nowhere else. For $r > 0$ the potentials and their derivatives are continuous and so there are no surface charges or currents for $r > 0$. Because the divergence of the right hand side in \eqref{eq:efield} is 0 away from any poles, there is no charge density for $r > 0$. Because the curl of the right hand side in \eqref{eq:bfield} is 0 away from any poles, there is no current density for $r > 0$. Consequently, the source for the $\sqrt{\text{Kerr}}$ particle must only have support for $r = 0$. For $r = 0$ the oblate spheroidal coordinates reduce to the disk in the $xy$ plane with center at the origin and radius $a$. In order to approach this disk from above we must take $r \to 0$ with $\theta < \frac{\pi}{2}$ and in order to approach from below we must take $r\to 0$ with $\theta > \frac{\pi}{2}$. It is useful to define the coordinate $\chi = \theta$ for $z > 0$ and $\chi = \pi - \theta$ for $z < 0$. This way, two points approaching the disk, one from above and one from below, will limit to the same point in space when their limiting values of $\chi$ and $\varphi$ coincide. The disk is then parameterized by the coordinates $\chi$ and $\varphi$ by:
	\begin{align}
		x = a\sin\chi\cos\varphi, \p y = a\sin\chi\sin\varphi, \p z = 0.
	\end{align}
	We can also use the cylindrical radial coordinate $\mc{r} = \sqrt{x^2+y^2} = a\sin\chi$. 
	
	The surface charge density inside the disk is given by:
	\begin{equation}
		\sigma_\text{disk} = \vec e_z\cdot(\vec E(0, \chi, \varphi) - \vec E(0, \pi-\chi, \varphi)) = -\frac{qa}{2\pi(a^2-\mc{r}^2)^\frac{3}{2}}.
	\end{equation}
	We can integrate this from $\chi = 0$ to $\chi = \frac{\pi}{2} - \eps$ to get the total charge inside the disk:
	\begin{equation}
		Q_\text{disk} = \int_{\text{disk}} \sigma_{\text{disk}} d\bb{A}  = q\(1 - \frac{1}{\sin\eps}\).
	\end{equation}	
	The surface charge density cannot be extended all the way to the ring at $\chi = \frac{\pi}{2}$ as it has a nonintegrable divergence. We know from before that the total charge of the distribution is $q$ and so there must be a charged ring at $\chi = \frac{\pi}{2}$ so that the total charge between this ring and the disk is $q$. This ring is precisely the ring on which $r^2 + a^2\cos^2\theta = 0$ and so where the potentials and fields are divergent. The charge density on the ring must be azimuthally symmetric because the fields are. So:
	\begin{equation}
		q = Q_{\text{disk}} + Q_{\text{ring}} = q - \frac{q}{\sin\eps} + \int_0^{2\pi} \lambda_\text{ring} a d\varphi \implies \lambda_\text{ring} = \frac{q}{2\pi a \sin\eps}.
	\end{equation}
	Thus the total charge density of the stationary $\sqrt{\text{Kerr}}$ particle is:
	\begin{equation}
		\rho = -\frac{qa}{2\pi(a^2 - \mc{r}^2)^\frac{3}{2}}\delta(z) \vartheta(a\cos\eps - \mc{r}) + \frac{q}{2\pi a\sin\eps}\delta(z)\delta(\mc{r} - a). \label{eq:kerrdense}
	\end{equation}
	The surface current density inside the disk is:
	\begin{align}
		\vec K_\text{disk} &= \vec e_z\times(\vec B(0, \chi, \varphi) - \vec B(0, \pi-\chi, \varphi))  = \sigma_\text{disk}\frac{\mc{r}}{a}\vec\iota_\varphi
	\end{align}
	where $\vec\iota_\varphi$ is the unit vector in the direction of $\vec e_\varphi$. This is precisely the surface current density produced by rigidly rotating the surface charge density of the disk with angular velocity $\vec\omega$ given by:
	\begin{equation}
		\vec\omega = \frac{1}{a}\vec e_z.
	\end{equation}
	Rotating the ring with this same angular velocity produces a current in the ring of:
	\begin{equation}
		\vec I_\text{ring} = \lambda_{\text{ring}}\vec\iota_\varphi = \frac{q}{2\pi a\sin\eps}\vec\iota_\varphi.
	\end{equation}
	These currents produce a magnetic dipole moment:
	\begin{align}
		\vec\mu &= \frac{1}{2}\int_{\text{disk}} \vec x\times\vec K_\text{disk} d\bb{A} + \frac{1}{2}\int_{\text{ring}} \vec x\times\vec I_\text{ring} ds = q\vec a\(1 - \frac{1}{2}\sin\eps\).
	\end{align}
	In the $\eps\to 0$ limit this produces exactly the magnetic dipole moment we found before. Therefore, the current density for the stationary $\sqrt{\text{Kerr}}$ distribution is:
	\begin{equation}
		\vec J = \rho\frac{\vec a\times\vec x}{a^2}	\label{eq:rigidcurrent}
	\end{equation}
	and the stationary $\sqrt{\text{Kerr}}$ particle is exactly a charged disk and ring with the charge distribution given by equation \eqref{eq:kerrdense} and rotating with the angular velocity $\frac{1}{a}$ about the central axis of the disk. These charge and current densities are precisely analogous to the mass density and energy-momentum tensor found in Ref.~\cite{israel:1970} for a stationary Kerr black hole.
	
	\subsection{Stationary Multipole Moments}
	
	We can now compute the $K_n$ moments explicitly for a stationary \rk particle:
	\begin{equation}
		K^{a_1...a_n 0}_n = \int_{\bb{S}} x^{a_1}...x^{a_n} \rho(\vec x) d^3\vec x, \p K^{a_1...a_nb}_n = \int_{\bb{S}} x^{a_1}...x^{a_n} J^b(\vec x) d^3\vec x.
	\end{equation}
	By using \eqref{eq:rigidcurrent}, we find immediately:
	\begin{equation}
		K^{a_1...a_nb}_n = \frac{1}{a^2}{\epsilon^b}_{cd} a^c K^{da_1...a_n0}_{n+1}	\label{eq:Kreduce}
	\end{equation}
	and so only the multipole moments of $\rho$ need to be computed in order to determine the moments of the full distribution. Further, due to the full symmetrization of $K_n^{a_1...a_n0}$, for an arbitrary 3-vector $k_a$:
	\begin{equation}
		K^{a_1...a_n0}_n = \frac{1}{n!} \frac{\d^n}{\d k_{a_1}...\d k_{a_n}}\( K_n^{b_1...b_n 0} k_{b_1}...k_{b_n}\).
	\end{equation}
	
	The charge density of the $\sqrt{\text{Kerr}}$ particle is localized to the $xy$ plane and is rotationally symmetric about the $z$ axis. So, it can be written as:
	\begin{equation}
		\rho(\vec x) = \sigma(\mc{r})\delta(z) \label{eq:symdense}
	\end{equation}
	where in particular for \rk:
	\begin{equation}
		\sigma(\mc{r}) = -\frac{qa}{2\pi(a^2 - \mc{r}^2)^\frac{3}{2}} \vartheta(a\cos\eps - \mc{r}) + \frac{q}{2\pi a\sin\eps}\delta(\mc{r} - a). \label{eq:kerrsurf}
	\end{equation}
	Then:
	\begin{equation}
		K_n^{a_1...a_n 0} k_{a_1}...k_{a_n} = (k_x^2+k_y^2)^{\frac{n}{2}}\int_0^{2\pi}\cos^n \varphi d\varphi \int_0^\8 \sigma(\mc{r})\mc{r}^{n+1} d\mc{r}.
	\end{equation}
	If $n$ is odd the azimuthal integral gives 0, so we only need to consider even $n$. For even $n$ the azimuthal integral is:
	\begin{equation}
		\int_0^{2\pi}\cos^{2n} \varphi d\varphi = 2\pi \frac{(2n)!}{4^n n!^2}.
	\end{equation}
	With the integration variable $x = \frac{1}{a}\sqrt{a^2-\mc{r}^2}$ the radial integral becomes:
	\begin{equation}
		\int_0^\8 \sigma(\mc{r})\mc{r}^{2n+1}d\mc{r} = \frac{qa^{2n}}{2\pi}\(1 + \int_{\sin\eps}^1 \frac{1- (1-x^2)^n}{x^2} dx\).
	\end{equation}
	Here it is safe to take the $\eps\to 0$ limit to give:
	\begin{equation}
		\int_0^\8 \sigma(\mc{r})\mc{r}^{2n+1}d\mc{r} = \frac{qa^{2n}}{2\pi}\frac{4^n n!^2}{(2n)!}.
	\end{equation}
	Thus the nonzero $K_n$ moments become:
	\begin{align}
		K_{2n}^{a_1...a_{2n}0} v_{a_1}...v_{a_{2n}} &= q|\vec a\times\vec v|^{2n} \\
		K_{2n+1}^{a_1...a_{2n+1}b} v_{a_1}...v_{a_{2n+1}} &= q|\vec a\times\vec v|^{2n}(\vec a\times\vec v)^b.
	\end{align}
	
	\subsection{Dynamical Multipole Moments}	
	
	Now we return the $K_n$ moments to the definition of the reduced multipole moments. For a generic body allowed to respond to external fields, the multipole moments will in general depend on those external fields and so the general moments in external fields cannot be fully determined from the stationary moments. It is unclear at this time what is the appropriate response for a \rk body to external fields. So, for an exact dynamical \rk body:
    \begin{equation}
        K_{n}^{a_1...a_n b} = K_{n\text{ stat}}^{a_1...a_n b} + \OO(\fc{F})
    \end{equation}
    where $K_{n\text{ stat}}^{a_1...a_n b}$ are the $K_n$ moments we found for a stationary \rk particle. To express the reduced moments it is useful to define the projector $\perp^\mu{}_\nu$ which projects 4-vectors into the plane of the disk:
    \begin{equation}
        \perp^\mu{}_\nu = \delta^\mu_\nu + u^\mu u_\nu - \hat a^\mu \hat a_\nu.
    \end{equation}
    We also write $k_\perp^\mu$ as a shorthand for $k_\perp^\mu = \perp^\mu{}_\nu k^\nu$ and $k_\perp$ without an index as a shorthand for $\sqrt{|k_\mu\perp^\mu{}_\nu k^\nu|}$. Using the definition of the reduced multipole moments for arbitrary 4-vectors $k_\mu$ and $v_\mu$ we find:
	\begin{align}
		k_{\lambda_1}...k_{\lambda_{2n}} v_\mu m^{\lambda_1...\lambda_{2n} \mu}_{2n} &= q a^{2n} k_\perp^{2n} u\cdot v - q a^{2n} k_\perp^{2n-2}(u\cdot k)(k_\perp\cdot v) + \OO(\fc{F})\\
		k_{\lambda_1}...k_{\lambda_{2n+1}} v_\mu m^{\lambda_1...\lambda_{2n+1}\mu}_{2n+1} &= qa^{2n}k^{2n}_\perp \epsilon_{\mu\nu\rho\sigma} u^\mu a^\nu k^\rho v^\sigma + \OO(\fc{F}).
	\end{align}
    Returning these to \eqref{eq:emmomrecovery}, we find:
    \begin{align}
        \fc{Q}_{2n}^{\rho_1...\rho_{2n}\mu\nu} \d^{2n}_{\rho_1...\rho_{2n}} \fc{F}_{\mu\nu} &= \frac{q}{(2n+1)!} a^{2n} \d_\perp^{2n} {}^\star\fc{F}_{\mu\nu}a^\mu u^\nu + \OO(\fc{F}^2) \\
        \fc{Q}_{2n+1}^{\rho_1...\rho_{2n+1}\mu\nu}\d^{2n+1}_{\rho_1...\rho_{2n+1}}\fc{F}_{\mu\nu} &= -\frac{q}{(2n+2)!} a^{2n+2} \d_\perp^{2n} \d_\perp^\mu \fc{F}_{\mu\nu} u^\nu + \OO(\fc{F}^2).
    \end{align}
    With the definition of $\fc{Q}_n$, these produce the dynamical mass function:
    \begin{equation}
        \fc{M} = m + q \frac{\sinh(a\d_\perp)}{a\d_\perp} {}^\star \fc{F}_{\mu\nu}a^\mu u^\nu + q \frac{1-\cosh(a\d_\perp)}{\d_\perp^2} \d_\perp^\mu \fc{F}_{\mu\nu} u^\nu + \OO(\fc{F}^2).
    \end{equation}
    There is no subtlety in defining the square root or inverse of the differential operator here because once the trigonometric functions are series expanded only positive even powers of $\d_\perp$ survive. We will see later that for \rk particles, at least at low orders in spin, the squared dynamical mass function $\fc{M}^2$ may be a simpler object when $\OO(\fc{F})^2$ operators are considered. So, going forward we will express the dynamical mass function as:
    \begin{equation}
        \fc{M}^2 = m^2 + 2m q \frac{\sinh(a\d_\perp)}{a\d_\perp} {}^\star \fc{F}_{\mu\nu}a^\mu u^\nu + 2m q \frac{1-\cosh(a\d_\perp)}{\d_\perp^2} \d_\perp^\mu \fc{F}_{\mu\nu} u^\nu + \OO(\fc{F}^2). \label{eq:fulldmf}
    \end{equation}
    This \rk dynamical mass function is our primary result for electromagnetism. When acting on a field strength which in the neighborhood of the body is a vacuum solution ($\d_\nu \fc{F}^{\mu\nu} = 0$) of Maxwell's equations, this reduces to:
	\begin{align}
		\fc{M}^2 = m^2 + 2mq\frac{\sin(a\Delta)}{a \Delta} {}^\star{\fc{F}}_{\mu\nu} a^\mu u^\nu + 2mq\frac{1-\cos(a\Delta)}{(a\Delta)^2}(a\cdot \d)\fc{F}_{\mu\nu} a^\mu u^\nu + \OO(\fc{F}^2)	\label{eq:simpledmf}
	\end{align}
	where the $\Delta$ differential operator is defined by:
	\begin{equation}
		a\Delta = \sqrt{(a\cdot\d)^2 - a^2(u\cdot\d)^2}.
	\end{equation}
	Again there is no subtlety in defining the square root or inverse of the differential operator. In general \eqref{eq:simpledmf} and \eqref{eq:fulldmf} are not equivalent, however we show that for computing Compton amplitudes they are interchangeable.
	
	The dynamical mass function in \eqref{eq:simpledmf} very nearly matches the couplings in Ref.~\cite{Levi:2015msa} (if the analysis done there for gravity is done for electromagnetism). In particular, phrased as a dynamical mass function, the couplings in Ref.~\cite{Levi:2015msa} determine:
	\begin{equation}
		\fc{M}^2 = m^2 + 2mq\frac{\sin(a\cdot \d)}{a\cdot \d}{}^\star{\fc{F}}_{\mu\nu} a^\mu u^\nu + 2mq\frac{1-\cos(a\cdot\d)}{a\cdot \d} \fc{F}_{\mu\nu} a^\mu u^\nu + \OO(\fc{F}^2)    \label{eq:lsm}
	\end{equation}
	Equation \eqref{eq:simpledmf} becomes this result with the replacement:
	\begin{equation}
		a\Delta \rightarrow a\cdot \d.
	\end{equation}
	The $(u\cdot \d)$ terms present in $a\Delta$ do not contribute to the three point amplitude and so $a\Delta$ and $a\cdot \d$ are indistinguishable in a three point matching calculation. For this reason, the analyses of Refs.~\cite{Vines:2017hyw, Bern:2020buy} were insensitive to the couplings on terms of the form $S^{2n}(u\cdot\d)^{2n} \fc{F}_{\cdot\cdot}$. The first of such terms is present in Ref.~\cite{Siemonsen:2019dsu} with an undetermined Wilson coefficient. Similarly, in Ref.~\cite{Levi:2015msa} $(u\cdot \d)$ are not considered because if one uses the equations of motion in the action, they can be shuffled into order $\fc{F}^2$ terms. One can see this quickly on the lowest order such term, for example:
    \begin{align}
        \ec a^2(u\cdot \d)^2{}^\star\fc{F}_{\mu\nu}a^\mu u^\nu &= a^2 a^\mu u^\nu u^\rho \dot z^\sigma \d_\sigma \d_\rho {}^\star\fc{F}_{\mu\nu} + \OO(\fc{F}^2) \\
        &= \frac{d}{d\lambda}\(a^2 a^\mu u^\nu u^\rho \d_\rho {}^\star\fc{F}_{\mu\nu}\) - \frac{d}{d\lambda}(a^2 a^\mu u^\nu u^\rho) \d_\rho {}^\star\fc{F}_{\mu\nu} + \OO(\fc{F}^2) \\
        &= \frac{d}{d\lambda}\(a^2 a^\mu u^\nu u^\rho \d_\rho {}^\star\fc{F}_{\mu\nu}\) + \OO(\fc{F}^2).
    \end{align}
    For three-point amplitudes, \eqref{eq:fulldmf}, \eqref{eq:simpledmf}, and \eqref{eq:lsm} are interchangeable. For Compton amplitudes, \eqref{eq:fulldmf} and \eqref{eq:simpledmf} are interchangeable but distinct from \eqref{eq:lsm}. The worldline evolution of all three are distinct. Importantly, by the definition of Dixon's moments \eqref{eq:intcurrent} holds for any test function vector field $A_\alpha(X)$, not only the physically relevant vector potential and that it holds without using the solution to the equations of motion. We have shown that for a \rk particle the only dynamical mass function for which \eqref{eq:intcurrent} holds without using the electromagnetic MPD equations is \eqref{eq:fulldmf}. If one shuffles away $(u\cdot \d)$ operators in favor of $\fc{F}^2$  so that the linear in $\fc{F}$ dynamical mass function is given by \eqref{eq:lsm}, then \eqref{eq:intcurrent} will not be true identically.  In this sense, the advantage of \eqref{eq:fulldmf} is that it uniquely provides at each $\lambda$ the physically correct multipole moments for \rk.

\section{Electromagnetic Compton Amplitude} \label{sec:emcompton}
	
	In this section we formally compute the \rk Compton amplitude to all orders in spin up to contact terms. Those contact terms are determined by $\fc{F}^2$ operators in the dynamical mass function which the multipole analysis is insensitive to. We begin by computing a formal expression for the all orders in spin Compton amplitude for a generic charged spinning body in terms of the dynamical mass function. Next we consider that generic Compton amplitude explicitly to order $S^3$. At $\OO(S^1)$ there is a single Wilson coefficient in the action and it can be determined by matching the $\OO(S^1)$ three-point amplitude or Compton amplitude. At $\OO(S^2)$ there are 5 new Wilson coefficients in the action. One of them can be determined by matching the $\OO(S^2)$ three-point amplitude while the other 4 can be determined uniquely by matching the $\OO(S^2)$ Compton amplitude. At $\OO(S^3)$ there are 8 new Wilson coefficients in the action. One of them can be determined by matching the $\OO(S^3)$ three-point amplitude while the other 7 appear only as contact terms in the Compton amplitude. Among those 7, there are only 6 linearly independent structures appearing in the $\OO(S^3)$ Compton amplitude, and so there is one linear combination of Wilson coefficients that the Compton amplitude is independent of at this order in spin. 

    Once we have the results for a generic body through $\OO(S^3)$, we specialize our interest to \rk. Requiring the exponentiation of spin structure as found in Ref.~\cite{Guevara:2018wpp} through $\OO(S^2)$ for the helicity-preserving amplitude (which develops a spurious pole starting at $\OO(S^3)$) and through $\OO(S^3)$ for the helicity-reversing amplitude (which has no such spurious pole) fixes all Wilson coefficients through $\OO(S^2)$ and 4 of the 8 new operators at $\OO(S^3)$. Alternatively, requiring the helicity-preserving amplitude to have the shift-symmetry described in Refs.~\cite{Aoude:2022trd, Bern:2022kto, Haddad:2023ylx,  Aoude:2023vdk} through $\OO(S^3)$ fixes 3 of the 8 new operators at $\OO(S^3)$. The spin-exponentiation and shift-symmetry are consistent with each other, and together fix 6 of the 8 new operators at $\OO(S^3)$ (as they share one redundant condition). The dynamical mass function \eqref{eq:simpledmf} determines 2 Wilson coefficients at $\OO(S^3)$, one of which is fixed by the three-point amplitude and the other of which is independent of and consistent with both spin-exponentiation and shift-symmetry. Together then Dixon's multipole moments, spin-exponentiation, and shift-symmetry fix 7 of the 8 Wilson coefficients at $\OO(S^3)$, which is the maximum amount possible by using the Compton amplitude (due to the presence of a linearly independent combination of Wilson coefficients that the amplitude is independent of).

    \subsection{Formal Classical Compton}\label{sec:FormalClassicalCompton}

    We consider Compton scattering of an incoming photon with polarization $\fc{E}_1$ and momentum $k_1$ off of a massive spinning charged body with initial momentum $m v$ ($v\cdot v = -1$) and initial spin $s$ to an outgoing photon with polarization $\fc{E}_2$ and momentum $k_2$ and perturbed massive body. For a plane wave vector potential with strength $\epsilon$:
    \begin{equation}
		A_\mu(X) = \epsilon\fc{E}_\mu e^{ik\cdot x}, \p \fc{F}_{\mu\nu}(X) = \epsilon f_{\mu\nu} e^{ik\cdot x}, \p f_{\mu\nu} = ik_\mu \fc{E}_\nu - ik_\nu \fc{E}_\mu.
	\end{equation}

    Because the tree level Compton amplitude is $\OO(q^2)$, it depends only on the $\OO(q)$ and $\OO(q^2)$ pieces of the dynamical mass function. Consequently, we will only be concerned with operators in $\fc{M}$ which are linear or quadratic in $\fc{F}$ and so we consider an $\fc{M}$ of the form:
	\begin{equation}
		\fc{M}^2(z, u, S) = m^2 + q\delta\fc{M}_1^2(z, u, S) + q^2\delta\fc{M}_2^2(z, u, S) + \OO(q^3\fc{F}^3)
	\end{equation}
	where $\delta \fc{M}_1^2$ is of the form:
	\begin{equation}
		\delta\fc{M}_1^2(z, u, S) = \sum_{n=0}^\8 T_n^{\rho_1...\rho_n\mu\nu}(u, S) \d^n_{\rho_1...\rho_n} \fc{F}_{\mu\nu}(z)
	\end{equation}
	for some functions $T_n^{\rho_1...\rho_n\mu\nu}(u, S)$ satisfying:
	\begin{equation}
		T_n^{\rho_1...\rho_n\mu\nu} = T_n^{(\rho_1...\rho_n)[\mu\nu]}
	\end{equation}
	and $\delta \fc{M}_2^2$ is of the form:
	\begin{equation}
		\delta\fc{M}_2^2(z, u, S) = \sum_{n=0}^\8\sum_{l=0}^\8 V_{nl}^{\rho_1...\rho_n\mu\nu | \kappa_1...\kappa_l \tau\omega}(u, S) \d^n_{\rho_1...\rho_n} \fc{F}_{\mu\nu}(z)\d^l_{\kappa_1...\kappa_l}\fc{F}_{\tau\omega}(z)
	\end{equation}
	for some functions $V_{nl}^{...}(u, S)$ satisfying:
	\begin{align}
		V_{nl}^{\rho_1...\rho_n\mu\nu | \kappa_1...\kappa_l \tau\omega} &= V_{nl}^{(\rho_1...\rho_n)[\mu\nu] | (\kappa_1...\kappa_l) [\tau\omega]} \\
		V_{nl}^{\rho_1...\rho_n\mu\nu | \kappa_1...\kappa_l \tau\omega} &= V_{ln}^{\kappa_1...\kappa_l \tau\omega|\rho_1...\rho_n\mu\nu}
	\end{align}

    Maxwell's equations together with the flat space electromagnetic MPD equations with the described initial conditions will produce solutions of the form:
    \begin{align}
		z^\mu(\lambda) &= v^\mu \lambda + q\epsilon\delta z^\mu(\lambda) + \OO(q^2)\\
		p_{\mu}(\lambda) &= mv_\mu + q\epsilon\delta p_\mu(\lambda) + \OO(q^2) \\
		S^\mu(\lambda) &= s^\mu + q\epsilon\delta S^\mu(\lambda) + \OO(q^2) \\
        J^\mu(X) &= qJ^\mu_{\stat}(X) + q^2\epsilon \delta J(X) + \OO(q^3) \\
		A_{\mu}(X) &= \epsilon\fc{E}_{1\mu} e^{ik\cdot x} + qA^\stat_\mu(X) + q^2\epsilon \delta A_\mu(X) + \OO(q^3)
	\end{align}
    where $qJ^\mu_\stat$ is the current density produced by the stationary spinning body in the absence of the incoming photon $(\epsilon\to 0)$ and $qA^\stat_\mu$ is the (Lorenz gauge) vector potential produced by that current density. The equation of motion perturbations will be oscillatory from solving the electromagnetic MPD equations. In particular, their solutions take the form:
    \begin{equation}
		\delta z^\mu = \delta\twid{z}^\mu e^{i k_1\cdot v \lambda}, \p \delta u^\mu = \delta\twid{u}^\mu e^{ik_1\cdot v \lambda}, \p \delta S^\mu = \delta\twid{S}^\mu e^{i k_1\cdot v\lambda}
	\end{equation}
    for constant vectors $\delta\twid{z}, \delta\twid{u}, \delta\twid{S}$. ($\delta p$ is determined from $\delta u$ and $\delta\fc{M
    }_1^2$ because $p^\mu = \fc{M} u^\mu$.)
    
    The $\OO(q^2\epsilon)$ piece of the vector potential, $\delta A_\mu$, determines the linear in $\epsilon$ outgoing electromagnetic field and thus determines the tree level Compton amplitude. From the Lorenz-gauge Maxwell equation, $\delta A_\mu$ satisfies:
    \begin{equation}
        -\d^2 \delta A^\mu = \delta J^\mu
    \end{equation}
    and limits to 0 in the asymptotic past (when the appropriate $i\eps$ is used for the oscillatory solutions). Therefore, it is determined by the delayed Green's function solution to the wave equation and given by:
    \begin{equation}
        \delta A^\mu(X) = \int_{\bb{S}} \frac{\delta J^\mu(t-|\vec x-\pvec x|,\pvec x)}{4\pi|\vec x-\pvec x|} d^3\pvec x
    \end{equation}
    where:
    \begin{equation}
        t = -v\cdot x, \p \vec x^\mu = x^\mu + v^\mu v\cdot x.
    \end{equation}
    Thus, $\delta A$ is determined by $\delta J$. The current perturbation is determined by the equation of motion perturbations $\delta z, \delta p, \delta S$. From \eqref{eq:Qmoment} we may identify:
	\begin{equation}
		\fc{Q}_n^{\rho_1...\rho_n\mu\nu} = \frac{q}{2m}\(1-\frac{q\delta\fc{M}_1^2}{2m^2}\)T_n^{\rho_1...\rho_n\mu\nu} + \frac{q^2}{m}\sum_{l=0}^\8 V^{\rho_1...\rho_n \mu\nu | \kappa_1...\kappa_l \tau \omega}_{nl} \d^l_{\kappa_1...\kappa_l}\fc{F}_{\tau\omega} + \OO(\fc{F}^2).
	\end{equation}
    Define the Fourier transform of the current:
	\begin{equation}
	   \twid{J}^\mu(k) = \int \frac{e^{-ik\cdot x}}{(2\pi)^2} J^\mu(X) d^4 X.
	\end{equation}
	In terms of the $\fc{Q}_n$ the exact Fourier transform of $J^\mu(X)$ is:
	\begin{equation}
		\twid{J}^\mu(k_2) = \int_{-\8}^\8\(q\dot z^\mu + 2\sum_{n=0}^\8 (-i)^{n+1} \fc{Q}_n^{\rho_1...\rho_n\mu\nu} k_{2\rho_1}...k_{2\rho_n} k_{2\nu}\) \frac{e^{-ik_2\cdot z}}{(2\pi)^2} d\lambda
	\end{equation}
	For a plane wave vector potential it is useful to introduce the function:
	\begin{equation}
		\fc{N}(u, S, k, \fc{E}) = \sum_{n=0}^\8 \frac{i^n}{2m} T_n^{\rho_1...\rho_n\mu\nu}(u, S) k_{\rho_1}...k_{\rho_n} f_{\mu\nu}
	\end{equation}
	so that if we evaluate $\delta\fc{M}_1^2$ on a plane wave vector potential:
	\begin{equation}
		\left.\delta\fc{M}_1^2\right|_{\text{plane wave}} = 2m\epsilon\fc{N}(u, S, k, \fc{E}) e^{ik\cdot z}.
	\end{equation}
	Similarly, for a pair of plane waves it is useful to introduce the function:
	\begin{equation}
		\fc{P}(u, S, k, \fc{E}, k', \fc{E}') = \sum_{n=0}^\8\sum_{l=0}^\8\frac{i^{l-n}}{m} V_{nl}^{\rho_1...\rho_n\mu\nu|\kappa_1...\kappa_l\tau\omega}(u,S)k'_{\rho_1}...k'_{\rho_n} f'^*_{\mu\nu} k_{\kappa_1}...k_{\kappa_l} f_{\tau\omega}.
	\end{equation}
    As a shorthand, we write:
	\begin{equation}
		\fc{N}_1 = \fc{N}(v, s, k_1, \fc{E}_1), \p \fc{N}_2 = \fc{N}(v, s, k_2, \fc{E}_2), \p \fc{P}_{12} = \fc{P}(v, s, k_1, \fc{E}_1, k_2, \fc{E}_2).
	\end{equation}
	Expanding the definition of the $\fc{Q}_n$ moments and returning the result to the Fourier transform of the current produces:
    \begin{equation}
        \delta\twid{J}^\mu(k_2) = H^\mu(k_2) \frac{\delta(k_1\cdot v-k_2\cdot v)}{2\pi}
    \end{equation}
    where:
    \begin{align}
        H^\mu(k_2) =& -ik_2\cdot\delta\twid{z} v^\mu + ik_2\cdot v\delta\twid{z}^\mu + 2\sum_{n=0}^\8(-i)^{n+1} \delta\twid{\fc{Q}}_n^{\rho_1...\rho_n\mu\nu}k_{2\rho_1}...k_{2\rho_n}k_{2\nu} \\
        \delta\twid{\fc{Q}}^{\rho_1...\rho_n\mu\nu}_n =& \ \frac{1}{2m}\(-\frac{\fc{N}_1}{m} + \delta\twid{u}^\sigma\frac{\d}{\d v^\sigma} + \delta\twid{S}^\sigma\frac{\d}{\d s^\sigma}\)T_n^{\rho_1...\rho_n\mu\nu}(v,s) \nonumber \\
        &\p \p + \sum_{l=0}^\8 \frac{i^l}{m}V_{nl}^{\rho_1...\rho_n\mu\nu|\kappa_1...\kappa_l\tau\omega}k_{1\kappa_1}...k_{1\kappa_l}f_{1\tau\omega}.
    \end{align}

    Returning our current perturbation to $\delta A$ produces:
    \begin{equation}
        \delta A^\mu = \frac{e^{i\omega(r-t)}}{4\pi r}H^\mu(k_2) + \OO\(\frac{1}{r^2}\)
    \end{equation}
    where
    \begin{equation}
		r = \sqrt{x^2 + (x\cdot v)^2}, \p n^\mu = v^\mu + \frac{x^\mu + (x\cdot v)v^\mu}{r}, \p \omega = -k_1\cdot v, \p k_2^\mu = \omega n^\nu.
	\end{equation}
    Therefore, the canonically normalized Compton amplitude is:
    \begin{equation}
        \fc{A}_\text{canonical} = \fc{E}^*_{2\mu} H^\mu(k_2).
    \end{equation}
    The covariant (Feynman) normalized Compton amplitude can be obtained by multiplying by the usual factor of $\sqrt{(2\ee_1)(2\ee_2)}$. Because the calculation is performed in the classical limit and in a frame in which the initial body is at rest, this normalization factor simply becomes $2m$. Thus:
    \begin{equation}
        \fc{A} = 2m \fc{E}^*_{2\mu} H^\mu(k_2).
    \end{equation}

    Expanding the electromagnetic MPD equations to linear order in $\epsilon$ produces the solutions:
    \begin{align}
		\delta\twid{u}^\mu =& \frac{-i f^{\mu\nu}_1 v_\nu }{m k_1\cdot v} - \frac{\fc{N}_1}{m}\(v^\mu + \frac{k_1^\mu}{k_1\cdot v}\) \label{eq:emcomptonsol1}\\
		\delta\twid{S}^\mu =& v^\mu \delta\twid{u}\cdot s - \frac{i}{k_1\cdot v} \epsilon^{\mu\nu\rho\sigma} v_\nu s_\rho \frac{\d\fc{N}_1}{\d s^\sigma} \\
		\delta\twid{z}^\mu =& \frac{-i}{k_1\cdot v}\(\delta\twid{u}^\mu + \frac{\delta^\mu_\alpha + v^\mu v_\alpha}{m^2} {}^\star{f}_{1}^{\alpha\beta}s_\beta + \frac{\eta^{\mu\nu} + v^\mu v^\nu}{m}\frac{\d\fc{N}_1}{\d v^\nu} \right.\nonumber \\
         &\p \p \p \p\left.+ \frac{s^\mu}{m} v^\nu \frac{\d\fc{N}_1}{\d s^\nu} + \frac{i}{m^2}\fc{N}_1\epsilon^{\mu\nu\rho\sigma}v_\nu s_\rho k_{1\sigma}\).\label{eq:emcomptonsol2}
	\end{align}
    Simplifying the Compton amplitude allows it to be expressed in terms of these solutions as:
    \begin{equation}
		\fc{A} = 2m\(f^*_{2\mu\nu} \delta\twid{z}^\mu v^\nu + ik_2\cdot \delta\twid{z} \fc{N}_2^* - \delta\twid{u}^\sigma \frac{\d\fc{N}_2^*}{\d v^\sigma} - \delta \twid{S}^\sigma \frac{\d\fc{N}_2^*}{\d s^\sigma} + \frac{\fc{N}_1\fc{N}_2^*}{m} -\fc{P}_{12}\). \label{eq:emcompton}  
    \end{equation}
    This gives the formal tree-level electromagnetic Compton amplitude for an arbitrary dynamical mass function.

    Using either the dynamical mass function in \eqref{eq:fulldmf} or in \eqref{eq:simpledmf} determines the $\fc{N}$ function to be:
	\begin{align}
		\fc{N}(u, S, k, \fc{E}) = &\ \frac{\sin\(\frac{1}{m}\sqrt{S^2(u\cdot k)^2 - (S\cdot k)^2}\)}{\sqrt{S^2(u\cdot k)^2 - (S\cdot k)^2}} {}^\star{f}_{\mu\nu} S^\mu u^\nu \nonumber \\
		&\ \ \ + ik\cdot S\frac{1-\cos\(\frac{1}{m}\sqrt{S^2(u\cdot k)^2 - (S\cdot k)^2}\)}{S^2(u\cdot k)^2 - (S\cdot k)^2} f_{\mu\nu} S^\mu u^\nu
	\end{align}
    however the multipole analysis is unable to determine the $\fc{P}$ function. Thus, this determines the Compton amplitude through equations \eqref{eq:emcomptonsol1}-\eqref{eq:emcompton} up to contact terms. Because the electromagnetic Compton amplitude only depends on the linear in $\fc{F}$ dynamical mass function through the $\fc{N}$ function and \eqref{eq:fulldmf} and \eqref{eq:simpledmf} determine the same $\fc{N}$ function, they are interchangeable for Compton amplitudes.

	\subsection{Compton Amplitude through Spin Cubed}

    In this subsection we explicitly compute the Compton amplitude for a generic dynamical mass function through order $\OO(S^3)$. Requiring a match to the spin-exponentiated result for a \rk particle fixes all Wilson coefficients on $\fc{F}^1$ and $\fc{F}^2$ operators through $\OO(S^2)$. The implied values of the Wilson coefficients on $\fc{F}^1$ operators match those determined by \eqref{eq:simpledmf}. For the helicity-conserving Compton amplitude, the spin-exponentiation cannot be continued past $\OO(S^2)$ due to spurious poles. However, the helicity-reversing Compton amplitude has a perfectly healthy exponentiation at $\OO(S^3)$. The helicity-conserving amplitude can instead be required to satisfy the shift symmetry at $\OO(S^3)$ (which is automatic for lower orders which satisfy spin-exponentiation). We find that at $\OO(S^3)$ requiring spin-exponentiation for the helicity-reversing Compton and shift symmetry for the helicity-conserving Compton are consistent with each other and consistent with \eqref{eq:simpledmf}. However, we find that these three requirements still leave one remaining free parameter in the $\OO(\fc{F}^2S^3)$ piece of the dynamical mass function.
    
	We consider only effects in the dynamical mass function which introduce no additional length scales beyond $\frac{S}{m}$. Consequently, we only consider terms for which the number of powers of spin equals the number of derivatives on the vector potential(s). As well, we only consider terms which are parity symmetric and not proportional to the field equations (no $\d_\nu \fc{F}^{\mu\nu}$ or $\d^2 \fc{F}^{\mu\nu}$ terms). (Terms proportional to $\d_\nu \fc{F}^{\mu\nu}$ or $\d^2\fc{F}_{\mu\nu}$ do not contribute to the electromagnetic Compton amplitude because they evaluate to 0 in the $\fc{N}$ function.) The most general such $\delta\fc{M}_1^2$ to order $S^3$ is:
	\begin{align}
		\delta\fc{M}_1^2 =&\ 2C_1 {}^\star\fc{F}_{\mu\nu} S^\mu u^\nu + \frac{C_2}{m} (S\cdot \d)\fc{F}_{\mu\nu} S^\mu u^\nu \nonumber \\
            &\ \ - \frac{C_3}{3m^2}(S\cdot\d)^2 {}^\star\fc{F}_{\mu\nu}S^\mu u^\nu + \frac{E_3}{3m^2}S^2(u\cdot\d)^2 {}^\star\fc{F}_{\mu\nu}S^\mu u^\nu + \OO(S^4)
	\end{align}
	for some Wilson coefficients $C_1$, $C_2$, $C_3$, $E_3$. The most general such $\delta\fc{M}_2^2$ to order $S^3$ is:
	\begin{align}
		\delta\fc{M}_2^2 =& \ \frac{D_{2a}}{m^2} (\fc{F}_{\mu\nu} S^\mu u^\nu)^2 + \frac{D_{2b}}{m^2} ({}^\star\fc{F}_{\mu\nu} S^\mu u^\nu)^2 + \frac{D_{2c}}{m^2} S^2\fc{F}^{\mu\nu}\fc{F}_{\mu\nu} + \frac{D_{2d}}{m^2} S^2 u^\mu \fc{F}_{\mu\nu}\fc{F}^{\nu\rho}u_\rho \nonumber \\
            &\ \ +\frac{D_{3a}}{m^3}{}^\star\fc{F}_{\mu\nu}S^\mu u^\nu (S\cdot\d)\fc{F}_{\rho\sigma}S^\rho u^\sigma+\frac{D_{3b}}{m^3}S^2{}^\star\fc{F}^{\mu\nu}u_\mu \d_\nu \fc{F}_{\rho\sigma}S^\rho u^\sigma \nonumber \\
            &\ \ + \frac{D_{3c}}{m^3}S^2{}^\star\fc{F}^{\mu\nu}u_\mu (S\cdot \d)\fc{F}_{\nu\rho}u^\rho+\frac{D_{3d}}{m^3}S^2{}^\star\fc{F}^{\mu\nu}S_\mu (u\cdot\d)\fc{F}_{\nu\rho}u^\rho \nonumber \\
            &\ \ + \frac{D_{3e}}{m^3}{}^\star\fc{F}^{\mu\nu}S_\mu(S\cdot\d)\fc{F}_{\nu\rho}S^\rho+\frac{D_{3f}}{m^3}S^2{}^\star\fc{F}^{\mu\nu}(S\cdot\d)\fc{F}_{\mu\nu} + \OO(S^4)
	\end{align}
	for some Wilson coefficients $D_{2a}$, $D_{2b}$, $D_{2c}$, $D_{2d}$ for quadratic-in-spin terms and $D_{3a}$, $D_{3b}$, $D_{3c}$, $D_{3d}$, $D_{3e}$, $D_{3f}$ for cubic in spin.
	 These lead to the $\fc{N}$ and $\fc{P}$ functions:
	\begin{align}
		\fc{N} =&\ \frac{C_1}{m}{}^\star f_{\mu\nu} S^\mu u^\nu + \frac{iC_2}{2m^2}k\cdot S f_{\mu\nu} S^\mu u^\nu + \frac{C_3}{6m^3}(k\cdot S)^2 {}^\star f_{\mu\nu}S^\mu u^\nu - \frac{E_3}{6m^3}S^2(k\cdot u)^2 {}^\star f_{\mu\nu}S^\mu u^\nu ,
    \label{eq:NFunctionEM}
        \\
		\fc{P} =&\ \frac{D_{2a}}{m^3} f'^*_{\mu\nu} S^\mu u^\nu  f_{\rho\sigma} S^\rho u^\sigma + \frac{D_{2b}}{m^3} {}^\star f'^*_{\mu\nu} S^\mu u^\nu {}^\star f_{\rho\sigma} S^\rho u^\sigma + \frac{D_{2c}}{m^3} S^2 f'^*_{\mu\nu}f^{\mu\nu} + \frac{D_{2d}}{m^3} S^2 u^\mu f'^*_{\mu\nu} f^{\nu\rho} u_\rho \nonumber \\
            &\ \ + \frac{i D_{3a}}{2m^4}{}^\star f'^*_{\mu\nu}S^\mu u^\nu (S\cdot k) f_{\rho\sigma}S^\rho u^\sigma - \frac{iD_{3a}}{2m^4}{}^\star f_{\mu\nu}S^\mu u^\nu (S\cdot k')f'^*_{\rho\sigma}S^\rho u^\sigma \nonumber \\
            &\ \ + \frac{i D_{3b}}{2m^4}S^2{}^\star f'^*_{\mu\nu}u^\mu k^\nu f_{\rho\sigma}S^\rho u^\sigma - \frac{i D_{3b}}{2m^4}S^2 {}^\star f_{\mu\nu}u^\mu k'^\nu f'^*_{\rho\sigma}S^\rho u^\sigma \nonumber \\
            &\ \ + \frac{i D_{3c}}{2m^4}S^2{}^\star f'^{*\mu\nu}u_\mu (S\cdot k) f_{\nu\rho}u^\rho - \frac{iD_{3c}}{2m^4} S^2 {}^\star f^{\mu\nu}u_\mu (S\cdot k')f'^*_{\nu\rho}u^\rho \nonumber \\
            &\ \ + \frac{i D_{3d}}{2m^4}S^2{}^\star f'^{*\mu\nu}S_\mu (u\cdot k)f_{\nu\rho}u^\rho - \frac{iD_{3d}}{2m^4}S^2{}^\star f^{\mu\nu}S_\mu (u\cdot k')f'^*_{\nu\rho}u^\rho \nonumber \\
            &\ \ + \frac{i D_{3e}}{2m^4}{}^\star f'^{*\mu\nu}S_\mu (S\cdot k)f_{\nu\rho}S^\rho - \frac{iD_{3e}}{2m^4}{}^\star f^{\mu\nu}S_\mu(S\cdot k')f'^*_{\nu\rho}S^\rho \nonumber \\
            &\ \ + \frac{i D_{3f}}{2m^4}S^2{}^\star f'^{*\mu\nu}(S\cdot k)f_{\mu\nu} - \frac{iD_{3f}}{2m^4}S^2{}^\star f^{\mu\nu}(S\cdot k')f'^*_{\mu\nu}.
	\end{align}

    We find it advantageous to express our results for the Compton amplitude in a basis of definite-helicity/circularly polarized/(anti-)self-dual states for the incoming and outgoing electromagnetic waves, or ``photons,'' while manifesting Lorentz covariance and gauge invariance (ultimately).
    For concreteness (initially), we can work in a particular Lorentz frame, associated to inertial Cartesian coordinates $x^\mu=(t,x,y,z)$, such that the charged massive spinning particle's initial velocity $v$ and the two photon wavevectors, $k_1$ incoming and $k_2$ outgoing (both future-pointing), are given by 
	\begin{equation}
		v^\mu = (1, 0, 0, 0),               \qquad 
        k_1^\mu = \omega(1, 0, 0, 1),
        \qquad
        k_2^\mu =\omega (1, \sin\theta, 0, \cos\theta).
    \label{eq:uk1k2frame}
    \end{equation}
    Then $\theta$ is the photon scattering angle in the $z$-$x$-plane, and $\omega=-v\cdot k_1=-v\cdot k_2$ is the waves' angular frequency.  The ``momentum transfer'' (per $\hbar$) $q=k_2-k_1$ squares to
    \begin{equation}
        q^2=(k_2-k_1)^2=-2k_1\cdot k_2=2\omega^2(1-\cos\theta)=4\omega^2\sin^2\frac{\theta}{2},
    \label{eq:qsqoftheta}
    \end{equation}
    vanishing at forward scattering, $\theta=0$.  
    In choosing a particular basis of definite-helicity (complex null) polarization vectors, $\fc E_{1\pm}$ incoming, $\fc E_{2\pm}$ outgoing, with $k_n\cdot\fc E_{n\pm}=0=\fc E_{n\pm}^2$, it is natural to fix the gauge freedom $\fc E_{n}\to\fc E_{n}+\alpha k_n$ by imposing $v\cdot\fc E=0$. Up to little group transformations ($\fc E\to e^{ 2i\varphi}\fc E$), this determines 
    \begin{equation}
        \fc{E}_{1\sigma_1}^\mu =\frac{1}{\sqrt{2}} \(0, 1,  i\sigma_1, 0\), \qquad
        \fc{E}_{2\sigma_2}^\mu =\frac{1}{\sqrt{2}} \(0, \cos\theta,  i\sigma_2, -\sin\theta\),
    \label{eq:helbasis}
	\end{equation}
    for helicities $\sigma_1=\pm1$ and $\sigma_2=\pm1$, with complex conjugates $\fc E_{n\pm}^{*\mu}=\fc E_{n\mp}^{\mu}$, normalized as $\fc E_{n\sigma_n}\cdot\fc E_{n\sigma_n}^*=1$.  This frame (\ref{eq:uk1k2frame}) and polarization basis (or gauge) (\ref{eq:helbasis}) are just as in \cite{Saketh:2022wap} and in \cite{Bautista:2022wjf}.
    The spinless Compton amplitude is given simply by the contraction of the ingoing and conjugate-outgoing polarization vectors (only) in this $v\cdot\fc E=0$ gauge:
	\begin{equation}
        \fc{A}^{(0)}_{\sigma_1\sigma_2} =-2\fc E_{1\sigma_1}\cdot\fc E_{2\sigma_2}^*= -\sigma_1\sigma_2 - \cos\theta.	\label{eq:spinlessem}
	\end{equation}

    In analyzing the helicity-preserving amplitudes  $\fc A_{\pm\pm}\propto\fc E_{1\pm}^\mu\fc E_{2\pm}^{*\nu}$, it is useful to define as in \cite{Aoude:2020onz}\footnote{In \cite{Aoude:2020onz} as in most Amplitudes literature, our \emph{helicity-preserving} amplitudes $\fc A_{\pm\pm}\propto\fc E_{1+}^\mu\fc E_{2+}^{\pm\nu}$ are called ``opposite-helicity amplitudes $\mathsf A_{\pm\mp}$'',  while our \emph{helicity-reversing} amplitudes $\fc A_{\pm\mp}\propto\fc E_{1\pm}^\mu\fc E_{2\mp}^{*\nu}$ are called ``same-helicity amplitudes $\mathsf A_{\pm\pm}$'',  due to differing conventions (essentially $k_n\leftrightarrow-k_n$ entailing $\sigma_n\leftrightarrow-\sigma_n$).  
    } a complex null vector $w$ orthogonal to both $k_1$ and $k_2$. The conditions $w^2=k_1\cdot w=k_2\cdot w=0$ determine $w$ up to an overall normalization, which we fix by setting $v\cdot w=-\omega$, and up to a binary choice of branch ($w\propto|1\rangle[ 2|$ or $w\propto|2\rangle[ 1|$) to be correlated with the photons' helicities.  For our $++$ (or $--$) case, the appropriate $w$ is given by
    \begin{alignat}{3}
        w^\mu&=\frac{\omega}{4\omega^2-q^2}\Big(2\omega(k_1^\mu+k_2^\mu)-q^2v^\mu-2i\epsilon^\mu{}_{\nu\rho\sigma}v^\nu k_1^\rho k_2^\sigma\Big)
    \label{eq:def_w}
    \end{alignat}
    (or the complex conjugate $w^{*\mu}$). In the frame of (\ref{eq:uk1k2frame}), 
    \begin{equation}
        w^\mu = \omega\Big(1, \tan\frac{\theta}{2}, i\tan\frac{\theta}{2}, 1\Big),
    \end{equation}
    noting $4\omega^2-q^2=4\omega^2\cos^2\frac{\theta}{2}$ along with (\ref{eq:qsqoftheta}). 
    The complex null direction $\propto w$ provides an alternative $(+)$-helicity polarization direction for the incoming $k_1$ photon, as well as an alternative conjugate $(+)$-helicity polarization direction for the outgoing $k_2$ photon. We see that we can recover the normalized orthogonal-to-$v$ polarization vectors $\fc E_{1+}$ and $\fc E_{2+}^*$ of (\ref{eq:helbasis}) from $w$ via $\fc E_n\to\fc E_n+\alpha k_n$ shifts and rescalings:
    \begin{equation}
        \fc E_{1+}^\mu=\frac{w^\mu-k_1^\mu}{\sqrt{2}\omega\tan\frac{\theta}{2}},
        \qquad
        \fc E_{2+}^{*\mu}=\frac{k_2^\mu-w^\mu}{\sqrt{2}\omega\tan\frac{\theta}{2}},
    \label{eq:E1plus}
    \end{equation}
    As in \cite{Haddad:2023ylx} (modulo conventions), let us define the vectors
    \begin{equation}
        \check k_1^\mu=k_1^\mu-w^\mu,
        \qquad
        \check k_2^\mu=k_2^\mu-w^\mu,
    \end{equation}
    proportional to those in (\ref{eq:E1plus}), which, along with $w$ as in (\ref{eq:def_w}), provide a relatively compact way to express the $++$ Compton amplitude.

    The complex field-strength amplitudes $ f^{\pm}_{\mu\nu}=2i k^{\phantom{\pm}}_{[\mu}\fc E^{\pm}_{\nu]}$ are invariant under $\fc E\to\fc E+\alpha k$, they transform under the little groups like the $\fc E$s, and they are self-dual (or anti-self-dual), \begin{equation}
    {}^\star f^{\pm}_{\mu\nu}=\frac{1}{2}\epsilon_{\mu\nu}{}^{\kappa\lambda}f^{\pm}_{\kappa\lambda}=\pm i f^{\pm}_{\mu\nu},
    \end{equation}
    for states of helicity $+1$ (or $-1$), while the complex conjugates are reversed: ${}^\star f_{\mu\nu}^{{\pm}{*}}=\mp i f_{\mu\nu}^{{\pm}{*}}$. For one way to see this, we can construct $f^{+}_{1\mu\nu}$ from the $\fc E^{+}_{1\mu}$ of (\ref{eq:helbasis}), using (\ref{eq:E1plus}) with (\ref{eq:def_w}),
    \begin{alignat}{3}
        f^{+}_{1\mu\nu}&=2ik^{\phantom{\pm}}_{1[\mu}\fc E_{1\nu]}^+=\frac{\sqrt{2}i}{\omega\tan\frac{\theta}{2}}k_{1[\mu}w_{\nu]}
    \label{eq:f1plus}
        \\\nonumber
        &=\frac{\sqrt{2}i}{\omega^2\sin\theta}k_{1[\mu}\Big(
        v_{\nu]}(k_1\cdot k_2)+\omega k_{2\nu]}-i\epsilon_{\nu]\rho\alpha\beta}v^\rho k_1^\alpha k_2^\beta\Big)
        \\\nonumber
        &=\frac{\sqrt{2}i}{\omega^2\sin\theta}\Big(\delta_{[\mu}{}^{\alpha}\delta_{\nu]}{}^\beta
        -\frac{i}{2}\epsilon_{\mu\nu}{}^{\alpha\beta}\Big)k_{1\alpha}\big(v_{\beta}(k_1\cdot k_2)+\omega k_{2\beta}\big)
        \\\nonumber
        &=\frac{\sqrt{2}i}{\omega^2\sin\theta}2{}^{\,+}\fc G_{\mu\nu}{}^{\alpha\beta}k_{1\alpha}2v_{[\beta}k_{2\gamma]}k_1^\gamma,
    \end{alignat}
    where we have used $0=5k_{1[\mu}\epsilon_{\nu\rho\alpha\beta]}v^\rho k_1^\alpha k_2^\beta=2 k_{1[\mu}\epsilon_{\nu]\rho\alpha\beta}v^\rho k_1^\alpha k_2^\beta-\epsilon_{\mu\nu\alpha\beta}k_1^\alpha(v^\beta(k_1\cdot k_2)+\omega k_2^\beta)=2k_{1[\mu}\epsilon_{\nu]vk_1k_2}-\epsilon_{\mu\nu\alpha\beta}k_1^\alpha \,2v^{[\beta}k_2^{\gamma]}k_{1\gamma}$ and $2\tan\frac{\theta}{2}\cos^2\frac{\theta}{2}=\sin\theta$, and where we recognize the \mbox{(anti-)}self-dual ((A)SD) projector,
    \begin{equation}
        {}^\pm\fc G_{\mu\nu}{}^{\kappa\lambda}=\frac{1}{2}\delta_{[\mu}{}^\kappa\delta_{\nu]}{}^\lambda\mp \frac{i}{4}\epsilon_{\mu\nu}{}^{\kappa\lambda}={}^\pm\fc G_{\mu\nu}{}^{\alpha\beta}{}^\pm\fc G_{\alpha\beta}{}^{\kappa\lambda}=\mp i\,{}^\star{}^\pm\fc G_{\mu\nu}{}^{\kappa\lambda},
    \label{eq:ASD_proj}
    \end{equation}
    which maps a 2-form (or any tensor) $A_{\mu\nu}$ onto its (A)SD part ${}^\pm\! A_{\mu\nu}$:
    \begin{equation}
        {}^\pm\! A_{\mu\nu}={}^\pm\fc G_{\mu\nu}{}^{\kappa\lambda} A_{\kappa\lambda}=\frac{1}{2}\big(A_{[\mu\nu]}\mp i\,{}^\star \! A_{[\mu\nu]}\big),
        \qquad
        {}^\star{}^\pm \!A_{\mu\nu}=\pm i {}^\pm\! A_{\mu\nu}.
    \end{equation}
    Note the useful identity
    \begin{equation}
        {}^\pm\fc G_{\mu\nu}{}^{(\alpha}{}_\rho {}^\pm\fc G_{\kappa\lambda}{}^{\beta)\rho}=\frac{1}{4}{}^\pm\fc G_{\mu\nu\kappa\lambda}g^{\alpha\beta},
    \label{eq:GGmagic}
    \end{equation}
    or ${}^\pm\fc G_{\mu\nu(\alpha}{}^\gamma{}^{\,\pm}\! A_{\beta)\gamma}=\frac{1}{4}{}^\pm\! A_{\mu\nu}g_{\alpha\beta}$ and thus ${}^\pm\fc G_{\mu\nu}{}^{\rho\sigma} p_\sigma {}^{\pm}\!A_{\rho\tau}p^\tau=\frac{1}{4}p^2{}^{\,\pm}\!A_{\mu\nu}$, following from $\epsilon_{\alpha\beta\gamma\delta}\epsilon^{\mu\nu\rho\sigma}$ $=$ $-24\delta_\alpha{}^{[\mu}\delta_\beta{}^\nu\delta_\gamma{}^\rho\delta_\delta{}^{\sigma]}$ and $\delta_{\omega}{}^{[\lambda}\delta_\alpha{}^{\mu}\delta_\beta{}^\nu\delta_\gamma{}^\rho\delta_\delta{}^{\sigma]}=0$.  Collecting (\ref{eq:f1plus}) along with its conjugate outgoing versions, also following from (\ref{eq:def_w}) and (\ref{eq:E1plus}), we have
    \begin{alignat}{3}
        f^{+}_{1\mu\nu}&=+2ik^{\phantom{+}}_{1[\mu}\fc E_{1\nu]}^{+\phantom{*}}=\frac{4\sqrt{2}i}{\omega^2\sin\theta}{}^{\,+}\fc G_{\mu\nu}{}^{\alpha\beta}k_{1\alpha}v_{[\beta}k_{2\gamma]}k_1^\gamma=\frac{\sqrt{2}i}{\omega\tan\frac{\theta}{2}}k_{1[\mu}w_{\nu]\,}
        ,
        \nonumber\\
        f^{+*}_{2\mu\nu} &=-2ik^{\phantom{+}}_{2[\mu}\fc E_{2\nu]}^{+*}=\frac{4\sqrt{2}i}{\omega^2\sin\theta}{}^{\,-}\fc G_{\mu\nu}{}^{\alpha\beta}k_{2\alpha}v_{[\beta}k_{1\gamma]}k_2^\gamma=\frac{\sqrt{2}i}{\omega\tan\frac{\theta}{2}}k_{2[\mu}w_{\nu]\,}
        ,
        \nonumber\\
        f^{-*}_{2\mu\nu} &=-2ik^{\phantom{+}}_{2[\mu}\fc E_{2\nu]}^{-*}=\frac{4\sqrt{2}i}{\omega^2\sin\theta}{}^{\,+}\fc G_{\mu\nu}{}^{\alpha\beta}k_{2\alpha}v_{[\beta}k_{1\gamma]}k_2^\gamma=\frac{\sqrt{2}i}{\omega\tan\frac{\theta}{2}}k^{\phantom{*}}_{2[\mu}w^*_{\nu]\,},
    \label{eq:f2plusconj}
    \end{alignat}
    noting $w^{*\mu}=-w^\mu+\omega\dfrac{2\omega(k_1+k_2)^\mu-q^2 v^\mu}{4\omega^2-q^2}$ and 
    \begin{equation}
        \omega^2\sin\theta=\sqrt{4\omega^4\sin^2\frac{\theta}{2}\cos^2\frac{\theta}{2}=\frac{q^2}{4}(4\omega^2-q^2)=-\epsilon^\mu{}_{vk_1k_2}\epsilon_{\mu}{}^{ vk_1k_2}\propto[12]\langle 12\rangle[2|v|1\rangle[1|v|2\rangle}.
    \end{equation}
    
    In simplifying the helicity-basis Compton amplitudes $\fc A_{+\pm}\propto f_{1\mu\nu}^+f_{2\alpha\beta}^{\pm*}$, the (A)SD properties ${}^\star f_{1\mu\nu}^{+}=+if_{1\mu\nu}$ and ${}^\star f_{2\mu\nu}^{\pm *}=\mp if_{2\mu\nu}^{\pm *}$ can be used immediately within (before differentiation of) the $\fc N(f)$ and $\fc P(f,f')$ functions in (\ref{eq:NFunctionEM}). These functions completely determine the amplitudes via (\ref{eq:emcomptonsol1})--(\ref{eq:emcompton}).  They can finally be evaluated directly in terms of the complex null $w^\mu(k_1,k_2,v)$ by using the extreme equalities of (\ref{eq:f2plusconj}), noting e.g.\ $\epsilon_{vak_1w}=i\omega\check k_1\cdot a$, $\epsilon_{vak_2w}=-i\omega\check k_2\cdot a$,  $\epsilon_{vk_1k_2w}=i\omega q^2/2$ following from (\ref{eq:def_w}), recalling $\check k_1=k_1-w$ and $\check k_2=k_2-w$; they otherwise depend only on $k_1^\mu$, $k_2^\mu$, $v^\mu$ and the initial spin $s^\mu=ma^\mu$.
    
    For the ++ amplitudes $\fc A^{(n)}_{++}\propto f_1^+ f_2^{+*}a^n$ at $n$th order in spin, we find
    \begin{subequations}
    \label{eq:AppEM_kcheck}
    \begin{alignat}{3}
        \fc{A}_{++}^{(0)}
        &=-\frac{4\omega^2-q^2}{2\omega^2},
    \end{alignat}
    \begin{alignat}{3}
        \fc{A}_{++}^{(1)}
        &=-\frac{4\omega^2-q^2}{2\omega^2}\bigg\{C_1(\check k_1+\check k_2)\cdot a-(C_1-1)^2w\cdot a\bigg\},
    \end{alignat}
    \begin{alignat}{3}
        \fc A_{++}^{(2)}&=-\frac{4\omega^2-q^2}{2\omega^2}\Bigg\{\frac{C_2}{2}[(\check k_1+\check k_2)\cdot a]^2-\Big((C_1-1) C_2+D_{2d}\Big) (w\cdot a)^2
        \\\nonumber
        &\qquad\qquad\qquad\quad+(C_1-1) (C_1-C_2)\,w\cdot a\,(\check k_1+\check k_2)\cdot a
        \\\nonumber
        &\quad+\bigg[\Big((C_1-1)(2C_1-C_2)+2D_{2d}-D_{2a}-D_{2b}\Big)\frac{2\omega^2}{q^2}+C_1^2-C_2\bigg]\check k_1\cdot a\,\check k_2\cdot a
        \Bigg\},
    \end{alignat}
    \begin{alignat}{3}
        \fc A_{++}^{(3)}
        &=-\frac{4\omega^2-q^2}{2\omega^2}\Bigg\{\frac{C_3}{6}[(\check k_1+\check k_2)\cdot a]^3+ \hat{\fc{A}}_{++}^{(3)(C_2-1, C_1-1)}
        \nonumber \\
        &\qquad+\bigg(\,^{\!}\frac{4}{3}+2F_\text{i}-\frac{F_\text{ii}}{4}+\frac{F_\text{iii}}{2}\bigg)(w\cdot a)^3+\bigg(F_\text{i}-\frac{C_3-1}{6}\bigg)(w\cdot a)^2(\check k_1+\check k_2)\cdot a
        \nonumber\\
        &\qquad+\bigg[F_\text{ii}\frac{\omega^2}{q^2}w\cdot a+\bigg(F_\text{iii}\frac{\omega^2}{q^2}-\frac{C_3-1}{2}\bigg)(\check k_1+\check k_2)\cdot a\bigg]\check k_1\cdot a\,\check k_2\cdot a\Bigg\},
    \label{eq:App_w}
    \end{alignat}
    where
    \begin{alignat}{7}
        F_\text{i}&=\frac{E_3-1}{6}-\frac{D_{3b}+D_{3c}+D_{3e}}{2},
        \qquad\quad&
        F_\text{ii}&=3-2D_{3a}+4D_{3c}+4D_{3d},
        \nonumber\\
        F_\text{iii}&=-1-\frac{2}{3}E_3-D_{3a}+2D_{3b}+2D_{3c},
    \end{alignat}
    \end{subequations}
    and $\hat{\fc{A}}_{++}^{(3)(C_2-1, C_1-1)}$ vanishes when $C_2 = C_1 = 1$ (which for \rk are fixed by lower-order-in-spin pieces of the Compton amplitude, or by the three-point amplitude). Here we have used
    \begin{equation}
        \omega^2a^2=(w\cdot a)^2-\frac{4\omega^2}{q^2}\check k_1\cdot a\,\check k_2\cdot a,
    \label{eq:asq_wasq}
    \end{equation}
    resulting from $0=(u^{[\lambda}k_1^\mu k_2^\nu w^\rho a^{\sigma]})^2$, to eliminate $a^2$ in favor of $(w\cdot a)^2$.  This leads to the relatively compact expressions (\ref{eq:AppEM_kcheck}) ----- paralleling the parametrization \eqref{eq:hatApp5kcheck} of the black hole--graviton $++$ Compton amplitude at fifth order in spin as formulated in \cite{Bautista:2022wjf} ----- but it makes the amplitude appear to have (in addition to the physical poles at $v\cdot k_1=v\cdot k_2$ $=$ $-\omega=0$) unphysical poles: firstly, explicitly, at $q^2=4\omega^2\sin^2\frac{\theta}{2}=0$ --- at forward scattering, $\theta=0$ --- the would-be physical pole corresponding to an internal photon of momentum $q=k_2-k_1$ going on shell, but on which the residue must be zero because the three-photon amplitude vanishes; and secondly, hidden inside $w$ (and $\check k_1$ and $\check k_2$) in (\ref{eq:def_w}), at $4\omega^2-q^2=4\omega^2\cos^2\frac{\theta}{2}=0$ --- at back-scattering, $\theta=\pi$ --- ``the spurious pole.''  However, no unphysical poles are actually present, for arbitrary values of the Wilson coefficients, as can be made manifest by using (\ref{eq:asq_wasq}) to eliminate factors of $(w\cdot a)^2$ in favor of $a^2$.

    The helicity-preserving amplitude $\fc A_{++}$ is well expressed in terms of the spin component $w\cdot a$ along the complex null $w^\mu(k_1,k_2,u)$ of (\ref{eq:def_w}) because of its symmetry $w\leftrightarrow w^*$ under $k_1\leftrightarrow k_2$ [with $-2\omega= (k_1+k_2)\cdot v$].

    Turning to the helicity-reversing amplitude $\fc A_{+-}$, the appropriate symmetry is reflected by a vector $x^\mu(k_1,k_2,v)$ with $x\leftrightarrow-x$ under $k_1\leftrightarrow k_2$ (modulo any component along $v$). An apt choice is $x(k_1,k_2,v)\cdot a\propto w(k_2,-k_1,v)\cdot a$:
    \begin{alignat}{3}
        w\cdot a\,&=\Big(\omega(k_1+k_2)\cdot a+i\epsilon_{\mu\nu\rho\sigma}v^\mu k_1^\nu k_2^\rho a^\sigma\Big)\frac{2\omega}{4\omega^2-q^2}\,,
        \nonumber\\
        x\cdot a\,&=\Big(\omega(k_2-k_1)\cdot a+i\epsilon_{\mu\nu\rho\sigma}v^\mu k_1^\nu k_2^\rho a^\sigma\Big)\,\frac{2\omega}{q^2}=w\cdot a +\frac{4\omega^2}{q^2}(k_1-w)\cdot a
        \nonumber\\
        &=\frac{2\omega}{q^2}\Big(\omega\, q\cdot a+ik_1\times k_2\cdot a\Big)\,,
    \end{alignat}
	coinciding with ``$w_O\cdot a$'' from \cite{Saketh:2022wap} or $\propto$ ``$w_{++}\cdot a$'' from \cite{Haddad:2023ylx}, with $x\cdot a\propto \langle1|av|1\rangle$, in contrast to $w\cdot a\propto[2|a|1\rangle$. The identity (\ref{eq:asq_wasq}) for $(w\cdot a)^2$ translates into
    \begin{alignat}{3}
        \frac{q^2}{4\omega^2}\Big((x\cdot a)^2-\omega^2a^2\Big)&=\dfrac{-q^2}{4\omega^2-q^2}(k_1\cdot a-x\cdot a)(k_2\cdot a+x\cdot a)
        \\\nonumber
        &=(k_1\cdot  a)(k_2\cdot  a)-(x\cdot  a)(q\cdot  a)-\omega^2 a^2=(aya)
    \label{eq:def:aya}
    \end{alignat}
    for $(x\cdot a)^2$ and defines a convenient quadratic $(aya)$ in the spin.  With this, we find
    \begin{alignat}{3}
        \fc A_{+-}&=\frac{q^2}{2\omega^2}\Bigg\{1-C_1\,q\cdot a+(C_1^2-1)x\cdot a
        \\
        &\qquad\;\;\,+\frac{C_2}{2}(q\cdot a)^2-(C_1-1)(C_1+C_2)k_1\cdot a\,k_2\cdot a
        \nonumber\\
        &\qquad\qquad+\bigg[\Big(D_{2b}-D_{2a}-(C_1-1)(2C_1+C_2)\Big)\frac{2\omega^2}{q^2}+(C_2-1)C_1\bigg](aya)
        \nonumber\\
        &\qquad\qquad -\Big(4D_{2c}+D_{2d}+(C_1-1)C_1\Big)\omega^2a^2
        \nonumber\\
        &\qquad\;\;\,-\frac{C_3}{6}(q\cdot a)^3+(C_3-1)x\cdot a\,k_1\cdot a\,k_2\cdot a +\hat{\fc A}_{+-}^{(3)(C_2-1,C_1-1)}\phantom{\bigg|}
        \nonumber\\
        &\qquad\qquad -\bigg((1+D_{3a})\frac{\omega^2}{q^2}+\frac{C_3-1}{2}\bigg)(aya)q\cdot a
        -\bigg(\frac{3}{4}+E_3+D_{3b}+D_{3d}\bigg)\omega^2a^2x\cdot a
        \nonumber\\\nonumber
        &\qquad\qquad-\bigg(\frac{1+E_3}{6}+\frac{D_{3b}+D_{3c}-D_{3e}}{2}+2D_{3f}+\frac{C_3-1}{6}\bigg)\omega^2 a^2 q\cdot a+\fc O(a^4)\Bigg\},
    \end{alignat}
    where again $\hat{\fc{A}}_{+-}^{(3)(C_2-1, C_1-1)}$ vanishes when $C_2 = C_1 = 1$.
    
    At linear order in spin, $C_1$ is determined by the $\OO(S^1)$ amplitude. At $\OO(S^2)$, all 5 new operators, $C_2, D_{2a,b,c,d}$ contribute linearly independent structures to the amplitude. At $\OO(S^3)$ there are 8 new operators $C_3, D_{3a,b,c,d,e,f}, E_{3}$ but only 7 linearly independent structures in the Compton. In particular, the Compton amplitude is independent of the value of the linear combination:
    \begin{equation}
        Z = -6 E_3 + D_{3f} + 2 C_1 D_{3b} - 2 (1+ C_1) D_{3c} + 4 D_{3d}
    \end{equation}
 
	These amplitudes produce the spin-exponentiation of Ref.~\cite{Guevara:2018wpp} through order $\OO(S^2)$ if and only if:
	\begin{equation}
		C_1 = C_2 = 1, \p D_{2a} = D_{2b} = D_{2c} = D_{2d} = 0.
	\end{equation}
	For these values of the Wilson coefficients, we recover the spin-exponentiated amplitudes:
	\begin{align}
		\fc{A}_{++}&= \fc{A}^{(0)}_{++}\exp\(a\cdot(\check k_1+\check k_2)\) + \OO(S^3) \\
		\fc{A}_{+-}&= \fc{A}^{(0)}_{+-}\exp\(a\cdot(k_1-k_2)\) + \OO(S^3) \label{eq:oppositehel}
	\end{align}

    For the equal helicity amplitude the spin-exponentiation cannot continue past $\OO(S^2)$ due to the spurious pole in $\cos(\theta/2)$. There is no such trouble for the opposite helicity amplitude. If one demands the continuance of the spin-exponentiation for the opposite helicity amplitude through $\OO(S^3)$, it fixes $C_3$ and 3 of the $D_{3a,b,c,d,e,f}$ coefficients. In particular it determines:
    \begin{gather}
        C_3 = 1, \p D_{3a} = -1, \p D_{3d} = -D_{3b} - E_3 - \frac{3}{4}, \nonumber \\
        D_{3f} = -\frac{1}{12} - \frac{E_3}{12} - \frac{D_{3b}}{4} - \frac{D_{3c}}{4} + \frac{D_{3e}}{4}.
    \end{gather}
    The value $C_3 = 1$ is consistent with \eqref{eq:simpledmf}. Equation \eqref{eq:simpledmf} fixes the value of $E_3 = 1$ which is possible while continuing the spin-exponentiation but not demanded by it.

    It is also interesting to study the shift symmetry condition identified in Refs.~\cite{Aoude:2022trd, Bern:2022kto, Haddad:2023ylx,  Aoude:2023vdk}. In order for the same helicity Compton amplitude to maintain shift symmetry through $\OO(S^3)$ according to the criteria of Ref.~\cite{Aoude:2022trd}, some of the $C_3$, $E_3$, $D_{3a,b,c,d,e,f}$ are fixed. In particular:
    \begin{gather}
        C_3 = 1, \p D_{3c} = \frac{E_3}{3} + \frac{1+D_{3a}}{2} - D_{3b},  \p D_{3d} = D_{3b} -\frac{E_3}{3} - \frac{5}{4}.
    \end{gather}
    Thus, at this order the shift symmetry is also consistent with \eqref{eq:simpledmf}. 

    The conditions necessary to maintain opposite helicity spin-exponentiation, shift symmetry, and match \eqref{eq:simpledmf} are consistent with each other at $\OO(S^3)$ and lead to the combined set of conditions:
    \begin{gather}
        C_3 = 1, \p E_3 = 1, \p D_{3a} = -1, \p D_{3b} = -\frac{1}{12} \nonumber \\
        D_{3c} = \frac{5}{12}, \p D_{3d} = -\frac{5}{3}, \p D_{3f} = \frac{D_{3e}-1}{4}.
    \end{gather}
    Thus, at $\OO(S^3)$ there is a one parameter family of dynamical mass functions (as $D_{3e}$ is fully undetermined) satisfying all of these constraints. 

    Following the decomposition of Ref.~\cite{Aoude:2022trd}, the same helicity amplitude in terms of $D_{3e}$ through $\OO(S^3)$ may be written as:
    \begin{equation}
        \fc{A}_{++} = e^{a\cdot(k_1+k_2)} \sum_{n=0}^3 \frac{\bar{I}_n}{n!} + \OO(S^4)
    \end{equation}
    with:
    \begin{align}
        &\bar{I}_0 = -2\cos^2\frac{\theta}{2},& &\bar{I}_1 = -2 a\cdot w \bar{I}_0,& \nonumber\\
        &\bar{I}_2 = \(2a\cdot w\)^2 \bar{I}_0,& &\bar{I}_3 = -(3D_{3e}+1) (a\cdot w)^2 a\cdot (k_1+k_2) \bar{I}_0.&
    \end{align}
    Thus we can identify $c_0^{(3)}$ in equation (3.9b) of Ref.~\cite{Aoude:2022trd} with $-2-6 D_{3e}$.

    We can also compare to the recent work of \cite{Cangemi:2023ysz} on \rk amplitudes from higher-spin gauge interactions. To that end, following the lead of \cite{Cangemi:2023ysz}, instead of using \eqref{eq:asq_wasq} to eliminate $a^2\omega^2$ leaving $(w\cdot a)^2$ as we did in \eqref{eq:AppEM_kcheck}, we can express the ++ amplitude in terms of both $(w\cdot a)^2$ and $\omega^2a^2$ while eliminating $q^2/\omega^2$ using \eqref{eq:asq_wasq}.  Defining $k_\pm=k_2\pm k_1$,
    \begin{alignat}{7}
        k_+&\,=\,k_1+k_{2\,},
        \qquad\;\,
         \check k_1+\check k_2\,=\,k_{+\!}-2w_{\,},
        \\\nonumber
        q\,=\,k_-&\,=\,k_2-k_1\;=\; \check k_2-\check k_{1\,},
        \quad\;\;\;\;\;\;
        [(k_{+\!}-2w)\cdot a]^2-(q\cdot a)^2=4_{\,} \check k_1\cdot a\,\check k_2\cdot a_{\,},
    \end{alignat}
    and replacing $\omega^2/q^2$ with $[(w\cdot a)^2-\omega^2a^2]/(4_{\,}\check k_1\cdot a\,\check k_2\cdot a)$, our amplitude \eqref{eq:AppEM_kcheck} becomes
    \begin{alignat}{3}
        \frac{\fc A_{++}}{\fc A^{(0)}_{++}}&=1+C_1(k_{+\!}-2w)\cdot a-(C_1-1)^2w\cdot a
        \nonumber\\
        &\qquad+\frac{1}{2}[(k_{+\!}-2w)\cdot a]^2\frac{C_1^2+C_2}{2}+\frac{C_2-C_1^2}{4}(q\cdot a)^2
        \nonumber\\
        &\qquad\qquad\qquad\qquad\;\;\,+(C_1-1)\bigg[(C_1-C_2)k_{+\!}\cdot a\,w\cdot a+\frac{C_2-2C_1}{2}\Big(\omega^2a^2+(w\cdot a)^2\Big)\bigg]
        \nonumber\\
        &\qquad\qquad -\frac{D_{2a}+D_{2b}}{2}(w\cdot a)^2+\frac{D_{2a}+D_{2b} -2D_{2d}}{2}\omega^2a^2 \phantom{\bigg|}
        \nonumber\\
        &\qquad+\frac{1}{6}(k_{+\!}-2w)\cdot a\Big[(k_{+\!}-4w)\cdot a\,k_{+\!}\cdot a\frac{3+C_3}{4}+3\frac{C_3-1}{4}(q\cdot a)^2\Big]+ \hat{\fc{A}}_{++}^{(3)(C_2-1, C_1-1)}
        \nonumber\\
        &\qquad\qquad+\frac{1-D_{3a}-2D_{3e}}{4}(w\cdot a)^2k_{+\!}\cdot a+\frac{-15+12(D_{3b}-D_{3d})-4E_3}{12}\omega^2 a^2 w\cdot a \phantom{\Bigg|}
        \nonumber\\
        &\qquad\qquad +\frac{3+3D_{3a}-6(D_{3b}+D_{3c})+2E_3}{12}\omega^2 a^2k_{+\!}\cdot a+\fc O(a^4).
    \label{eq:App_noqsq}
    \end{alignat}
    Similarly, using \eqref{eq:def:aya},
    \begin{alignat}{3}
        \frac{\fc A_{+-}}{\fc A_{+-}^{(0)}}&=1-C_{1\,}q\cdot a+(C_1^2-1)x\cdot a
        \nonumber\\
        &\qquad+\frac{1}{2}(q\cdot a)^2\frac{C_1^2+C_2}{2}+\frac{C_2-C_1^2}{4}(k_{+\!}\cdot a)^2-(C_2-1)C_{1\,}q\cdot a\,x\cdot a
        \nonumber\\
        &\qquad\qquad+\frac{1}{2}\Big((1-C_1)(C_2+2C_1)-D_{2a}+D_{2b}\Big)(x\cdot a)^2
        \nonumber\\
        &\qquad\qquad+\frac{1}{2}\Big(2C_1-C_2-C_1C_2+D_{2a}-D_{2b}-8 D_{2c}-2D_{2d}\Big)\omega^2a^2
        \nonumber\\
        &\qquad-\frac{1}{6}(q\cdot a)^3\frac{3+C_3}{4}+\frac{C_3-1}{4}\Big[(q\cdot a)^2x\cdot a-\frac{1}{2}(k_{+\!}\cdot a)^2(q-2x)\cdot a+\frac{4}{3}\omega^2a^2q\cdot a\Big]\phantom{\bigg|}
        \nonumber\\
        &\qquad\qquad-\frac{1+D_{3a}}{4}(x\cdot a)^2q\cdot a-\frac{3+4(D_{3b}+D_{3d}+E_3)}{4}\omega^2a^2x\cdot a+\hat{\fc A}_{+-}^{(3)(C_2-1,C_1-1)}
        \nonumber\\
        &\qquad\qquad
        +\frac{1+3D_{3a} -6(D_{3b}+D_{3c}-D_{3e}) -24D_{3f}-2E_3}{12}\omega^2a^2q\cdot a+\fc O(a^4).\phantom{\bigg|}
    \label{eq:Apm_noqsq}
    \end{alignat}
    With $C_1=C_2=C_3=1$,
    \begin{alignat}{3}
        &\frac{\fc A_{++}}{\fc A^{(0)}_{++}}=1+(k_{+\!}-2w)\cdot a+\frac{1}{2}[(k_{+\!}-2w)\cdot a]^2
        +\frac{D_{2a}+D_{2b}}{2}\big[\omega^2a^2-(w\cdot a)^2\big]-D_{2d\,}\omega^2a^2
        \nonumber\\\nonumber
        &\qquad\qquad\;\,+\frac{1}{6}(k_{+\!}-4w)\cdot a\,(k_{+\!}-2w)\cdot a\,k_{+\!}\cdot a+\frac{1-D_{3a}-2D_{3e}}{4}(w\cdot a)^2k_{+\!}\cdot a\phantom{\bigg|}
        \\\nonumber
        &\quad+\frac{-15+12(D_{3b} -D_{3d})-4E_3}{12}\omega^2 a^2 w\cdot a +\frac{3+3D_{3a}-6(D_{3b}+D_{3c})+2E_3}{12}\omega^2 a^2k_{+\!}\cdot a
        \nonumber\\
        &+\fc O(a^4),
    \label{eq:App_Ceq1}
    \end{alignat}
    and
    \begin{alignat}{3}
        \frac{\fc A_{+-}}{\fc A_{+-}^{(0)}}&=1-q\cdot a+\frac{1}{2}(q\cdot a)^2
        +\frac{D_{2a}-D_{2b}}{2}\big[\omega^2a^2-(x\cdot a)^2\big]
        +(4 D_{2c}
        +D_{2d})\omega^2a^2
        \nonumber\\
        &\quad-\frac{1}{6}(q\cdot a)^3
        -\frac{1+D_{3a}}{4}(x\cdot a)^2q\cdot a-\frac{3+4(D_{3b}+D_{3d}+E_3)}{4}\omega^2a^2x\cdot a
        \nonumber\\
        &\qquad \quad+\frac{1+3D_{3a}-6(D_{3b}+D_{3c}-D_{3e})-24D_{3f}-2E_3}{12}\omega^2a^2q\cdot a+\fc O(a^4).\phantom{\Bigg|}
    \label{eq:Apm_Ceq1}
    \end{alignat}
    This matches (6.76) of \cite{Cangemi:2023ysz},
    \begin{alignat}{3}
        \frac{\fc A_{++}}{\fc A^{(0)}_{++}}&=1+(k_{+\!}-2w)\cdot a+\frac{1}{2}[(k_{+\!}-2w)\cdot a]^2+2\delta\big[\omega^2a^2-(w\cdot a)^2\big]
        \\\nonumber
        &\ \ +\frac{1}{6}(k_{+\!}-2w)\cdot a\Big[(k_{+\!}-4w)\cdot a\, k_{+\!}\cdot a+4\omega^2a^2\Big]+\frac{4}{3}\delta\big[\omega^2a^2-(w\cdot a)^2\big]k_{+\!}\cdot a+\fc O(a^4),
    \end{alignat}
    and $\fc A_{+-}/\fc A_{+-}^{(0)}=e^{-q\cdot a}+\fc O(a^4)$ if
    \begin{alignat}{7}
        D_{2a}&=D_{2b} =2\delta, 
        \qquad &
        D_{2c}&=D_{2d}=0,
        \\\nonumber
        D_{3a}&=-1, \qquad &
        D_{3b}&=-\frac{5+4E_3}{12}, \qquad &
        D_{3c}&=\frac{-11+8E_3-32\delta}{12}, 
        \\\nonumber
        D_{3d}&=-\frac{1+2E_3}{3}, \qquad &
        D_{3e}&=1+\frac{8}{3}\delta, \qquad &
        D_{3f}&=\frac{3-E_3+8\delta}{6}.
    \end{alignat}

\section{Gravitational MPD Equations} \label{sec:grmpd}
	
	We now turn our attention to gravity. We find that the line of analysis is directly analogous to that of electromagnetism and the resulting dynamical mass function is very similar. The motion of a generic spinning body in general relativity is described by the MPD equations~\cite{Mathisson:1937zz, Papapetrou:1951pa, Tulczyjew, Dixon:1970mpd, dixon2, dixon3}. It is well established~\cite{dixon3, Ehlers:1977rud, bailey1975, Vines:2017hyw, Vines:2016unv, JanSteinhoff:2015ist, Marsat:2014xea, Levi:2015msa} that the MPD equations can be derived from a variational principle through an action $\fc{S}$ of the form:
	\begin{equation}
		\fc{S}[z, p, \Lambda, S, \alpha, \beta] = \int_{-\8}^\8\(p_\mu\dot z^\mu + \frac{1}{2}\twid{\epsilon}_{\mu\nu\rho\sigma} u^\mu S^\nu \Omega^{\rho\sigma} - \frac{\alpha}{2}(p^2 + \fc{M}^2) + \beta p\cdot S\)d\lambda	\label{eq:grmpdact}
	\end{equation}
	where the pseudotensor Levi-Civita symbol is defined by:
	\begin{equation}
		\twid{\epsilon}_{\mu\nu\rho\sigma} = \epsilon_{\mu\nu\rho\sigma}\sqrt{-\det g}
	\end{equation}
	in terms of the purely numerical antisymmetric Levi-Civita symbol $\epsilon_{\mu\nu\rho\sigma}$ which has $\epsilon_{0123} = 1$. In curved spacetime, the ${\Lambda^\mu}_A(\lambda)$ tetrad satisfies:
	\begin{equation}
		g^{\mu\nu}(Z) = \Lambda^\mu{}_A\Lambda^\nu{}_B \eta^{AB}, \p \eta_{AB} = g_{\mu\nu}(Z)\Lambda^\mu{}_A\Lambda^\nu{}_B.
	\end{equation}
	Just as in the case of electromagnetism, we take:
	\begin{equation}
		{\Lambda^\mu}_0(\lambda) = u^\mu(\lambda), \p {\Lambda^\mu}_3(\lambda) = \hat S^\mu(\lambda).
	\end{equation}
	The angular velocity tensor of the body is defined by:
	\begin{equation}
		\Omega^{\mu\nu} = \eta^{AB}{\Lambda^\mu}_A\frac{D{\Lambda^\nu}_B}{D\lambda}	\label{eq:angvelgr}
	\end{equation}
	where $\frac{D}{D\lambda}$ indicates covariant $\lambda$ differentiation:
	\begin{equation}
		\frac{D{\Lambda^\mu}_A}{D\lambda} = \frac{d{\Lambda^\mu}_A}{d\lambda} + \Gamma^\mu{}_{\rho\sigma} \dot z^\rho \Lambda^\sigma{}_A.
	\end{equation}
	The dynamical mass function $\fc{M}(z, u, S)$ now encodes the mass of the body and all of its nonminimal couplings to gravity and in particular takes the form:
	\begin{equation}
		\fc{M}^2(z, u, S) = m^2 + \OO(R)
	\end{equation}
	where $m$ is the mass of the body in vacuum and $R_{\mu\nu\rho\sigma}$ is the Riemann tensor.
	
	For variations of the action, it is useful to define the covariant variations:
	\begin{align}
		\Delta p_\mu &= \delta p_\mu - {\Gamma^\rho}_{\sigma \mu} p_\rho \delta z^\sigma \\
		\Delta S^\mu &= \delta S^\mu + {\Gamma^\mu}_{\rho\sigma} S^\rho \delta z^\sigma \\
		\Delta {\Lambda^\mu}_A &= \delta {\Lambda^\mu}_A + \Gamma^\mu{}_{\rho\sigma} \delta z^\rho \Lambda^\sigma{}_A
	\end{align}
	and the antisymmetric tensor:
	\begin{equation}
		\Delta\theta^{\mu\nu} = \eta^{AB}{\Lambda^{[\mu}}_A \Delta\Lambda^{\nu]}{}_B.
	\end{equation}
	This definition leads to the identity:
	\begin{equation}
		\Delta\Omega^{\mu\nu} = \frac{D}{D\lambda}\Delta\theta^{\mu\nu} + {\Omega^\mu}_\rho\Delta\theta^{\rho\nu} - \Omega^\nu{}_\rho\Delta\theta^{\rho\mu} + R^{\mu\nu}{}_{\rho\sigma}\dot z^\rho \delta z^\sigma.
	\end{equation}
	Then, the variation of the above action gives:
	\begin{align}
		\delta \fc{S} = \int_{-\8}^\8&\(\delta z^\mu\( -\frac{Dp_\mu}{D\lambda} - R^\star_{\mu\nu\rho\sigma} \dot z^\nu u^\rho S^\sigma - \ec \nabla_\mu\fc{M}\)\right.\nonumber \\
		&+\Delta p_\mu\(\dot z^\mu - \ec u^\mu - \frac{\ec}{|p|}\frac{\d\fc{M}}{\d u^\nu}(g^{\mu\nu}+u^\mu u^\nu) + \beta S^\mu\right. \nonumber \\
        &\p \p \p\left.+ \frac{1}{2|p|}\twid{\epsilon}^{\mu\nu\rho\sigma}S_\nu\Omega_{\rho\sigma} + \frac{u^\mu}{2|p|}\twid{\epsilon}_{\alpha\beta\rho\sigma} u^\alpha S^\beta \Omega^{\rho\sigma}\) \nonumber \\
		& +\frac{1}{2}\Delta\theta^{\rho\sigma}\(-\frac{D}{D\lambda}\(\twid{\epsilon}_{\mu\nu\rho\sigma} u^\mu S^\nu\) + \twid{\epsilon}_{\mu\nu\rho\alpha} u^\mu S^\nu \Omega^\alpha{}_\sigma - \twid{\epsilon}_{\mu\nu\sigma\alpha} u^\mu S^\nu \Omega^\alpha{}_\rho\) \nonumber \\
		& \left.+\Delta S^\mu\(-\frac{1}{2}\twid{\epsilon}_{\mu\nu\rho\sigma} u^\nu \Omega^{\rho\sigma} - \ec\frac{\d\fc{M}}{\d S^\mu} + \beta p_\mu\) - \frac{\delta\alpha}{2}\(p^2+\fc{M}^2\) + \delta\beta p\cdot S\) d\lambda
	\end{align}
	where the right-dual of the Riemann tensor is defined by:
	\begin{equation}
		R^\star_{\mu\nu\rho\sigma} = \frac{1}{2}\twid{\epsilon}_{\rho\sigma}{}^{\alpha\beta} R_{\mu\nu\alpha\beta}.
	\end{equation}
	Using the $\delta S^\mu$ variation to solve for the angular velocity tensor, one can then determine the value of $\beta$. That value of $\beta$ can then be used to simplify the spin and trajectory equations of motion. Explicitly, these give:
	\begin{align}
		\Omega^{\mu\nu} &= \frac{Du^\mu}{D\lambda} u^\nu - u^\mu \frac{Du^\nu}{D\lambda} + \ec \twid{\epsilon}^{\mu\nu\rho\sigma} u_\rho \frac{\d\fc{M}}{\d S^\sigma} \label{eq:angveleomgr} \\
		\beta &= -\frac{\ec}{\fc{M}}u^\mu \frac{\d\fc{M}}{\d S^\mu}\\
		\frac{DS^\mu}{D\lambda} &= u^\mu \frac{Du^\nu}{D\lambda} S_\nu + \ec \epsilon^{\mu\nu\rho\sigma} u_\nu S_\rho \frac{\d\fc{M}}{\d S^\sigma} \\
		\dot z^\mu &= \ec u^\mu + \frac{\ec}{\fc{M}}(g^{\mu\nu} + u^\mu u^\nu)\frac{\d\fc{M}}{\d u^\nu} + \frac{\ec}{\fc{M}} S^\mu u^\nu \frac{\d\fc{M}}{\d S^\nu} + \frac{1}{\fc{M}^2} \twid{\epsilon}^{\mu\nu\rho\sigma} S_\nu u_\rho \frac{Dp_\sigma}{D\lambda}.	\label{eq:velsol1gr}
	\end{align}
	In order to determine the trajectory evolution explicitly we must insert the momentum equation of motion into \eqref{eq:velsol1gr}. To simplify, it is useful to introduce the two sided dual Riemann tensor:
	\begin{equation}
		{}^\star R^\star_{\mu\nu\rho\sigma} = \frac{1}{2}\twid{\epsilon}_{\mu\nu\alpha\beta} R^{\star\alpha\beta}{}_{\rho\sigma}
	\end{equation}
	Simplifying finally gives the gravitational MPD equations of motion for the spinning body:
	\begin{align}
		\(1 + \frac{{}^\star R^\star_{uSuS}}{\fc{M}^2}\)\frac{\dot z^\mu}{\ec} =&\ u^\mu + \frac{g^{\mu\nu} + u^\mu u^\nu}{\fc{M}}\frac{\d\fc{M}}{\d u^\nu} + S^\mu\frac{u^\nu}{\fc{M}}\frac{\d\fc{M}}{\d S^\nu} + \frac{1}{\fc{M}^2} \twid{\epsilon}^{\mu\nu\rho\sigma} u_\nu S_\rho\nabla_\sigma\fc{M} \nonumber \\
		& - \frac{{}^\star R^{\star\mu S u S}}{\fc{M}^2} - \frac{1}{\fc{M}^3}\(S^\nu \frac{\d\fc{M}}{\d u^\nu} + S^2 u^\nu \frac{\d\fc{M}}{\d S^\nu}\) {}^\star R^{\star\mu u u S}\\
		\frac{Dp_\mu}{D\lambda} =&\ -R^\star_{\mu\nu\rho\sigma} \dot z^\nu u^\rho S^\sigma - \ec \nabla_\mu\fc{M} \\
		\frac{DS^\mu}{D\lambda} =&\ u^\mu S_\nu\frac{Du^\nu}{D\lambda} + \ec \twid{\epsilon}^{\mu\nu\rho\sigma} u_\nu S_\rho \frac{\d\fc{M}}{\d S^\sigma}.
	\end{align}
    Vectors such as $u$ and $S$ are used as indices to indicate contractions with them (${}^\star R^{\star\mu uu S} = {}^\star R^{\star\mu\nu\rho\sigma} u_\nu u_\rho S_\sigma$). For solving these equations of motion we will always choose $\lambda$ so that $\ec = 1$. 
	
	To understand how the dynamical mass function relates to the multipole moments of the body, we will need to study the energy-momentum tensor produced by our action. The energy-momentum tensor will be given by:
	\begin{equation}
		T_{\mu\nu} = -\frac{2}{\sqrt{-\det g}} \frac{\delta \fc{S}}{\delta g^{\mu\nu}}.
	\end{equation}
	In varying the metric, we have $\delta z^\mu = 0$, $\delta p_\mu = 0$, $\delta S^\mu = 0$, and $\Delta \theta^{\mu\nu} = 0$. ${\delta \Lambda}^\mu{}_A$ cannot be made 0 as its variation is related to the metric variation on the worldline. It is useful to introduce the DeWitt index shuffling operator~\cite{JanSteinhoff:2015ist} ${\hat G}^\alpha{}_\beta$, which acts on tensors according to:
	\begin{align}
		\hat{G}^{\alpha}{}_\beta F^{\mu_1...\mu_m}{}_{\nu_1...\nu_n} =& \delta^{\mu_1}_\beta F^{\alpha \mu_2...\mu_m}{}_{\nu_1...\nu_n} + ... + \delta^{\mu_m}_\beta F^{\mu_1...\mu_{m-1}\alpha} {}_{\nu_1...\nu_n} \nonumber \\
		&\ \ -\delta^{\alpha}_{\nu_1} F^{\mu_1...\mu_m}{}_{\beta \nu_2...\nu_n} - ... - \delta^\alpha_{\nu_n} F^{\mu_1...\mu_m}{}_{\nu_1...\nu_{n-1} \beta}
	\end{align}
	so that:
	\begin{equation}
		\nabla_\rho F^{\mu_1...\mu_m}{}_{\nu_1...\nu_n} = \d_\rho F^{\mu_1...\mu_m}{}_{\nu_1...\nu_n} + \Gamma^\beta{}_{\alpha\rho} \hat G^\alpha{}_\beta F^{\mu_1...\mu_m}{}_{\nu_1...\nu_n}.
	\end{equation}
	For expressing the energy-momentum tensor simply it is important to use the scalarity of the dynamical mass function. In particular, requiring it to be a scalar function of $u_\mu$, $S^\mu$, $g_{\mu\nu}$, and symmetric covariant derivatives of $R_{\mu\nu\rho\sigma}$ implies it is invariant under a small diffeomorphism. This is only true if:
	\begin{equation}
		\frac{\d\fc{M}}{\d g^{\alpha\beta}} = \frac{1}{2}\frac{\d\fc{M}}{\d u^\alpha} u_\beta - \frac{1}{2}S_\alpha\frac{\d\fc{M}}{\d S^\beta} - \frac{1}{2}\sum_{n=0}^\8 \frac{\d\fc{M}}{\d \nabla^n_{(\lambda_1...\lambda_n)} R_{\mu\nu\rho\sigma}} \hat G_{\alpha\beta} \nabla^n_{(\lambda_1...\lambda_n)} R_{\mu\nu\rho\sigma}.
	\end{equation}
	With these ingredients, the variation of the action with respect to the metric becomes:
	\begin{align}
		\delta \fc{S} &= \int_{-\8}^\8\(-\frac{1}{2} p_\mu \dot z_\nu \delta g^{\mu\nu} + \frac{1}{2}\twid{\epsilon}_{\mu\nu\rho\sigma} u^\mu S^\nu \dot z_\alpha \nabla^\sigma \delta g^{\rho\alpha}\right. \nonumber \\
		&\p \left.+\ec \sum_{n=0}^\8 \frac{\d\fc{M}}{\d\nabla^n_{(\lambda_1...\lambda_n)} R_{\mu\nu\rho\sigma}}\(\frac{1}{2}\delta g^{\alpha\beta}\hat G_{\alpha\beta} \nabla^n_{(\lambda_1...\lambda_n)} R_{\mu\nu\rho\sigma} - \delta(\nabla^n_{(\lambda_1...\lambda_n)}R_{\mu\nu\rho\sigma})\)\) d\lambda
	\end{align}

    For computing variations of derivatives of the Riemann curvature, a strategy from Ref.~\cite{wald846} is helpful. Consider a tensor field $F^{\lambda_1...\lambda_n \mu\nu\rho\sigma}$ with the same index symmetries as $\nabla^n_{(\lambda_1...\lambda_n)} R_{\mu\nu\rho\sigma}$ which decays to 0 sufficiently quickly at infinity for no surface terms to be necessary upon the relevant integrations by parts we will perform. Then, a short calculation gives:
    \begin{align}
        &\int_{\bb{T}} F^{\lambda_1...\lambda_n\mu\nu\rho\sigma}\delta(\nabla^n_{\lambda_1...\lambda_n} R_{\mu\nu\rho\sigma}) D^4 x \nonumber \\
        &\ = \int_{\bb{T}}\(F^{\lambda_1...\lambda_n\mu\nu\rho\sigma}\delta\Gamma^\beta{}_{\alpha\lambda_n}\hat G^\alpha{}_\beta\nabla^{n-1}_{\lambda_1...\lambda_{n-1}}R_{\mu\nu\rho\sigma} - \nabla_{\lambda_n} F^{\lambda_1...\lambda_n\mu\nu\rho\sigma} \delta(\nabla^{n-1}_{\lambda_1...\lambda_{n-1}}R_{\mu\nu\rho\sigma})\) D^4 x.
    \end{align}
    The final term is now in the same form as the initial variational problem, but of a lower rank. Applying this formula iteratively allows all derivatives on the Riemann tensor to eventually be pushed past the variation, resulting in:
    \begin{align}
        &\int_{\bb{T}} F^{\lambda_1...\lambda_n\mu\nu\rho\sigma}\delta(\nabla^n_{\lambda_1...\lambda_n} R_{\mu\nu\rho\sigma}) D^4 x \nonumber \\
        &\p = \int_{\bb{T}}\(\sum_{k=0}^{n-1}(-1)^k\nabla^k_{\lambda_n...\lambda_{n-k+1}} F^{\lambda_1...\lambda_n\mu\nu\rho\sigma} \delta \Gamma^\beta{}_{\alpha\lambda_{n-k}}\hat G^\alpha{}_\beta\nabla^{n-k-1}_{\lambda_1...\lambda_{n-k-1}} R_{\mu\nu\rho\sigma}\right.\nonumber \\
        &\p \p \p \left.+ (-1)^n \nabla^n_{\lambda_1...\lambda_n}F^{\lambda_1...\lambda_n\mu\nu\rho\sigma}\delta R_{\mu\nu\rho\sigma}\)D^4 x.
    \end{align}
    Now using:
    \begin{align}
        \delta \Gamma^\rho{}_{\mu\nu} &= \frac{1}{2}g_{\mu\alpha}g_{\nu\beta}\nabla^\rho \delta g^{\alpha\beta} - \frac{1}{2}g_{\mu\sigma}\nabla_\nu \delta g^{\rho\sigma} - \frac{1}{2}g_{\nu\sigma}\nabla_\mu \delta g^{\rho\sigma} \\
        \delta R^\rho{}_{\mu\sigma\nu} &= \nabla_\sigma\delta \Gamma^\rho{}_{\nu\mu} - \nabla_\nu\delta \Gamma^\rho{}_{\sigma\mu}
    \end{align}
    we are able to arrive at:
	\begin{align}
        &\int F^{\lambda_1...\lambda_n\mu\nu\rho\sigma}\delta(\nabla^n_{(\lambda_1...\lambda_n)} R_{\mu\nu\rho\sigma}) D^4 x \nonumber \\
        &\ = \int\(\sum_{k=0}^{n-1} (-1)^k \delta g^{\alpha\beta} g_{\alpha\lambda_{n-k}}\nabla^\gamma\nabla^k_{(\lambda_{n-k+1}...\lambda_n)} F^{\lambda_1...\lambda_n \mu\nu\rho\sigma}\hat G_{[\gamma\beta]}(\nabla^{n-k-1}_{(\lambda_1...\lambda_{n-k-1})} R_{\mu\nu\rho\sigma})\right. \nonumber \\
        &\ps\ps +\sum_{k=0}^{n-1}(-1)^k \delta g^{\alpha\beta}g_{\alpha\lambda_{n-k}}\nabla^k_{(\lambda_{n-k+1}...\lambda_n)} F^{\lambda_1...\lambda_n\mu\nu\rho\sigma} g^{\gamma\tau}\hat G_{[\tau\beta]}(\nabla_\gamma \nabla^{n-k-1}_{(\lambda_1...\lambda_{n-k-1})} R_{\mu\nu\rho\sigma}) \nonumber \\
        &\ps\ps +\sum_{k=0}^{n-1}2(-1)^k \delta g^{\alpha\beta}g_{\alpha\lambda_{n-k}}\nabla^k_{(\lambda_{n-k+1}...\lambda_n)} F^{\lambda_1...\lambda_n\mu\nu\rho\sigma} \nabla_\beta\nabla^{n-k-1}_{(\lambda_1...\lambda_{n-k-1})}R_{\mu\nu\rho\sigma} 
        \nonumber \\
        &\ps\ps +(-1)^n \delta g^{\alpha\beta}\nabla^n_{(\lambda_1...\lambda_n)} F^{\lambda_1...\lambda_n\mu\nu\rho\sigma}g_{\alpha\mu} R_{\beta\nu\rho\sigma} + 2(-1)^n\nabla^2_{(\mu\nu)} \nabla^n_{(\lambda_1...\lambda_n)} F^{\lambda_1...\lambda_n\mu}{}_{\alpha}{}^\nu{}_\beta\delta g^{\alpha\beta} 
        \nonumber \\
        &\ps\ps \left. + \frac{1}{2}\delta g^{\alpha\beta} F^{\lambda_1...\lambda_n\mu\nu\rho\sigma}\hat G_{\alpha\beta}(\nabla^n_{\lambda_1...\lambda_n}R_{\mu\nu\rho\sigma})\) D^4x.
    \end{align}
    Define the gravitational $\fc{Q}_n$ moments:
    \begin{equation}
        \fc{Q}_n^{\lambda_1...\lambda_n\mu\nu\rho\sigma} = \frac{\d\fc{M}}{\d \nabla^n_{(\lambda_1...\lambda_n)}R_{\mu\nu\rho\sigma}}.
    \end{equation}
    As well, define the scalar Dirac delta distribution:
    \begin{equation}
        \delta(X,Z) = \frac{\delta(x-z)}{\sqrt{-\det g}}
    \end{equation}
    and the $\Phi_n$ fields:
    \begin{equation}
        \Phi_n^{\lambda_1...\lambda_n\mu\nu\rho\sigma}(X) = \int_{-\8}^\8 \fc{Q}_n^{\lambda_1...\lambda_n\mu\nu\rho\sigma} \delta(X,Z) D\lambda.
    \end{equation}
    Then, formally as a distributional expression the energy-momentum tensor becomes:
    \begin{equation}
        T_{\alpha\beta} = \int_{-\8}^\8 \(\dot z_{(\alpha} p_{\beta)} \delta(X,Z) + \nabla^\gamma(\dot z_{(\alpha} \twid{\epsilon}_{\beta)\gamma\rho\sigma} u^\rho S^\sigma\delta(X,Z))\)d\lambda + \sum_{n=0}^\8 \Psi^{(n)}_{\alpha\beta}  \label{eq:stress}
    \end{equation}
    with:
    \begin{align}
        \Psi^{(n)}_{\alpha\beta} =&\ \sum_{k=0}^{n-1}(-1)^k \nabla^k_{\lambda_{n-k+1}...\lambda_n}\Phi_n^{\lambda_1...\lambda_n\mu\nu\rho\sigma} g_{\lambda_{n-k}(\alpha}\nabla_{\beta)}\nabla^{n-k-1}_{\lambda_1...\lambda_{n-k-1}} R_{\mu\nu\rho\sigma} \nonumber \\
        & + \sum_{k=0}^{n-1}2(-1)^k \nabla_\gamma\(\nabla^k_{\lambda_{n-k+1}...\lambda_n}\Phi_n^{\lambda_1...\lambda_n\mu\nu\rho\sigma} g_{\lambda_{n-k}(\alpha}g_{\beta)\delta} \hat G^{[\gamma\delta]}(\nabla^{n-k-1}_{\lambda_1...\lambda_{n-k-1}}R_{\mu\nu\rho\sigma})\) \nonumber \\
        &+2(-1)^n \nabla^n_{\lambda_1...\lambda_n} \Phi_n^{\lambda_1...\lambda_n\mu\nu\rho\sigma} g_{\mu(\alpha} R_{\beta)\nu\rho\sigma} + 4(-1)^n \nabla^2_{\mu\nu} \nabla^n_{\lambda_1...\lambda_n} \Phi^{\lambda_1...\lambda_n}_n{}^\mu{}_{(\alpha}{}^\nu{}_{\beta)}.
    \end{align}

\section{Kerr Multipole Moments} \label{sec:kerrmoments}

    In this section we perform the analysis of sections \ref{sec:dixon} and \ref{sec:rkmoments} but for the Kerr metric in gravity instead of the \rk solution in electromagnetism. We begin by describing Dixon's definition of the multipoles of the energy-momentum tensor. Then, following analysis done by Israel in Ref.~\cite{israel:1970} we identify the energy-momentum tensor which acts as the source of the Kerr metric in the causally maximal extension of the Kerr spacetime.
 
	\subsection{Moments of the Energy Momentum Tensor}

    For precisely the same reasons that cause the naive moments of the current density to be interdependent due to the continuity equation, the naive moments of the energy-momentum tensor are interdependent due to its covariant conservation. Define the quantities:
    \begin{align}
        \Theta^{\kappa\lambda\mu\nu}_n(Z,X) &= (n-1)\int_0^1 \sigma^{\kappa\alpha}\sigma^{(\mu}{}_\alpha \sigma^{\nu)}{}_\beta \sigma^{\lambda\beta} t^{n-2}dt, \ps (n\ge 2) \\
        \mc{p}^{\kappa_1...\kappa_n\lambda\mu\nu}_n &= 2(-1)^n\int_\Sigma \sigma^{\kappa_1}...\sigma^{\kappa_n} \Theta^{\rho\nu\lambda\mu}_n (-\sigma^{-1}_{\alpha\rho})T^{\alpha\beta} d\Sigma_\beta, \ps (n \ge 2) \\
        \mc{t}^{\kappa_1...\kappa_n\lambda\mu}_n &= (-1)^n\int_\Sigma \sigma^{\kappa_1}...\sigma^{\kappa_n}{\sigma^\lambda}{}_\alpha \sigma^\mu{}_\beta T^{\alpha\beta} w^\gamma d\Sigma_\gamma, \ps (n \ge 2) \\
        \fc{J}^{\kappa_1...\kappa_n\lambda\mu\nu\rho}_n &= \mc{t}^{\kappa_1...\kappa_n[\lambda['\nu\mu]\rho]'}_{n+2} + \frac{1}{n+1}\mc{p}^{\kappa_1...\kappa_n[\lambda['\nu\mu]\rho]'\tau}_{n+2} \frac{\dot z_\tau}{\ec} \\
        I^{\lambda_1...\lambda_n\mu\nu}_n &= \frac{4(n-1)}{n+1} \fc{J}^{(\lambda_1...\lambda_{n-1}|\mu|\lambda_n)\nu}_{n-2}, \ps (n \ge 2).
    \end{align}
    These definitions are precisely analogous to equations \eqref{eq:Thetadef} through \eqref{eq:reducedef} for electromagnetism, in precisely the same order. The $I_n$ moments will serve as the interdependence-free reduced multipole moments of the energy-momentum tensor. The other quantities defined are useful intermediate pieces for calculation. The four-index antisymmetrization $[\lambda['\nu\mu]\rho]'$ is defined so that $\lambda,\mu$ are antisymmetrized and $\nu,\rho$ are independently antisymmetrized:
    \begin{equation}
        I_2^{[\lambda['\nu\mu]\rho]'} = \frac{1}{4}\(I_2^{\lambda\nu\mu\rho}-I_2^{\mu\nu\lambda\rho}-I_2^{\lambda\rho\mu\nu}+I_2^{\mu\rho\lambda\nu}\).
    \end{equation}
    As well, the inclusion of vertical bars $|\mu|$ around indices in the midst of a symmetrization indicates that those indices should be skipped over when performing the symmetrization. For example:
    \begin{equation}
        I_1^{(\lambda|\mu|\nu)} = \frac{1}{2}(I_1^{\lambda\mu\nu} + I_1^{\nu\mu\lambda}).
    \end{equation}
    
    Dixon's reduced moments automatically satisfy:
	\begin{gather}
		I_n^{\lambda_1...\lambda_n\mu\nu} = I_n^{(\lambda_1...\lambda_n)(\mu\nu)}, \p I^{(\lambda_1...\lambda_n\mu)\nu}_n = 0, \label{eq:Inindex1}\\
		u_{\lambda_1} I^{\lambda_1...\lambda_{n-2}[\lambda_{n-1}['\lambda_n \mu] \nu]'}_n = 0, \p (n\ge 3).	\label{eq:Inindex2}
	\end{gather}
	Dixon finds that beyond these conditions, the reduced moments are not restricted by the covariant conservation of $T^{\alpha\beta}$ and that they are independent of each other for different values of $n$. It is useful to define the $0^\text{th}$ and $1^\text{st}$ moments:
	\begin{equation}
        I_{0}^{\lambda\mu} = \frac{p^{(\lambda}\dot z^{\mu)}}{\ec}, \p I_{1}^{\kappa\lambda\mu} = \frac{S^{\kappa(\lambda} \dot z^{\mu)}}{\ec}
	\end{equation}
    where $p^\lambda$ is the total linear momentum of the body defined by:
    \begin{equation}
        p^\kappa = \int_\Sigma (-\sigma^{-1}_{\alpha\lambda})\sigma^{\lambda\kappa} T^{\alpha\beta}d\Sigma_\beta    \label{eq:grmomdef}
    \end{equation}
    and $S^{\lambda\mu}$ is the total spin tensor of the body defined by:
    \begin{equation}
        S^{\kappa\lambda} = 2\int_\Sigma \sigma^{[\kappa}(\sigma^{-1})^{\lambda]}{}_\alpha T^{\alpha\beta} d\Sigma_\beta.    \label{eq:grspindef}
    \end{equation}
    We will always choose our definition of the worldline and Cauchy slicing so that $p^\mu(\lambda)$ is orthogonal to all tangent vectors to $\Sigma(\lambda)$. With this choice we can require:
    \begin{equation}
        p^\mu = |p| u^\mu, \p S^{\mu\nu}p_\nu = 0
    \end{equation}
    and thus we can introduce the spin vector $S^\mu$ defined so that:
    \begin{equation}
        S^\mu = -\frac{1}{2}\twid{\epsilon}^{\mu\nu\rho\sigma}u_\nu S_{\rho\sigma} \implies S_{\mu\nu} = \twid{\epsilon}_{\mu\nu\rho\sigma}u^\rho S^\sigma.
    \end{equation}
	The conservation of $T^{\alpha\beta}$ determines the time evolution of $p^\mu$ and $S^\mu$ (through the MPD equations) but determines nothing about the time evolution of the higher moments. Dixon finds also that the reduced moments are independent of the $0^\text{th}$ and $1^\text{st}$ moments. Then, define the reduced moment generating function:
	\begin{equation}
		I^{\mu\nu}(\lambda, k) = \sum_{n=0}^\8 \frac{(-i)^n}{n!} I^{\lambda_1...\lambda_n \mu\nu}_n k_{\lambda_1}...k_{\lambda_n}.
	\end{equation}
	Like the naive moments, the reduced moment generating functions determine the behavior of $T^{\alpha\beta}$ against test functions. In particular, for an arbitrary symmetric tensor field $h_{\alpha\beta}(X)$:
	\begin{equation}
		\int_\Sigma h^*_{\alpha\beta}(X) T^{\alpha\beta}(X) w^\gamma d\Sigma_\gamma = \int \twid{h}^*_{\mu\nu}(Z, k) I^{\mu\nu}(\lambda, k) \frac{D^4k}{(2\pi)^2}.	  \label{eq:intstress}
	\end{equation}
	The moment generating function automatically satisfies:
	\begin{equation}
        I^{\lambda\mu} k_\lambda = I^{\lambda\mu}_0 k_\lambda - i k_\kappa k_\lambda I^{\kappa\lambda\mu}_1.
	\end{equation}
	Dixon proved~\cite{dixon2} that these reduced moments are the unique set of moments which are independent of each other for different $n$, have only $I_0$ and $I_1$ restricted by the conservation law, and satisfy the index symmetry conditions in equations \eqref{eq:Inindex1} and \eqref{eq:Inindex2}.

    Through \eqref{eq:intstress}, the energy-momentum tensor is determined in terms of the reduced multipole moments. Explicitly comparing that behavior against test functions to \eqref{eq:stress} and using crucially that the reduced multipole moments are unique and contain no interdependencies, we can identify:
    \begin{equation}
        I_n^{\rho_1...\rho_n\mu\nu} = 4 n! \fc{Q}_{n-2}^{(\rho_1...\rho_{n-1}|\mu|\rho_{n})\nu} + \OO(R)
    \end{equation}
    which gives the reduced multipole moments from the couplings in the action. Alternatively, this can be nicely inverted using the index symmetry conditions of both quantities to find:
    \begin{equation}
        \fc{Q}_n^{\lambda_1...\lambda_n\rho\mu\sigma\nu} = \frac{n+1}{(n+3)!} I_{n+2}^{\lambda_1...\lambda_n [\rho['\sigma\mu]\nu]'} + \OO(R)   \label{eq:grmomrecovery}
    \end{equation}
    This allows the direct determination of the coupling of the body to the Riemann tensor in the action from its reduced multipole moments. 
	
	\subsection{Source of the Kerr Metric}

    Here we summarize the analysis of Ref.~\cite{israel:1970} to identify the energy-momentum tensor which produces the Kerr metric. The Kerr metric has no intrinsic singularities away from $r = 0$ and everywhere away from $r = 0$ it is a solution to the vacuum Einstein equations. Therefore, the source of the Kerr metric can only have support on the surface $r = 0$. This surface is a disk and we introduce the same coordinate $\chi = \theta$ above the disk and $\chi = \pi - \theta$ below the disk as we did for \rk so that $t, \chi, \varphi$ provide an intrinsic coordinate system. Restricting the Kerr solution to the surface $r = 0$ produces the flat metric $\gamma_{ij}$ on the disk:
    \begin{align}
        &\gamma_{tt} = -1,& &\gamma_{t\chi} = 0,& &\gamma_{t\varphi} = 0, \nonumber \\
        &\gamma_{\chi\chi} = a^2\cos^2\chi,& &\gamma_{\chi\varphi} = 0,& &\gamma_{\varphi\varphi} = a^2\sin^2\chi.
    \end{align}
    The extrinsic curvature $K_{ij}$ of the disk, when approached from above toward $r=0$, is determined by the first $r$ derivative of the Kerr solution and produces:
    \begin{align}
        &K_{tt} = \frac{G m}{a^2 \cos^3\chi},& &K_{t\chi} = 0,& &K_{t\varphi} = -\frac{Gm}{a} \frac{\sin^2\chi}{\cos^3\chi}, \nonumber \\
        &K_{\chi\chi} = 0,& &K_{\chi\varphi} = 0,& &K_{\varphi \varphi} = Gm\frac{\sin^4\chi}{\cos^3\chi}.
    \end{align}
    The extrinsic curvature tensor of the disk when approached from below is simply the negative of the extrinsic curvature when approached from above. We may now use Israel's junction conditions to determine the surface energy-momentum tensor on the disk $\text{S}^{\text{disk}}_{ij}$:
    \begin{equation}
        \text{S}^{\text{disk}}_{ij} = -\frac{1}{8\pi G}\left.\(K_{ij} - K \gamma_{ij}\)\right|^{\text{above}}_{\text{below}}.
    \end{equation}
    The resulting surface energy-momentum tensor is:
    \begin{equation}
        \text{S}^{\text{disk}}_{ij} = \frac{\sigma_\text{disk}}{2}(\zeta_i\zeta_j+\xi_i\xi_j)
    \end{equation}
    where:
    \begin{align}
        & & &\sigma_\text{disk} = - \frac{m}{2\pi a^2\cos^3\chi}& & \\
        &\zeta_t = 0, & &\zeta_\chi = a\cos^2 \chi,& & \zeta_\varphi = 0 \\
        &\xi_t = -\sin\chi,& &\xi_\chi = 0,& &\xi_\varphi = a\sin\chi.
    \end{align}
    As a distribution the disk energy-momentum tensor is:
    \begin{equation}
        T^{\text{disk}}_{\mu\nu} = \frac{\sigma_{\text{disk}}}{2}(\zeta_\mu\zeta_\nu + \xi_\mu\xi_\nu) \delta(r\cos\chi).
    \end{equation}
    Taking the worldline which passes through the center of the disk as the worldline of the metric, we can use equations \eqref{eq:grmomdef} and \eqref{eq:grspindef} to compute the total momentum and spin of the Kerr solution. With only the given surface energy-momentum tensor, the resultant linear momentum and spin are not $m u^\mu$ and $m\epsilon_{\mu\nu\rho\sigma} u^\rho a^\sigma$. Instead, the integrals diverge as $\chi\to \frac{\pi}{2}$ in precisely the same way as occurred for \rk. In order to produce the correct total momentum and spin it is necessary to have a linear energy-momentum tensor density on the ring singularity at $r =0, \chi = \frac{\pi}{2}$. In particular, the necessary effective mass density is:
    \begin{equation}
        \rho = -\frac{m}{2\pi a^2\cos^4\chi}\delta(r)\vartheta\(\frac{\pi}{2}-\eps-\chi\) + \frac{m}{2\pi a^2\sin\eps\cos^2\chi}\delta(r)\delta\(\chi-\frac{\pi}{2}\)
    \end{equation}
    which is the same density as the \rk solution. The resulting energy-momentum tensor is:
    \begin{equation}
        T_{\mu\nu} = \frac{\rho}{2}(\zeta_\mu\zeta_\nu+\xi_\mu\xi_\nu).
    \end{equation}
    With this energy-momentum tensor using \eqref{eq:grmomdef} and \eqref{eq:grspindef} we have precisely:
    \begin{equation}
        p^\mu = m u^\mu, \p S^\mu = m a^\mu.
    \end{equation}

	\subsection{Stationary Multipole Moments of Kerr}

    We now consider the Minkowski space limit of Dixon's moments for the energy-momentum tensor. Using the same $y^a$ coordinates as before, we find:
    \begin{align}
        \mc{p}^{\kappa_1...\kappa_n\lambda\mu\nu}_n &= 2{\Lambda^{\kappa_1}}_{A_1}...{\Lambda^{\kappa_n}}_{A_n} \eta^{\nu(\mu}\int_\Sigma (-T^{\lambda)\rho} u_\rho) y^{A_1}...y^{A_n} d^3 y + \OO(R) \\
        \mc{t}^{\kappa_1...\kappa_n\lambda\mu}_n &= {\Lambda^{\kappa_1}}_{A_1}...{\Lambda^{\kappa_n}}_{A_n}\int_\Sigma y^{A_1}...y^{A_n} T^{\lambda\mu} d^3 y + \OO(R).
    \end{align}
    Just like with the current density, we now define the naive $K$ moments:
    \begin{equation}
        K_n^{a_1...a_nBC} = \int_\Sigma y^{a_1}...y^{a_n}{\Lambda_\alpha}^B{\Lambda_\beta}^C T^{\alpha\beta} d^3 y.
    \end{equation}
    In terms of these moments, we have:
    \begin{align}
        \mc{p}^{\kappa_1...\kappa_n\lambda\mu\nu}_n &= 2 {\Lambda^{\kappa_1}}_{a_1}...{\Lambda^{\kappa_n}}_{a_n} \eta^{\nu(\mu} {\Lambda^{\lambda)}}_B K_n^{a_1...a_n B0} + \OO(R)\\
        \mc{t}^{\kappa_1...\kappa_n\lambda\mu}_n &= {\Lambda^{\kappa_1}}_{a_1}...{\Lambda^{\kappa_n}}_{a_n} {\Lambda^\lambda}_B{\Lambda^\mu}_C K_n^{a_1...a_nBC} + \OO(R).
    \end{align}
    For arbitrary vectors on the internal Lorentz indices, $k^A$ and $v^A$ we have:
    \begin{equation}
        K_n^{a_1...a_n BC} k_{a_1}...k_{a_n} v_B v_C = \int_{\Sigma}(x^a k_a)^n \frac{\sigma(\mc{r})}{2}\delta(z)((\zeta^A v_A)^2+ (\xi^A v_A)^2) d^3 x.
    \end{equation}
    Following the same integrations as for \rk allows these moments to be computed with no additional complications giving:
    \begin{align}
        K_{2n}^{a_1...a_{2n} BC} k_{a_1}...k_{a_{2n}} v_B v_C &= m \frac{n+1}{2n+1} |\vec k\times\vec a|^{2n} v_0^2 + m\frac{n}{2n+1}|\vec k\times\vec a|^{2n-2}(\vec a\cdot(\vec k\times\vec v))^2 \\
        K_{2n+1}^{a_1...a_{2n+1}BC}k_{a_1}...k_{a_{2n+1}} v_B v_C &= m v_0 |\vec a\times\vec k|^{2n} \vec a\cdot(\vec k\times\vec v).
    \end{align}

	
	\subsection{Dynamical Multipole Moments of Kerr}

    Now that we have the stationary moments of the Kerr solution, we can use these to compute the dynamical moments of a spinning black hole, up to corrections of order of the Riemann tensor, exactly analogously to the calculation for \rk. We can compute the reduced moments of the energy-momentum tensor by returning the stationary results through the chain of definitions defining $I^{\rho_1...\rho_n\mu\nu}_n$. Using the same notation as for electromagnetism, we find:
    \begin{align}
        I_{2n}^{\rho_1...\rho_{2n}\mu\nu}k_{\rho_1}...k_{\rho_{2n}} v_{\mu}v_\nu =&\ \frac{n+1}{2n+1} ma^{2n}k_\perp^{2n-4}\(k_\perp^4 v_0^2 - 2k_\perp^2 (k_\perp\cdot v) k_0v_0 + \frac{k_\perp^2 v_\perp^2}{2n-1}k_0^2\right.\nonumber \\
        &\ \left. + 2 \frac{n-1}{2n-1}(k_\perp\cdot v)^2k_0^2 + \frac{n}{n+1}k_\perp^2\(\epsilon_{\mu\nu\rho\sigma}u^\mu \hat a^\nu k^\rho v^\sigma\)^2\) + \OO(R) \label{eq:evenmoments} \\
        I_{2n+1}^{\rho_1...\rho_{2n+1}\mu\nu}k_{\rho_1}...k_{\rho_{2n+1}}v_{\mu}v_{\nu} =&\ ma^{2n+1}k_\perp^{2n-2}\epsilon_{\mu\nu\rho\sigma}u^\mu\hat a^\nu k^\rho v^\sigma(k_\perp^2 v_0 - (k_\perp\cdot v)k_0) + \OO(R). \label{eq:oddmoments}
    \end{align}

    By returning these to \eqref{eq:grmomrecovery}, we find that:
    \begin{align}
        \fc{Q}_{2n}^{\lambda_1...\lambda_{2n}\rho\mu\sigma\nu}\nabla^{2n}_{(\lambda_1...\lambda_{2n})} R_{\rho\mu\sigma\nu} = \frac{m a^{2n+2} \nabla_\perp^{2n-2}}{(2n+3)!}&\((n+2)\nabla^2_\perp \perp^{\rho\sigma}R_{\rho u \sigma u}\right. \nonumber \\
        &\ \  + \frac{n+1}{2} \nabla_\perp^2 \perp^{\rho\sigma}\perp^{\mu\nu}R_{\rho\mu\sigma\nu} \nonumber \\
        &\left.\ \ - \frac{n(n+2)}{2n+1} (u\cdot\nabla)^2 \perp^{\rho\sigma}\perp^{\mu\nu}R_{\rho\mu\sigma\nu}\)\\
        \fc{Q}_{2n+1}^{\lambda_1...\lambda_{2n+1}\rho\mu\sigma\nu}\nabla^{2n+1}_{(\lambda_1...\lambda_{2n+1})} R_{\rho\mu\sigma\nu} = \frac{m a^{2n+2}}{(2n+3)!}\nabla^{2n}_\perp & R^\star_{\nabla_\perp u u a}
    \end{align}
    (up to terms which are quadratic in the Riemann tensor). $\nabla_\perp$ is consistent with our use of the $\perp$ symbol: $\nabla_\perp^\rho = \perp^{\rho\sigma}\nabla_\sigma$ and is used as an index to indicate contraction just as with $u$ and $a$. Now returning these to the dynamical mass function produces:
    \begin{align}
        \fc{M}^2 = m^2 & + 2 m^2 a^2 \scr{F}_1(a\nabla_\perp) \perp^{\rho\sigma}R_{\rho u \sigma u} + 2m^2 a^2 \scr{F}_2(a\nabla_\perp) \perp^{\rho\sigma}\perp^{\mu\nu}R_{\rho\mu\sigma\nu} \nonumber \\
        &+ 2 m^2 a^4 \scr{F}_3(a\nabla_\perp) (u\cdot\nabla)^2 \perp^{\rho\sigma}\perp^{\mu\nu} R_{\rho\mu\sigma\nu} + 2m^2 a^2 \scr{F}_4(a\nabla_\perp) R^\star_{\nabla_\perp u u a} + \OO(R^2) \label{eq:grdmf}
    \end{align}
    where:
    \begin{align}
        \scr{F}_1(x) &= \frac{\cosh x}{2x^2} + \frac{\sinh x}{2 x^3} - \frac{1}{x^2} \\
        \scr{F}_2(x) &= \frac{\cosh x}{4x^2} - \frac{\sinh x}{4 x^3} \\
        \scr{F}_3(x) &= \frac{1}{x^4} - \frac{5}{8} \frac{\cosh x}{x^4} - \frac{3}{8} \frac{\sinh x}{x^5} + \frac{3}{8 x^3}\int_0^x \frac{\sinh t}{t} dt \\
        \scr{F}_4(x) &= \frac{\sinh x - x}{x^3}.
    \end{align}
    
    The dynamical mass function in \eqref{eq:grdmf} is our principal result for spinning black holes, analogous to \eqref{eq:fulldmf}. If we neglect $u\cdot \nabla$ terms and consider only contributions to the dynamical mass function which are nonzero for a local vacuum solution of Einstein's equations $(R_{\mu\nu} = 0)$, then \eqref{eq:grdmf} simplifies to:
    \begin{align}
        \fc{M}^2 = m^2 - 2m^2 \frac{1-\cos(a\cdot\nabla)}{(a\cdot\nabla)^2} R_{uaua} + 2m^2 \frac{(a\cdot\nabla)-\sin(a\cdot\nabla)}{(a\cdot\nabla)^2} R^\star_{uaua} + \OO(R^2) \label{eq:lsdmf}
    \end{align}
    which are the equivalent couplings of Ref.~\cite{Levi:2015msa}. While \eqref{eq:grdmf} and \eqref{eq:lsdmf} produce the same three point amplitude (and so the same stationary energy-momentum tensor), they do not produce the same Compton amplitudes. Only \eqref{eq:grdmf} satisfies \eqref{eq:intstress} as an off-shell statement for black holes. In this way, \eqref{eq:grdmf} uniquely captures the physical multipole moments for a spinning black hole independent of its motion.

\section{Gravitational Compton Amplitude} \label{sec:grcompton}
	
	\subsection{Formal Classical Compton}

     To write Einstein's equations explicitly for the metric perturbation, it is useful to introduce a shorthand for the volume form Jacobian:
	\begin{equation}
		V = \frac{1}{\sqrt{-\det g}}.
	\end{equation}
    and to define the inverse metric tensor density $\mf{g}^{\mu\nu}$:
	\begin{equation}
		\mf{g}^{\mu\nu} = g^{\mu\nu}\sqrt{-\det g}.
	\end{equation}
	Then, derivatives of the metric can be expressed as:
	\begin{equation}
		\d_\rho g^{\mu\nu} = V\(\d_\rho \mf{g}^{\mu\nu} - \frac{1}{2}g^{\mu\nu}g_{\alpha\beta}\d_\rho \mf{g}^{\alpha\beta}\), \p \d_\rho V = -\frac{V^2}{2}g_{\mu\nu}\d_\rho \mf{g}^{\mu\nu}.
	\end{equation}
	With these definitions, we find that:
	\begin{align}
		\frac{2}{V^2}\(R^{\mu\nu} - \frac{1}{2}Rg^{\mu\nu}\) =&\ \mf{g}^{\rho\sigma}\d^2_{\rho\sigma}\mf{g}^{\mu\nu} + \mf{g}^{\mu\nu}\d^2_{\rho\sigma}\mf{g}^{\rho\sigma} - \mf{g}^{\mu\sigma}\d^2_{\rho\sigma}\mf{g}^{\rho\nu} - \mf{g}^{\nu\sigma}\d^2_{\rho\sigma}\mf{g}^{\rho\mu} \nonumber \\
		&\ -\d_\sigma\mf{g}^{\mu\rho} \d_\rho\mf{g}^{\nu\sigma} + \d_\rho\mf{g}^{\rho\sigma}\d_\sigma\mf{g}^{\mu\nu} - g_{\alpha\beta}g^{\rho\sigma}\d_\rho\mf{g}^{\mu\alpha}\d_\sigma\mf{g}^{\nu\beta} \nonumber \\
		&\ + g^{\mu\beta}g_{\rho\alpha}\d_\sigma\mf{g}^{\nu\alpha}\d_\beta\mf{g}^{\rho\sigma} + g^{\nu\beta}g_{\rho\alpha}\d_\sigma\mf{g}^{\mu\alpha}\d_\beta\mf{g}^{\rho\sigma} - \frac{1}{2}g^{\mu\nu}g_{\alpha\beta}\d_\sigma \mf{g}^{\alpha\beta}\d_\beta \mf{g}^{\rho\sigma} \nonumber \\
		&\ - \frac{1}{8}(2g^{\mu\tau}g^{\nu\omega} - g^{\mu\nu}g^{\tau\omega})(2g_{\alpha\rho}g_{\beta\sigma} - g_{\alpha\beta}g_{\rho\sigma})\d_\tau \mf{g}^{\alpha\beta}\d_\omega\mf{g}^{\rho\sigma}.
	\end{align}
	This expression is true in any coordinates. Going forward we will only use de Donder gauge, defined so that the coordinates are harmonic functions when viewed as scalars:
	\begin{equation}
		\nabla^2 x^\mu = -g^{\alpha\beta}\Gamma^\mu{}_{\alpha\beta} = V\d_\nu \mf{g}^{\mu\nu} \assert 0.
	\end{equation}
	Using this gauge, Einstein's equations can be written exactly in Landau-Lifshitz form as:
	\begin{align}
		-\mf{g}^{\rho\sigma}\d^2_{\rho\sigma}\mf{g}^{\mu\nu} =&\ -16\pi G T^{\mu\nu}|\det \mf{g}| - \d_\sigma \mf{g}^{\mu\rho}\d_\rho\mf{g}^{\nu\sigma} - g_{\alpha\beta}g^{\rho\sigma}\d_\rho \mf{g}^{\mu\alpha}\d_\sigma\mf{g}^{\nu\beta} \nonumber \\
		&\ + g^{\mu\beta}g_{\rho\alpha}\d_\sigma\mf{g}^{\nu\alpha}\d_\beta\mf{g}^{\rho\sigma} + g^{\nu\beta}g_{\rho\alpha}\d_\sigma \mf{g}^{\mu\alpha}\d_\beta\mf{g}^{\rho\sigma} - \frac{1}{2}g^{\mu\nu}g_{\alpha\rho} \d_\sigma\mf{g}^{\alpha\beta}\d_\beta\mf{g}^{\rho\sigma} \nonumber \\
		&\ -\frac{1}{8}(2g^{\mu\tau}g^{\nu\omega} - g^{\mu\nu}g^{\tau\omega})(2g_{\alpha\rho}g_{\beta\sigma}-g_{\alpha\beta}g_{\rho\sigma})\d_\tau \mf{g}^{\alpha\beta}\d_\omega\mf{g}^{\rho\sigma}
	\end{align}
	For studying gravitational waves we perturb about Minkowski space:
	\begin{equation}
		\mf{g}^{\mu\nu} = \eta^{\mu\nu} + \kappa h^{\mu\nu}, \p \kappa = \sqrt{32\pi G}.
	\end{equation}
	For this choice of coupling constant, the Einstein-Hilbert action is canonically normalized as a functional of the perturbation $h^{\mu\nu}$. When considering perturbations of Minkowski space, we raise and lower indices by using the Minkowski metric. In terms of $h^{\mu\nu}$, de Donder gauge is the requirement:
	\begin{equation}
		\d_\nu h^{\mu\nu} = 0.
	\end{equation}
	In terms of $h^{\mu\nu}$, Einstein's equations are exactly:
	\begin{align}
		-\d^2 h^{\mu\nu} = \kappa&\(-\frac{T^{\mu\nu}}{2}|\det \mf{g}| + h^{\rho\sigma}\d^2_{\rho\sigma}h^{\mu\nu} -\d_\sigma h^{\mu\rho}\d_\rho h^{\nu\sigma} - g_{\alpha\beta}g^{\rho\sigma}\d_\rho h^{\mu\alpha}\d_\sigma h^{\nu\beta}\right. \nonumber \\
		&\ \ \ \left.+ g^{\mu\beta}g_{\rho\alpha}\d_\sigma h^{\nu\alpha}\d_\beta h^{\rho\sigma} + g^{\nu\beta}g_{\rho\alpha}\d_\sigma h^{\mu\alpha}\d_\beta h^{\rho\sigma} - \frac{1}{2}g^{\mu\nu}g_{\alpha\rho}\d_\sigma h^{\alpha\beta}\d_\beta h^{\rho\sigma} \right.\nonumber \\
		&\ \ \ \left. \vphantom{\frac{T^{\mu\nu}}{2}}- \frac{1}{8}(2g^{\mu\tau}g^{\nu\omega}-g^{\mu\nu}g^{\tau\omega})(2g_{\alpha\rho}g_{\beta\sigma}-g_{\alpha\beta}g_{\rho\sigma})\d_\tau h^{\alpha\beta}\d_\omega h^{\rho\sigma}\)
	\end{align}
	
	With $\kappa$ acting as the coupling constant, $h^{\mu\nu}$ and $T^{\mu\nu}$ will have solutions in powers of $\kappa$:
	\begin{align}
		h^{\mu\nu} &= h^{\mu\nu}_{(0)} + \kappa h^{\mu\nu}_{(1)} + \kappa^2 h^{\mu\nu}_{(2)} + \OO(\kappa^3) \nonumber \\
		T^{\mu\nu} &= T^{\mu\nu}_{(0)} + \kappa T^{\mu\nu}_{(1)} + \OO(\kappa^2)
	\end{align}
	The $\kappa^0$ piece then satisfies the homogeneous wave equation from Einstein's equations:
	\begin{equation}
		\d^2 h^{\mu\nu}_{(0)} = 0.
	\end{equation}
    For Compton scattering, we will consider the incoming gravitational field to be a plane wave:
    \begin{equation}
        h^{\mu\nu}_{(0)} = \fc{E}_1^{\mu\nu} e^{ik_1\cdot x}
    \end{equation}
    for some polarization tensor $\fc{E}_1^{\mu\nu}$. We will further gauge fix within de Donder gauge so that $\fc{E}_1^{\mu\nu}$ is traceless, transverse to $k_{1\mu}$ and orthogonal to a vector $v^\mu$. Using the helicity polarization vectors from electromagnetism, the most general such tensor may be written as a linear combination:
	\begin{equation}
		\fc{E}^{\mu\nu}_1 = c_+ \fc{E}^\mu_{1+}\fc{E}^\nu_{1+} + c_- \fc{E}^\mu_{1-}\fc{E}^\nu_{1-}.
	\end{equation}
	Therefore for considering Compton scattering in the helicity basis we will take the incoming plane wave to be of the form:
	\begin{equation}
		h^{\mu\nu}_{(0)} = \epsilon \fc{E}^\mu_1\fc{E}^\nu_1 e^{i k_1\cdot x}.
	\end{equation}
	
	From Einstein's equations, the $\kappa^1$ piece $h^{\mu\nu}_{(1)}$ then satisfies:
	\begin{align}
		-\d^2 h^{\mu\nu}_{(1)} = &\ -\frac{T^{\mu\nu}_{(0)}}{2} + h^{\rho\sigma}_{(0)}\d^2_{\rho\sigma}h^{\mu\nu}_{(0)} -\d_\sigma h^{\mu\rho}_{(0)}\d_\rho h^{\nu\sigma}_{(0)} - \eta_{\alpha\beta}\eta^{\rho\sigma}\d_\rho h^{\mu\alpha}_{(0)}\d_\sigma h^{\nu\beta}_{(0)} \nonumber \\
		&\ \ \ + \eta^{\mu\beta}\eta_{\rho\alpha}\d_\sigma h^{\nu\alpha}_{(0)}\d_\beta h^{\rho\sigma}_{(0)} + \eta^{\nu\beta}\eta_{\rho\alpha}\d_\sigma h^{\mu\alpha}_{(0)}\d_\beta h^{\rho\sigma}_{(0)} - \frac{1}{2}\eta^{\mu\nu}\eta_{\alpha\rho}\d_\sigma h^{\alpha\beta}_{(0)}\d_\beta h^{\rho\sigma}_{(0)} \nonumber \\
		&\ \ \ - \frac{1}{8}(2\eta^{\mu\tau}\eta^{\nu\omega}-\eta^{\mu\nu}\eta^{\tau\omega})(2\eta_{\alpha\rho}\eta_{\beta\sigma}-\eta_{\alpha\beta}\eta_{\rho\sigma})\d_\tau h^{\alpha\beta}_{(0)}\d_\omega h^{\rho\sigma}_{(0)}.
	\end{align}
	For Compton scattering we are only concerned with the response of the system to linear order in the incoming field strength $\epsilon$ and so it is useful to define $h^{\mu\nu}_\stat$ as the stationary response of the metric perturbation to the unperturbed energy-momentum tensor. In the equation of motion for $h_{(1)}^{\mu\nu}$, all of the contributions from $h^{\mu\nu}_{(0)}$ are of order $\epsilon^2$ and so:
	\begin{equation}
		h^{\mu\nu}_{(1)} = h^{\mu\nu}_\stat + \OO(\epsilon^2), \p -\d^2 h^{\mu\nu}_\stat = -\frac{T^{\mu\nu}_{(0)}}{2}.
	\end{equation}

    The $\kappa^1$ correction to the metric perturbation can be decomposted into a statioanry piece from iterating corrections from $h^{\mu\nu}_\stat$, which is $\epsilon$ independent, a piece which is linear in $\epsilon$, and pieces which are of at least $\OO(\epsilon^2)$:
    \begin{equation}
		h^{\mu\nu}_{(2)} = h^{\mu\nu}_{\stat,2}+ \epsilon \delta h^{\mu\nu} + \OO(\epsilon^2).
	\end{equation}
    Because the lowest order stationary correction to the actual metric $g_{\mu\nu}$ is $\OO(\kappa^2)$, the $\kappa^1$ correction to the energy-momentum tensor has no $\epsilon^0$ piece and so may be written:
	\begin{equation}
		T^{\mu\nu}_{(1)} = \epsilon \delta T^{\mu\nu} + \OO(\epsilon^2).
	\end{equation}
	
	The quantity $\delta h^{\mu\nu}$ thus encodes the linearized response of gravity to the interaction of an incoming plane wave with the massive body and so determines the Compton amplitude. Returning these definitions to Einstein's equations, we find that $\delta h^{\mu\nu}$ satisfies: 
	\begin{align}
		-\epsilon\d^2 \delta h^{\mu\nu} =&\ -\frac{\epsilon}{2}\delta T^{\mu\nu} + h^{\rho\sigma}_\stat\d^2_{\rho\sigma} h^{\mu\nu}_{(0)} + h^{\rho\sigma}_{(0)} \d^2_{\rho\sigma} h^{\mu\nu}_\stat - \d_\sigma h^{\mu\rho}_\stat\d_\rho h^{\nu\sigma}_{(0)} - \d_\sigma h^{\mu\rho}_{(0)}\d_\rho h^{\nu\sigma}_\stat \nonumber \\
		&\ - \d^\rho h^{\mu\sigma}_\stat\d_\rho h^{\nu}_{(0)\sigma} - \d^\rho h^{\mu\sigma}_{(0)}\d_\rho h^\nu_{\stat \sigma} + \d^\mu h^{\rho\sigma}_\stat\d_\rho h^\nu_{(0)\sigma} + \d^\mu h^{\rho\sigma}_{(0)}\d_\rho h^\nu_{\stat\sigma} \nonumber \\
		&\ + \d^\nu h^{\rho\sigma}_\stat\d_\rho h^{\mu}_{(0)\sigma} + \d^\nu h^{\rho\sigma}_{(0)}\d_\rho h^{\mu}_{\stat\sigma} - \eta^{\mu\nu}\d_\lambda h^{\rho\sigma}_\stat\d_\rho h^{\lambda}_{(0)\sigma}\nonumber \\
		&\ - \frac{1}{4}\(\eta^{\mu\tau}\eta^{\nu\omega}+\eta^{\nu\tau}\eta^{\mu\omega} - \eta^{\mu\nu}\eta^{\tau\omega}\)\(2\eta_{\alpha\rho}\eta_{\beta\sigma}-\eta_{\alpha\beta}\eta_{\rho\sigma}\)\d_\tau h^{\alpha\beta}_\stat \d_\omega h^{\rho\sigma}_{(0)}.   \label{eq:flucth}
	\end{align}
	Define the source fluctuation $\delta\tau^{\mu\nu}$ so that:
	\begin{equation}
		-\d^2\delta h^{\mu\nu} = -\frac{\delta \tau^{\mu\nu}}{2}.
	\end{equation}

    Because the tree level Compton is $\OO(\kappa^2)$, it only depends on linear and quadratic in $R$ terms in the dynamical mass function. Consequently, we will consider the dynamical mass function to be of the form:
    \begin{equation}
        \fc{M}^2 = m^2 + \delta\fc{M}_1^2 + \delta\fc{M}_2^2 + \OO(R^3)
    \end{equation}
    where $\delta\fc{M}_1^2$ is of the form:
    \begin{equation}
        \delta\fc{M}_1^2 = \sum_{n=0}^\8 T_n^{\lambda_1...\lambda_n \rho\mu\sigma\nu}(g, u, S) \nabla^n_{(\lambda_1...\lambda_n)} R_{\rho\mu\sigma\nu}
    \end{equation}
    for some functions $T_n^{...}$ with the same index symmetries as $\nabla^n_{(\lambda_1...\lambda_n)} R_{\rho\mu\sigma\nu}$ and where $\delta\fc{M}_2^2$ is of the form:
    \begin{equation}
        \delta\fc{M}_2^2 = \sum_{n=0}^\8\sum_{l=0}^\8V_{nl}^{\lambda_1...\lambda_n\rho\mu\sigma\nu|\kappa_1...\kappa_l \gamma\alpha\delta\beta}(g, u, S) \nabla^n_{(\lambda_1...\lambda_n)}R_{\rho\mu\sigma\nu} \nabla^l_{(\kappa_1...\kappa_l)} R_{\gamma\alpha\delta\beta}
    \end{equation}
    for some functions $V_{nl}^{...}$ with the same index symmetries as $\nabla^n_{(\lambda_1...\lambda_n)}R_{\rho\mu\sigma\nu} \nabla^l_{(\kappa_1...\kappa_l)} R_{\gamma\alpha\delta\beta}$.

    Einstein's equations together with the MPD equations with the described initial conditions will produce solutions of the form:
    \begin{align}
		z^\mu(\lambda) &= v^\mu \lambda + \kappa\epsilon\delta z^\mu(\lambda) + \OO(\kappa^2)\\
		u_{\mu}(\lambda) &= v_\mu + \kappa\epsilon\delta u_\mu(\lambda) + \OO(\kappa^2) \\
		S^\mu(\lambda) &= s^\mu + \kappa\epsilon\delta S^\mu(\lambda) + \OO(\kappa^2) \\
        T^{\mu\nu}(X) &= T^{\mu\nu}_{(0)}(X) + \kappa \epsilon \delta T^{\mu\nu} + \OO(\kappa^2,\epsilon^2) \\
		h^{\mu\nu}(X) &= \epsilon\fc{E}_{1}^\mu\fc{E}_1^\nu e^{ik_1\cdot x} + \kappa h^{\mu\nu}_\text{stat}(X) + \kappa^2 h^{\mu\nu}_{\text{stat},2}(X) + \kappa^2\epsilon\delta h^{\mu\nu}(X) + \OO(\kappa^3,\epsilon^2).
	\end{align}
    The equation of motion perturbations will be oscillatory from solving the  MPD equations. In particular, their solutions take the form:
    \begin{equation}
		\delta z^\mu = \delta\twid{z}^\mu e^{i k_1\cdot v \lambda}, \p \delta u_\mu = \delta\twid{u}_\mu e^{ik_1\cdot v \lambda}, \p \delta S^\mu = \delta\twid{S}^\mu e^{i k_1\cdot v\lambda}
	\end{equation}
    for constant vectors $\delta\twid{z}, \delta\twid{u}, \delta\twid{S}$.

    It is useful to define the function $\fc{N}(u, S, k, \fc{E})$:
    \begin{equation}
        \fc{N}(u, S, k, \fc{E}) = -\sum_{n=0}^\8 \frac{i^n}{m} T_n^{\lambda_1...\lambda_n\rho\mu\sigma\nu}(\eta, u, S) k_{\lambda_1}...k_{\lambda_n} \fc{E}_\rho k_\mu \fc{E}_\sigma k_\nu
    \end{equation}
    so that for a plane-wave:
    \begin{equation}
        \left.\delta\fc{M}_1^2\right|_{\text{plane-wave}} = 2m\kappa\epsilon \fc{N}(u, S, k, \fc{E})e^{ik\cdot z} + \OO(\kappa^2).
    \end{equation}
    Similarly, we define the function $\fc{P}(u, S, k, \fc{E}, k', \fc{E}')$:
    \begin{equation}
        \fc{P} = \sum_{n=0}^\8\sum_{l=0}^\8 \frac{4i^{l-n}}{m}V_{nl}^{\lambda_1...\lambda_n\rho\mu\sigma\nu|\kappa_1...\kappa_l\gamma\alpha\delta\beta}k'_{\lambda_1}...k'_{\lambda_n}k'_\rho \fc{E}'^*_\mu k'_\sigma \fc{E}'^*_\nu k_{\kappa_1}...k_{\kappa_l}k_\gamma \fc{E}_\alpha k_\delta \fc{E}_\beta
    \end{equation}
    and adopt the shorthand:
    \begin{equation}
        \fc{N}_1 = \fc{N}(v,s,k_1, \fc{E}_1), \p \fc{N}_2 = \fc{N}(v,s,k_2,\fc{E}_2), \p \fc{P}_{12} = \fc{P}(v,s,k_1,\fc{E}_1,k_2,\fc{E}_2).
    \end{equation}
    
    Using the delayed Green's function solution to the wave equation to solve Einstein's equations for $\delta h^{\mu\nu}$, we find:
    \begin{equation}
        \delta h^{\mu\nu}(X) = -\int_{\bb{S}} \frac{\delta \tau^{\mu\nu}(t-|\vec x-\pvec x|,\pvec x)}{8\pi|\vec x-\pvec x|} d^3\pvec x
    \end{equation}
    where $t$ and $\vec x$ retain their definitions from electromagnetism. The source perturbation is determined by the equation of motion perturbations $\delta z, \delta u, \delta S$. Define the flat-space Fourier transform of the energy-momentum tensor:
    \begin{equation}
        \twid{T}^{\mu\nu}(k) = \int \frac{e^{-ik\cdot x}}{(2\pi)^2} T^{\mu\nu}(X) d^4 X.
    \end{equation}
    From \eqref{eq:stress} we can identify that the Fourier transform of the stationary stress tensor of the unperturbed body is:
    \begin{equation}
        \twid T_{(0)}^{\alpha\beta}(k) = M^{\alpha\beta}(k) \frac{\delta(k\cdot v)}{2\pi}
    \end{equation}
    where:
    \begin{equation}
        M^{\alpha\beta}(k) = mv^\alpha v^\beta + i v^{(\alpha} \epsilon^{\beta)\gamma\rho\sigma}k_\gamma v_\rho s_\sigma - \frac{2}{m}\sum_{n=0}^\8 (-i)^n T_n^{\lambda_1...\lambda_n\mu\alpha\nu\beta}(\eta, v, s) k_{\lambda_1}...k_{\lambda_n}k_\mu k_\nu.
    \end{equation}
    The resulting Fourier transform of the stationary metric perturbation is:
    \begin{equation}
        \twid h^{\alpha\beta}_\text{stat}(k) = -\frac{M^{\alpha\beta}(k)}{2k^2} \frac{\delta(k\cdot v)}{2\pi}.
    \end{equation}
    In terms of this solution, \eqref{eq:flucth} produces:
    \begin{align}
        \delta\twid \tau^{\mu\nu}(k_2) =&\ \delta\twid T^{\mu\nu}(k_2)
        + 2\fc{E}_1^\mu \fc{E}_1^\nu(\twid h^{\rho\sigma}_\text{stat} k_{1\rho} k_{1\sigma})
        + 2(\fc{E}_1\cdot k_2)^2 \twid h^{\mu\nu}_\text{stat}
        - 4(\fc{E}_1\cdot k_2) \fc{E}_1^{(\mu} \twid h^{\nu)\rho}_\text{stat} k_{1\rho} \nonumber \\
        &\ - 4(k_1\cdot k_2)\fc{E}_1^{(\mu} \twid h^{\nu)\rho}_\text{stat}\fc{E}_{1\rho}
        + 4\fc{E}_1^{(\mu}(k_2-k_1)^{\nu)}(\twid h^{\rho\sigma}_\text{stat}k_{1\rho}\fc{E}_{1\sigma})
        + 4(\fc{E}_1\cdot k_2) k_1^{(\mu} \twid h^{\nu)\rho}_\text{stat}\fc{E}_{1\rho} \nonumber \\
        &\ - 2\eta^{\mu\nu}(\fc{E}_1\cdot k_2)\twid h^{\rho\sigma}_\text{stat}k_{1\rho}\fc{E}_{1\sigma} - 2k_1^{(\mu}(k_2-k_1)^{\nu)}\twid h^{\alpha\beta}_\text{stat}\fc{E}_{1\alpha}\fc{E}_{1\beta} + \eta^{\mu\nu} k_1\cdot k_2 \twid h^{\alpha\beta}_{\text{stat}}\fc{E}_{1\alpha}\fc{E}_{1\beta}
    \end{align}
    where everywhere it appears in the above equation, $\twid h^{\rho\sigma}_\text{stat}$ is evaluated at $k_2-k_1$.

    From the definition of $\fc{Q}_n$, we find (up to terms of $\OO(R^2)$):
    \begin{equation}
        \fc{Q}_n^{\lambda_1...\lambda_n\rho\mu\sigma\nu} = \(1 - \frac{\delta\fc{M}_1^2}{2m^2}\)\frac{T_n^{\lambda_1...\lambda_n\rho\mu\sigma\nu}}{2m} + \frac{1}{m}\sum_{l=0}^\8 V_{nl}^{\lambda_1...\lambda_n\rho\mu\sigma\nu|\kappa_1...\kappa_l\gamma\alpha\delta\beta} \nabla^l_{(\kappa_1...\kappa_l)} R_{\gamma\alpha\delta\beta}.
    \end{equation}
    In terms of the solutions to the equations of motion:
    \begin{equation}
        \fc{Q}_n^{\lambda_1...\lambda_n\rho\mu\sigma\nu} = \frac{T_n^{\lambda_1...\lambda_n\rho\mu\sigma\nu}(\eta, v, s)}{2m} + \kappa \epsilon \delta \twid{\fc{Q}}_n^{\lambda_1...\lambda_n\rho\mu\sigma\nu} e^{ik_1\cdot v\lambda} + \OO(\kappa^2)
    \end{equation}
    where:
    \begin{align}
        \delta\twid{\fc{Q}}_n^{\lambda_1...\lambda_n\rho\mu\sigma\nu} =&\  \frac{1}{4m}\fc{E}_{1\alpha}\fc{E}_{1\beta}\hat G^{\alpha\beta} T_n^{\lambda_1...\lambda_n\rho\mu\sigma\nu} + \frac{1}{2m} \frac{\d T_n^{\lambda_1...\lambda_n\rho\mu\sigma\nu}}{\d v^\alpha} \delta\twid u^\alpha \nonumber \\
        &\ + \frac{1}{2m} \frac{\d T_n^{\lambda_1...\lambda_n\rho\mu\sigma\nu}}{\d s^\alpha}\(\delta\twid S^\alpha - \frac{1}{2}s\cdot\fc{E}_1 \fc{E}_1^\alpha\) - \frac{\fc{N}_1}{2m^2} T_n^{\lambda_1...\lambda_n\rho\mu\sigma\nu} \nonumber \\
        &\ - \frac{2}{m}\sum_{l=0}^\8 i^l V_{nl}^{\lambda_1...\lambda_n\rho\mu\sigma\nu|\kappa_1...\kappa_l \gamma\alpha\delta\beta} k_{1\kappa_1}...k_{1\kappa_l} k_{1\gamma}k_{1\delta}\fc{E}_{1\alpha}\fc{E}_{1\beta}.
    \end{align}

    Returning these to $\delta\tau$, we find that in Fourier space it takes the form:
    \begin{equation}
        \delta\twid\tau^{\mu\nu}(k_2) = H^{\mu\nu}(k_2)\frac{\delta(k_2\cdot v-k_1\cdot v)}{2\pi}
    \end{equation}
    and that at large distances, recycling the notation from electromagnetism. The metric perturbation becomes:
    \begin{equation}
        \delta h^{\mu\nu}(X) = -\frac{e^{i\omega(r-t)}}{8\pi r} H^{\mu\nu}(k_2) + \OO\(\frac{1}{r^2}\).
    \end{equation}
    Therefore, the covariantly normalized Compton amplitude is given by:
    \begin{equation}
        \fc{A} = -m \fc{E}_{2\mu}^* \fc{E}_{2\nu}^* H^{\mu\nu}(k_2).
    \end{equation}

    The solutions to the equations of motion are:
    \begin{align}
        \delta\twid{u}^\mu =&\ \frac{i}{2m} \fc{E}_1^\mu \epsilon^{v s k_1 \fc{E}_1} - \frac{\fc{N}_1}{m}\(v^\mu + \frac{k_1^\mu}{k_1\cdot v}\) \\
        \delta\twid{S}^\mu =&\ \frac{1}{2}(\fc{E}_1\cdot s)\fc{E}_1^\mu + v^\mu s\cdot\delta\twid u - \frac{i}{k_1\cdot v} \epsilon^{\mu\nu\rho\sigma} v_\nu s_\rho \frac{\d\fc{N}_1}{\d s^\sigma} \\
        \delta\twid{z}^\mu =&\  \frac{-i}{k_1\cdot v}\(\delta \twid{u}^\mu + \frac{\epsilon^{v s k_1\fc{E}_1}}{2m^2}\(\delta^\mu_\nu + v^\mu v_\nu\)\epsilon^{\nu s k_1\fc{E}_1} + \frac{\eta^{\mu\nu} + v^\mu v^\nu}{m}\frac{\d\fc{N}_1}{\d v^\nu} \right.\nonumber \\
        &\left.\vphantom{\frac{-i}{k_1}}\p\p + \frac{s^\mu v^\nu}{m}\frac{\d\fc{N}_1}{\d s^\nu} + \frac{i}{m^2}\fc{N}_1\epsilon^{\mu\nu\rho\sigma} v_\nu s_\rho k_{1\sigma}\).
    \end{align}
    Expressed in terms of these solutions, the gravitational Compton amplitude to all orders in spin is:
    \begin{align}
        \fc{A} =&\ 2m\(\frac{1}{2}(k_2\cdot v)(\fc{E}_2^*\cdot\delta \twid z) \epsilon^{\fc{E}_2^*k_2 v s} -\frac{i}{4}(k_2\cdot v) (\fc E_1\cdot\fc E_2^*) \epsilon^{\fc E_1\fc E_2^*vs} \right.\nonumber \\
        &\p \p\left.+ i (k_2\cdot\delta \twid z)\fc{N}_2^* -\delta\twid u^\alpha\frac{\d\fc{N}_2^*}{\d v^\alpha}-\(\delta\twid S^\alpha - \frac{1}{2}(s\cdot \fc{E}_1)\fc{E}_1^\alpha\)\frac{\d\fc{N}_2^*}{\d s^\alpha} + \frac{\fc{N}_1\fc{N}_2^*}{m} - \fc{P}_{12}\) \nonumber \\
        &-\frac{m(\fc{E}_1\cdot\fc{E}_2^*)^2}{2(k_1\cdot k_2)}(M_{\mu\nu}k_1^\mu k_1^\nu)
        - \frac{m(\fc{E}_1\cdot k_2)^2}{2(k_1\cdot k_2)}(M_{\mu\nu}\fc{E}_2^{*\mu}\fc{E}_2^{*\nu})
        + \frac{m(\fc{E}_1\cdot k_2)(\fc{E}_1\cdot \fc{E}_2^*)}{(k_1\cdot k_2)}(M_{\mu\nu}k_1^\mu \fc{E}_2^{*\nu})\nonumber \\
        &+ m(\fc{E}_1\cdot \fc{E}_2^*)(M_{\mu\nu}\fc{E}_1^\mu\fc{E}_2^{*\nu})
        + \frac{m(\fc{E}_1\cdot\fc{E}_2^*)(k_1\cdot\fc{E}_2^*)}{(k_1\cdot k_2)}(M_{\mu\nu}k_1^{\mu}\fc{E}_1^\nu) \nonumber \\
        &- \frac{m(\fc{E}_1\cdot k_2)(\fc{E}_2^*\cdot k_1)}{(k_1\cdot k_2)}(M_{\mu\nu}\fc{E}_1^\mu \fc{E}_2^{*\nu})
        -\frac{m(\fc{E}_2^*\cdot k_1)^2}{2(k_1\cdot k_2)}(M_{\mu\nu}\fc{E}_1^\mu\fc{E}_1^\nu) + \sum_{n=0}^\8 \delta\fc{A}_n
    \end{align}
    where:
    \begin{align}
        \delta\fc{A}_n =&\ -\frac{1}{2}\sum_{l=0}^{n-1} i^n (-1)^l (k_1\cdot \fc{E}_2^*) T_n^{\lambda_1...\lambda_n\rho\mu\sigma\nu}k_{1\lambda_1}...k_{1\lambda_{n-l-1}} \fc{E}^*_{2\lambda_{n-l}} q_{\lambda_{n-l+1}}...q_{\lambda_n} \delta\twid R_{\rho\mu\sigma\nu} \nonumber \\
        & + \sum_{l=0}^{n-1} i^n (-1)^l T_n^{\lambda_1...\lambda_n\rho\mu\sigma\nu} \fc{E}_2^{*[\tau} k_2^{\omega]} \hat G_{\tau\omega}(k_{1\lambda_1}...k_{1\lambda_{n-l-1}} \delta\twid R_{\rho\mu\sigma\nu}) \fc{E}^*_{2\lambda_{n-l}}q_{\lambda_{n-l+1}}...q_{\lambda_n} \nonumber \\
        &-(-i)^n T_n^{\lambda_1...\lambda_n\rho\mu\sigma\nu} q_{\lambda_1}...q_{\lambda_n} \fc{E}^*_{2\rho} \fc{E}^{*\alpha}_2\delta\twid R_{\alpha\mu\sigma\nu} \nonumber \\
        &+ 2(-1)^n\sum_{l=0}^{n-1} i^{n-1}\fc{E}^*_{2\alpha} \fc{E}^*_{2\beta} k_{2\mu}k_{2\nu}k_{2\lambda_{n-l+1}}...k_{2\lambda_n} \delta\twid \Gamma^\sigma{}_{\rho\lambda_{n-l}} \hat G^\rho{}_\sigma(T_n^{\lambda_1...\lambda_n\mu\alpha\nu\beta} q_{\lambda_1}...q_{\lambda_{n-l-1}}) \nonumber \\
        &-2i(-i)^n \fc{E}^*_{2\alpha} \fc{E}^*_{2\beta} k_{2\mu} \delta\twid\Gamma^\sigma{}_{\rho\nu} \hat G^\rho{}_\sigma (T_n^{\lambda_1...\lambda_n\mu\alpha\nu\beta} q_{\lambda_1}...q_{\lambda_n}) \nonumber \\
        &-2i(-i)^n\fc{E}^*_{2\alpha} \fc{E}^*_{2\beta} \delta\twid{\Gamma}^\sigma{}_{\rho\mu} \hat G^\rho{}_\sigma(T_n^{\lambda_1...\lambda_n\mu\alpha\nu\beta} q_\nu q_{\lambda_1}...q_{\lambda_n}) \nonumber \\
        &+(-i)^n k_{2\lambda_1}...k_{2\lambda_n} k_{2\rho} \fc{E}^*_{2\mu}k_{2\sigma}\fc{E}^*_{2\nu}\fc{E}_1^\alpha\fc{E}_1^\beta \hat G_{\alpha\beta}(T_n^{\lambda_1...\lambda_n\rho\mu\sigma\nu})
    \label{eq:deltaAn}
    \end{align}
    and where:
    \begin{align}
        q^\mu &= k_2^\mu - k_1^\mu \\
        \delta\twid \Gamma^\rho{}_{\mu\nu} &= \frac{i}{2}(k_1^\rho \fc{E}_{1\mu}\fc{E}_{1\nu} - \fc{E}_1^\rho k_{1\mu} \fc{E}_{1\nu} - \fc{E}_1^\rho k_{1\nu} \fc{E}_{1\mu}) \\
        \delta\twid R_{\rho\mu\sigma\nu} &= -\frac{1}{2}(k_{1\rho}\fc{E}_{1\mu}-k_{1\mu}\fc{E}_{1\rho})(k_{1\sigma}\fc{E}_{1\nu}-k_{1\nu}\fc{E}_{1\sigma}).
    \end{align}

	\subsection{Compton Amplitude through Spin to the Fifth}

    Similarly to electromagnetism, we consider only operators which have the same number of powers of spin as they do number of derivatives acting on the metric, so as not to include any length scales beyond $\frac{S}{m}$. For enumerating linear in Riemann operators, it is not safe to use Einstein's equations to simplify the possible terms when considering the Compton amplitude as the linear in Riemann piece of the dynamical mass function does not contribute solely in a perfectly on-shell way to the Compton amplitude in the way the linear in field strength piece did for electromagnetism (only through the $\fc{N}$ function). We find explicitly that some Ricci operators make independent contributions to the Compton amplitude. Through $\OO(S^5)$, the most general possible dynamical mass function with only length scale $\frac{S}{m}$ is:
    \begin{align}
        \delta \fc{M}_1^2 =&\ \delta\fc{M}_{1S^2}^2 + \delta\fc{M}_{1S^3}^2 + \delta\fc{M}_{1S^4}^2 + \delta\fc{M}_{1S^5}^2 + \OO(S^6) \\
        \delta\fc{M}_{1S^2}^2 =&\ -C_2 R_{uSuS}  -\frac{E_{2a}}{3} R_{SS} + E_{2b} S^2 R_{uu} + \frac{E_{2c}}{6} S^2 R, \\
        \delta\fc{M}_{1S^3}^2 =&\ \frac{C_3}{3m} (S\cdot\nabla) R^\star_{uSuS} + \frac{E_3}{3m} S^2 \nabla^\rho R^\star_{\rho uu S},
        \\
        \delta\fc{M}_{1S^4}^2 =&\ \frac{C_4}{12m^2} (S\cdot\nabla)^2 R_{uSuS} - \frac{E_{4a}}{20m^2} S^2(u\cdot\nabla)^2 R_{uSuS} - \frac{E_{4b}}{12m^2} S^2\nabla^2 R_{uSuS} \nonumber \\
        &\ + \frac{E_{4c}}{30m^2} (S\cdot\nabla)^2 R_{SS} - \frac{E_{4d}}{30m^2}S^2 (u\cdot\nabla)^2 R_{SS} - \frac{E_{4e}}{30m^2} S^2\nabla^2 R_{SS} \nonumber \\
        &\ - \frac{E_{4f}}{12m^2} S^2 (S\cdot\nabla)^2 R_{uu} + \frac{E_{4g}}{20 m^2} S^4 (u\cdot\nabla)^2 R_{uu} + \frac{E_{4h}}{12 m^2}S^4 \nabla^2 R_{uu} \nonumber \\
        &\ - \frac{E_{4i}}{60 m^2} S^2 (S\cdot\nabla)^2 R - \frac{E_{4j}}{60m^2} S^4 (u\cdot\nabla)^2 R + \frac{E_{4k}}{60 m^2} S^4 \nabla^2 R,  \\
        \delta\fc{M}_{1S^5}^2 =&\ -\frac{C_5}{60m^3} (S\cdot\nabla)^3 R^\star_{uSuS} + \frac{E_{5a}}{60m^3} S^2(u\cdot\nabla)^2 (S\cdot\nabla) R^\star_{uSuS} \nonumber \\
        &\ + \frac{E_{5b}}{60m^3}S^2(S\cdot\nabla)\nabla^2 R^\star_{uSuS} \nonumber - \frac{E_{5c}}{60m^3} S^2 (S\cdot\nabla)^2 \nabla^\rho R^\star_{\rho u u S} \nonumber \\
        &\ + \frac{E_{5d}}{60 m^3} S^4 (u\cdot\nabla)^2 \nabla^\rho R^\star_{\rho u u S} + \frac{E_{5e}}{60 m^3} S^4 \nabla^2 \nabla^\rho R_{\rho u u S}^\star.
    \end{align}
    Expanding \eqref{eq:grdmf} through order $S^5$ we find that it implies that all of these Wilson coefficients should be 1 except $E_{4d}$ and $E_{4j}$ which should both be 0:
    \begin{align}
        C_2 &= C_3 = C_4 = C_5 = E_{2a} = E_{2b} = E_{2c} = E_3 \nonumber \\
            &= E_{4a} = E_{4b} = E_{4c} = E_{4e} = E_{4f} = E_{4g} = E_{4h}\nonumber \\
            &= E_{4i} = E_{4k} = E_{5a} = E_{5b} = E_{5c} = E_{5d} = E_{5e} = 1 \nonumber \\
        E_{4d} &= E_{4j} = 0. \label{eq:dmfrequire}
    \end{align}

    While it is not valid to use Einstein's equations to simplify the linear in Riemann piece of the dynamical mass function when computing the Compton amplitude, because the quadratic in Riemann piece of the dynamical mass function only contributes to the Compton amplitude by being evaluated fully on-shell (through the $\fc{P}$ function) we are perfectly safe to only consider terms within it which are nonzero for vacuum solutions of Einstein's equations. If we assume vacuum, $R^{\rho}{}_{\mu\rho\nu}=0$, then the left and right duals of the Riemann tensor are equal, ${}^\star R= R^\star$, ${}^\star R^\star=-R$, and its (A)SD part 
    can be obtained by projecting with (\ref{eq:ASD_proj}) from either side, and it commutes through to the other,
    \begin{equation}
        {}^{\pm\!}R_{\rho\mu\sigma\nu}=\frac{1}{2}(R\mp i\,{}^\star R)_{\rho\mu\sigma\nu}={}^\pm\fc G_{\rho\mu}{}^{\alpha\beta}R_{\alpha\beta\sigma\nu}=R_{\rho\mu\gamma\delta}\,{}^\pm\fc G^{\gamma\delta}{}_{\sigma\nu}={}^\pm\fc G_{\rho\mu}{}^{\alpha\beta}R_{\alpha\beta\gamma\delta}\,{}^\pm\fc G^{\gamma\delta}{}_{\sigma\nu},
        \end{equation}
    satisfying ${}^\star {}^{\,\pm\!} R={}^{\,\pm\!} R^\star=\pm i {}^{\,\pm\!} R$ and ${}^{\,\pm\!} R^*={}^{\,\mp\!} R$. Defining the quadrupolar gravito-electric and -magnetic ``tidal'' curvature tensors with respect to a unit timelike direction $u$,
    \begin{equation}
        E_{\mu\nu}\mp i B_{\mu\nu}=(R_{\rho\mu\sigma\nu}\mp i{}^{\,\star} R_{\rho\mu\sigma\nu})u^\rho u^\sigma=2{}^{\,\pm\!} R_{\rho\mu\sigma\nu}{u^\rho u^\sigma},
    \end{equation}
    we see that the identities (\ref{eq:ASD_proj})--(\ref{eq:GGmagic}) allow us to reconstruct $R_{\rho\mu\sigma\nu}$ from its components $E_{\mu\nu}$ and $B_{\mu\nu}$ (and $u_\rho$), as the real part of
    \begin{equation}
        2{}^{\,\pm}\! R_{\rho\mu\sigma\nu}=(R\mp i{}^{\,\star} R)_{\rho\mu\sigma\nu}=16\,{}^\pm\fc G_{\rho\mu}{}^{\gamma\alpha}\,{}^\pm\fc G_{\sigma\nu}{}^{\delta\beta}(E_{\alpha\beta}\mp iB_{\alpha\beta}){u_\gamma u_\delta}.
    \end{equation}
    The tidal tensors $E_{\mu\nu}$ and $B_{\mu\nu}$ are symmetric and trace-free (STF) in vacuum, and orthogonal to $u^\mu$, forming irreducible representations of the SO(3) rotation little group of $u^\mu$. Moving on to the first derivative $\nabla_\lambda R_{\rho\mu\sigma\nu}$, the irreducible pieces w.r.t.\ $u^\mu$ are the fully STF-($\perp$$ u$) octupolar tidal tensors
    \begin{equation}
        E_{\lambda\mu\nu}\mp iB_{\lambda\mu\nu}=2\nabla_{\kappa}{}^\pm\! R^\rho{}_{(\lambda}{}^\sigma{}_{\mu}\big(\delta_{\nu)}{}^\kappa+u_{\nu)} u^\kappa\big)u_\rho u_\sigma,
    \end{equation}
    and the `time derivatives' of the quadrupolar curvature components,
    \begin{equation}
        \dot E_{\mu\nu}\mp i\dot B_{\mu\nu}=2u^\lambda\nabla_\lambda {}^\pm \!R_{\rho\mu\sigma\nu}u^\rho u^\sigma.
    \end{equation}

    In terms of the tidal tensors, the most general dynamical mass function through $S^5$ which is quadratic in the Riemann tensor and contains only pieces which are nonzero in vacuum is:
    \begin{align}
        \delta\fc{M}_2^2 =&\ \delta\fc{M}_{2 S^4}^2 + \delta\fc{M}_{2 S^5}^2 + \OO(S^6), \\
        \delta\fc{M}_{2 S^4}^2 =&\ \frac{D_{4a}}{m^2} (E_{SS})^2 + \frac{D_{4b}}{m^2} S^2 E^\mu{}_S E_{\mu S} + \frac{D_{4c}}{m^2} S^4 E^{\mu\nu} E_{\mu\nu} \nonumber \\
        &\ + \frac{D_{4d}}{m^2} (B_{SS})^2 + \frac{D_{4e}}{m^2} S^2 B^\mu{}_S B_{\mu S} + \frac{D_{4f}}{m^2} S^4 B^{\mu\nu}B_{\mu\nu}, 
        \\
        \delta\fc{M}_{2 S^5}^2 =&\ \frac{D_{5a}}{m^3} B_{SS} E_{SSS} + \frac{D_{5b}}{m^3} S^2 B^\mu{}_S E_{\mu SS} + \frac{D_{5c}}{m^3} S^4 B^{\mu\nu}E_{\mu\nu S} \nonumber \\
        &+ \frac{D_{5d}}{m^3} E_{SS}B_{SSS} + \frac{D_{5e}}{m^3} S^2 E^\mu{}_S B_{\mu SS} + \frac{D_{5f}}{m^3} S^4 E^{\mu\nu}B_{\mu\nu S}
        \\ \nonumber
        &+\(\frac{D_{5g}}{m^3}E_{S\mu}\dot E_{S\nu}+\frac{D_{5h}}{m^3}S^2E^{\lambda}{}_{\mu}\dot E_{\lambda\nu}+\frac{D_{5i}}{m^3}B_{S\mu}\dot B_{S\nu}+\frac{D_{5j}}{m^3}S^2B^{\lambda}{}_{\mu}\dot B_{\lambda\nu}\) \epsilon^{\mu\nu u S} S^2.
    \end{align}

    The spinless and linear in spin pieces of the amplitudes are universal (independent of the values of any Wilson coefficients). At quadratic order in spin, there are 4 Wilson coefficients in the dynamical mass function, $C_2$, $E_{2a,b,c}$. The amplitude for a given body at $\OO(S^2)$ determines $C_2$ and is independent of $E_{2a,b,c}$. In fact, the amplitude is independent of $E_{2a,b,c}$ to all orders in spin. At cubic order in spin, there are 2 Wilson coefficients in the dynamical mass function, $C_3$ and $E_3$. The amplitude for a given body at $\OO(S^3)$ determines $C_3$ and is independent of $E_3$. In fact, the amplitude is independent of $E_3$ to all orders in spin. At quartic order in spin, there are 18 Wilson coefficients in the dynamical mass function, $C_4$, $D_{4a,b,c,d,e,f}$, $E_{4a,b,c,d,e,f,g,h,i,j,k}$. The amplitude for a given body at $\OO(S^4)$ is independent of $E_{4c,d,e,f,g,h,i,j,k}$ and remains so at all orders in spin. Of the 9 remaining coefficients $C_4$, $D_{4a,b,c,d,e,f}$, $E_{4a,b}$, only 7 linear combinations can be determined from the $\OO(S^4)$ Compton amplitude. In particular, the $S^4$ amplitude is independent of the value of $E_{4a}$ (though the $S^5$ amplitude does depend on $E_{4a}$) and of the linear combination:
    \begin{equation}
        Z_4 = E_{4b} + \frac{D_{4b}}{2} - \frac{D_{4c}}{6} - \frac{D_{4e}}{2} + \frac{D_{4f}}{6}.
    \end{equation}
    We find agreement with the Compton amplitude computed by Ben-Shahar in Ref~\cite{Ben-Shahar:2023djm} to quartic order in spin.
    
    At quintic order in spin, there are 16 Wilson coefficients in the dynamical mass function, $C_5$, $D_{5a,b,c,d,e,f,g,h,i,j}$, $E_{5a,b,c,d,e}$. As well, $E_{4a}$ contributes to the $S^5$ amplitude. Of these 17 coefficients, 11 linearly combinations can be determined from the Compton amplitude. The amplitude is independent of the values of the 6 linear combinations:
    \begin{align}
        Z_{5a} &= E_{5a} + \frac{D_{5e}}{180} + \frac{D_{5g}}{60} - \frac{D_{5i}}{90} \\
        Z_{5b} &= E_{5b} -\frac{4}{15} D_{5b} + \frac{D_{5c}}{10} - \frac{4}{15}D_{5e} + \frac{D_{5f}}{10} + \frac{D_{5h}}{30} - \frac{D_{5j}}{30} \\
        Z_{5c} &= E_{5c} -\frac{D_{5b}}{12} + \frac{D_{5c}}{36} - \frac{17}{180}D_{5e} + \frac{D_{5f}}{30} - \frac{D_{5g}}{180} + \frac{D_{5h}}{180} - \frac{D_{5i}}{180} \\
        Z_{5d} &= E_{5d} -\frac{D_{5f}}{180} - \frac{D_{5h}}{90} + \frac{D_{5j}}{180} \\
        Z_{5e} &= E_{5e} + \frac{D_{5c}}{60} + \frac{D_{5f}}{60} + \frac{D_{5h}}{20} - \frac{D_{5j}}{20} \\
        Z_{5f} &= E_{4a} -\frac{D_{5b}}{20} - \frac{C_2}{10}D_{5g} + \frac{3}{20}D_{5i}.
    \end{align}
    There are precisely as many $Z$ combinations which the Compton amplitude is independent of as there are $E$ coefficients in the amplitude at this order, so if the amplitude is fully determined by some matching conditions those conditions can be expressed so that all of the $D$ coefficients are parameterized by the matched values and the undetermined $E$ coefficients. If \eqref{eq:dmfrequire} is used, all of these $E$ coefficients are set to 1, which then fully determines the $D$ coefficients and hence the dynamical mass function through this order. Because of these null $Z$ combinations, the values of the any values of $E$ coefficients can be made consistent with any matching conditions on the Compton amplitude.

    In order to match the spin-exponentiated amplitude of Ref.~\cite{Guevara:2018wpp} through $\OO(S^4)$, we must have:
    \begin{gather}
        C_2 = C_3 = C_4 = 1, \p D_{4a} = D_{4d} = 0, \nonumber \\
        D_{4b} = \frac{E_{4b}}{2}, \p D_{4c} = -\frac{E_{4b}}{6}, \p D_{4e} =-\frac{E_{4b}}{2}, \p D_{4f} = \frac{E_{4b}}{6}. \label{eq:gov4th}
    \end{gather}
    These conditions are consistent with \eqref{eq:dmfrequire}. For the opposite-helicity amplitude, it is possible to continue the spin-exponentiation through $S^5$. Doing so requires:
    \begin{align}
        C_5 &= 1, \nonumber \\ 
        D_{5d} &= \frac{1}{6} - D_{5a}, \nonumber \\
        D_{5e} &= -\frac{1}{15} - D_{5b} - \frac{E_{4a}}{20} + \frac{E_{5a}}{180} - \frac{8}{15}E_{5b} - \frac{8}{45}E_{5c}, \nonumber \\
        D_{5f} &= -D_{5c} + \frac{E_{5b}}{5} + \frac{11}{180}E_{5c} - \frac{E_{5d}}{180} + \frac{E_{5e}}{30}, \nonumber \\
        D_{5i} &= \frac{2}{9} + D_{5g} + \frac{E_{4a}}{4} - \frac{E_{5a}}{36} \nonumber \\
        D_{5j} &= D_{5h} - \frac{E_{5b}}{15} - \frac{E_{5c}}{180} + \frac{E_{5d}}{60} - \frac{E_{5e}}{10}
    \end{align}
    which are completely consistent with \eqref{eq:dmfrequire} (which simply sets $E_{4a} = E_{5a,b,c,d,e} = 1$ in these expressions). Therefore, if the dynamical mass function in \eqref{eq:grdmf} is used, opposite-helicity spin exponentiation can be maintained at $\OO(S^5)$. 

    It is also interesting to see what is required by shift-symmetry. In order for the same-helicity Compton amplitude to have shift-symmetry through $\OO(S^4)$ we find that it requires:
    \begin{gather}
        C_4 = 1, \p D_{4e} = -D_{4a} - D_{4b} - D_{4d}, \p D_{4f} = \frac{D_{4a}}{4} - D_{4c} + \frac{D_{4d}}{4}
    \end{gather}
    which are consistent with both \eqref{eq:dmfrequire} and \eqref{eq:gov4th}. It is possible to demand shift-symmetry at $S^5$ as well. Doing so requires:
    \begin{align}
        C_5 &= 1, \nonumber \\
        D_{5e} &= \frac{4}{5} + D_{5a} + D_{5b} - D_{5d} + \frac{E_{4a}}{20} + \frac{E_{5a}}{180} - \frac{E_{5c}}{90}, \nonumber \\
        D_{5f} &= -\frac{43}{120} - \frac{D_{5a}}{4} + D_{5c} + \frac{D_{5d}}{4} + \frac{E_{5c}}{180} - \frac{E_{5d}}{180}, \nonumber \\
        D_{5i} &= \frac{5}{18} - D_{5g} + \frac{E_{4a}}{20} + \frac{E_{5a}}{180} - \frac{E_{5c}}{90} \nonumber \\
        D_{5j} &= -D_{5h} + \frac{E_{5c}}{180} - \frac{E_{5d}}{180}
    \end{align}
    which are completely consistent with \eqref{eq:dmfrequire}. Therefore, if the dynamical mass function in \eqref{eq:grdmf} is used, shift-symmetry will may be continued at $\OO(S^5)$. The spin-exponentiation conditions and shift-symmetry conditions can be demanded simultaneously through $\OO(S^5)$.

    A final interesting case of comparison is to results from the Teukolsky equation. Following the analysis of Ref~\cite{Bautista:2022wjf}, a match to the analytically continued results of the Teukolsky equation depends on the combinations:
    \begin{alignat}{3}
        c_2^{(0)}&= -D_{5c} + D_{5f} - \frac{E_{5c}}{180} + \frac{E_{5d}}{180},
        \qquad\qquad\;\;
        c_3^{(0)}= -2 D_{5h} - 2 D_{5j} + \frac{E_{5c}}{90} - \frac{E_{5d}}{90},
        \nonumber\\
        c_2^{(1)}&= \frac{C_3}{6} - \frac{C_2 C_3}{3} + \frac{C_3^2}{6} - \frac{5 C_4}{24} + \frac{C_2 C_4}{4} + \frac{C_5}{120} - \frac{D_{5b}}{2} - 2 D_{5c} + \frac{D_{5e}}{2} + 2 D_{5f}  \nonumber \\
        &\ \ \qquad - \frac{E_{4a}}{40}- \frac{E_{5a}}{360} - \frac{E_{5c}}{180} + \frac{E_{5d}}{90}, \phantom{\bigg|}
        \nonumber\\
        c_3^{(1)}&= -\frac{C_3}{3} + \frac{2 C_2 C_3}{3} - \frac{2 C_3^2}{9} + \frac{C_4}{4} - \frac{C_2 C_4}{3} - \frac{C_5}{60} - D_{5g} - 4 D_{5h} - D_{5i} - 4 D_{5j} \nonumber \\
        &\ \ \qquad + \frac{3 E_{4a}}{20} - \frac{C_2 E_{4a}}{10} + \frac{E_{5a}}{180} + \frac{E_{5c}}{90} - \frac{E_{5d}}{45},
        \nonumber\\
        c_2^{(2)}&= -\frac{C_2^2}{8} - \frac{C_3}{12} + \frac{5 C_2 C_3}{24} - \frac{C_3^2}{24} + \frac{C_4}{24} - \frac{C_2 C_4}{24} - \frac{D_{5a}}{4} - \frac{D_{5b}}{2} - D_{5c} + \frac{D_{5d}}{4} + \frac{D_{5e}}{2} + D_{5f}\nonumber \\
        &\ \ \qquad - \frac{E_{4a}}{40} - \frac{E_{5a}}{360} + \frac{E_{5d}}{180}, \nonumber \\
        c_3^{(2)}&= \frac{C_2^2}{4} + \frac{C_3^2}{36} - D_{5g} - 2 D_{5h} - D_{5i} - 2 D_{5j} + \frac{3 E_{4a}}{20} - \frac{C_2 E_{4a}}{10} + \frac{E_{5a}}{180} - \frac{E_{5d}}{90}.
    \label{eq:littlecdef}
    \end{alignat}
    Matching Teukolsky at $\OO(S^5)$ requires:
    \begin{gather}
        c_2^{(0)} = c_2^{(1)} = c_2^{(2)} = 0, \p c_3^{(0)}=\dfrac{64}{15}\alpha, \p c_3^{(1)}=\dfrac{16}{3}\alpha, \p c_3^{(2)}=\dfrac{4}{15}(1+4\alpha),
    \end{gather}
    where $\alpha = 1$ if contributions from analytically continued digamma functions are to be kept or $\alpha = 0$ if such contributions are to be dropped. Matching to Teukolsky is inconsistent with shift-symmetry but consistent with continuing spin-exponentiation for the same helicity amplitude and with \eqref{eq:grdmf}. The combination of Teukolsky, spin-exponentiation, and \eqref{eq:dmfrequire} are consistent with each other and fully determine the $D$ type Wilson coefficients to be:
    \begin{align}
        &D_{5a} = -\frac{1}{10},&   &D_{5b} = -\frac{23}{60},&  &D_{5c} = \frac{13}{90},&   &D_{5d} = \frac{4}{15},& \nonumber \\
        &D_{5e} = -\frac{79}{180},&  &D_{5f} = \frac{13}{90},&   &D_{5g} = -\frac{7}{36}+\frac{8}{5}\alpha,&    &D_{5h} = \frac{7}{90} -\frac{16}{15}\alpha,& \nonumber \\
        & & &D_{5i} = \frac{1}{4} + \frac{8}{5}\alpha,& &D_{5j} = -\frac{7}{90} - \frac{16}{15}\alpha.& & &
    \end{align}

    For expressing the full amplitudes, recall $\check k_1=k_1-w$, $\check k_2=k_2-w$. Then, for the helicity-preserving amplitude we find:
    \begin{alignat}{3}
        \fc A_{++}=\frac{(4\omega^2-q^2)^2}{16 q^2\omega^2}\bigg\{&
        1+\check k_1\cdot a+\check k_2\cdot a
        \nonumber\\
        &+\frac{1}{2}(\check k_1\cdot a+\check k_2\cdot a)^2+\frac{C_2-1}{2}\Big((\check k_1\cdot a)^2+(\check k_2\cdot a)^2\Big)
        \nonumber \\
        &+\frac{1}{6}(\check k_1\cdot a+\check k_2\cdot a)^3+\frac{C_3-1}{6}\Big((\check k_1\cdot a)^3+(\check k_2\cdot a)^3\Big)
        \nonumber \\
        &\qquad\qquad +\frac{C_2-1}{2}( k_1+ k_2-2C_2w)\cdot a\,\check k_1\cdot a\,\check k_2\cdot a
        \nonumber\\
        &+\hat{\fc A}_{++}^{(4)}+\hat{\fc A}_{++}^{(5)}+\fc O(a^6)\bigg\},
    \end{alignat}
    with
    \begin{alignat}{3}
        \hat{\fc A}_{++}^{(4)}&=\frac{1}{24}(\check k_1\cdot a+\check k_2\cdot a)^4 - (D_{4a}+D_{4d}) \frac{2\omega^2}{q^2} (\check k_1\cdot a)^2(\check k_2\cdot a)^2\nonumber \\
        &\ \ \ 
        + (D_{4b}+D_{4e})\omega^2 a^2 (\check k_1\cdot a)(\check k_2\cdot a)- \frac{D_{4c}+D_{4f}}{2} q^2 \omega^2 a^4 \nonumber \\
        &\ \ \ 
        + \frac{C_2 -1 }{6}(\check k_1\cdot a)(\check k_2\cdot a)\Big(3(\check k_1\cdot a)(\check k_2\cdot a)-4(w\cdot a)^2+ \frac{8\omega^2}{q^2}(\check k_1\cdot a)(\check k_2\cdot a)\Big) \nonumber \\
        &\ \ \ 
        + \frac{(C_2-1)^2}{4}(\check k_1\cdot a)(\check k_2\cdot a)\Big((\check k_1\cdot a)(\check k_2\cdot a)+2(\check k_1+\check k_2)\cdot a(w\cdot a)\nonumber \\
        &\quad\quad\quad\quad\quad\quad\quad\quad\quad\quad\quad\quad\quad\quad\quad\quad+\frac{8\omega^2}{q^2}(\check k_1\cdot a)(\check k_2\cdot a)\Big) \nonumber \\
        &\ \ \ 
        + \frac{C_3-1}{6}(\check k_1\cdot a)(\check k_2\cdot a)((\check k_1\cdot a)^2+(\check k_2\cdot a)^2) +\frac{C_4 - 1}{24}((\check k_1\cdot a)^4 + (\check k_2\cdot a)^4) \nonumber \\
        &\ \ \ 
        - \frac{(C_2-1)(C_3-1)}{6}(\check k_1\cdot a)(\check k_2\cdot a)\Big(3(\check k_1+\check k_2)\cdot a(w\cdot a) + 4(w\cdot a)^2 \nonumber \\
        &\quad\quad\quad\quad\quad\quad\quad\quad\quad\quad\quad\quad\quad\quad\quad\quad+\frac{4\omega^2}{q^2}(\check k_1\cdot a)(\check k_2\cdot a)\Big)
    \end{alignat}
    noting
    \begin{equation}
        \omega^2a^2=(w\cdot a)^2-\frac{4\omega^2}{q^2}\check k_1\cdot a\,\check k_2\cdot a,
    \end{equation}
    and with
    \begin{alignat}{3}
        \hat{\fc A}_{++}^{(5)}&= \frac{1}{120}(\check k_1\cdot a + \check k_2\cdot a)^5 \nonumber \\
        &\ \ \ + (c_2^{(0)}(k_1+k_2)\cdot a + c_3^{(0)} w\cdot a)(w\cdot a)^4 \frac{q^2}{4\omega^2} \nonumber \\
        &\ \ \ - (c_2^{(1)}(k_1+k_2)\cdot a + c_3^{(1)} w\cdot a)(w\cdot a)^2 \check k_1\cdot a \check k_2\cdot a \nonumber \\
        &\ \ \ + (c_2^{(2)}(k_1+k_2)\cdot a + c_3^{(2)}w\cdot a)\frac{4\omega^2}{q^2}(\check k_1\cdot a)^2(\check k_2\cdot a)^2 \nonumber \\
        &\ \ \ + \frac{C_2 C_3 -1}{12} (\check k_1\cdot a)^2(\check k_2\cdot a)^2(\check k_1+\check k_2)\cdot a - \frac{(C_2-C_3)^2}{4} (\check k_1\cdot a)^2(\check k_2\cdot a)^2 (w\cdot a) \nonumber \\
        &\ \ \ + \frac{C_4-1}{24} (\check k_1\cdot a)(\check k_2\cdot a)((\check k_1\cdot a)^3+(\check k_2\cdot a)^3) + \frac{C_5-1}{120}((\check k_1\cdot a)^5+(\check k_2\cdot a)^5) \nonumber \\
        &\ \ \  + \frac{(C_2-1)(C_3-C_4)}{6}(\check k_1\cdot a)(\check k_2\cdot a)((\check k_1\cdot a)^2+(\check k_2\cdot a)^2)(w\cdot a),
    \label{eq:hatApp5kcheck}
    \end{alignat}
    using the previous definitions of the $c_i^{(j)}$ coefficients in \eqref{eq:littlecdef}. Because the $C$ coefficients are all determined by the three-point amplitude, for black holes they are all known to take the value 1.
    
    For the helicity reversing amplitude, it is useful to recall:
    \begin{alignat}{3}
        (aya)&:=(k_1\cdot  a)(k_2\cdot  a)-(x\cdot  a)(q\cdot  a)-\omega^2 a^2
        \nonumber\\
        &=\dfrac{-q^2}{4\omega^2-q^2}(k_1\cdot a-x\cdot a)(k_2\cdot a+x\cdot a)=\frac{q^2}{4\omega^2}\Big((x\cdot a)^2-\omega^2a^2\Big).
    \end{alignat}
    With this, we find:
    \begin{alignat}{3}
        \fc A_{+-}=\frac{q^2}{16\omega^2}\bigg\{
        &
        1-q\cdot a+\frac{(q\cdot a)^2}{2}C_2+(C_2-1)(aya)
        \nonumber\\\nonumber
        &-\frac{(q\cdot a)^3}{6}C_3+(aya)\Big((1-C_2-C_2^2+C_3)x+\frac{C_2-C_3}{2}q\Big)\cdot a
        \\
        &+\hat{\fc A}^{(4)}_{+-}+\hat{\fc A}^{(5)}_{+-}+\fc O(a^6)\bigg\},
    \end{alignat}
    with
    \begin{alignat}{3}
        \hat{\fc A}^{(4)}_{+-}
        &=\frac{(q\cdot a)^4}{24}C_4+(aya)^2\frac{3 C_2^2-4C_3+C_4}{12}+\frac{C_4-C_3}{6}(aya)q\cdot a(q-2x)\cdot a
        \nonumber\\\nonumber
        &\qquad\qquad \quad\;\;+(C_2-1)(aya)\bigg[\frac{C_3+3C_2}{3}\Big(2(aya)\frac{\omega^2}{q^2}+\omega^2 a^2\Big)+\frac{C_3+C_2}{2}q\cdot a\,x\cdot a\bigg]
        \\\nonumber
        &\quad +(aya)^2\frac{\omega^2}{q^2}2(D_{4d}-D_{4a})+q^2\omega^2a^4\bigg(\frac{D_{4f}-D_{4c}}{2}-\frac{E_{4b}}{6}\bigg)
        \\ 
        &\quad +(aya)\omega^2a^2(D_{4b}-D_{4e}-E_{4b}),
    \end{alignat}
    and
    \begin{alignat}{3}
        \hat{\fc A}^{(5)}_{+-}&= -\frac{C_5}{120}(q\cdot a)^5   
        +\Big(-\frac{C_2 C_3}{12} + \frac{C_4}{8} - \frac{C_5}{24}\Big)(q\cdot a)(aya)^2 
        +\frac{C_4-C_5}{24}(q\cdot a)^3 (aya) \nonumber \\
        &
        +\frac{-3 C_2^2 + 4C_3 + 3 C_3^2 - C_4 - 4 C_2 C_4 + C_5}{12} (x\cdot a)(aya)^2 \nonumber \\
        &
        +\frac{2 C_3 - 2 C_2 C_3 + C_4 -  2 C_2 C_4 + C_5}{12} (x\cdot a)(q\cdot a)^2 (aya) \nonumber \\
        &
        +\frac{-3 C_2^2 + 10 C_3 - 3 C_2 C_3 + 5 C_3^2 - C_4 - 9 C_2 C_4 + 2 C_5 - 6 D_{5a} - 6 D_{5d}}{6} \frac{\omega^2}{q^2}(q\cdot a)(aya)^2 \nonumber \\
        &
        + \Big(\frac{C_3}{2} - \frac{C_2 C_3}{3} + \frac{C_3^2}{6} + \frac{C_4}{24} - \frac{5 C_2 C_4}{12} + \frac{3 C_5}{40} \nonumber \\
        &\quad\quad\quad\quad\quad\quad\quad\quad + \frac{D_{5b}}{2} + \frac{D_{5e}}{2} + \frac{E_{4a}}{40} - \frac{E_{5a}}{360} + \frac{4 E_{5b}}{15} + \frac{4 E_{5c}}{45}\Big) (q\cdot a) (aya) \omega^2 a^2 \nonumber \\
        &
        + \Big(\frac{C_3}{3} + \frac{2 c_3^2}{9} - \frac{C_4}{12} - \frac{C_2 C_4}{3} + \frac{C_5}{12} \nonumber \\
        &\quad\quad\quad\quad\quad\quad\quad\quad + D_{5g} - D_{5i} + \frac{3 E_{4a}}{20} + \frac{C_2 E_{4a}}{10} - \frac{E_{5a}}{36}\Big) (x\cdot a)(aya) \omega^2 a^2 \nonumber \\
        & 
        + \Big(-\frac{D_{5c}}{4} - \frac{D_{5f}}{4} + \frac{E_{5b}}{20} + \frac{11 E_{5c}}{720} - \frac{E_{5d}}{720} + \frac{E_{5e}}{120}\Big) (q\cdot a) q^2\omega^2 a^4 \nonumber \\
        &
        + \Big(-\frac{D_{5h}}{2} + \frac{D_{5j}}{2} + \frac{E_{5b}}{30} + \frac{E_{5c}}{360} - \frac{E_{5d}}{120} + \frac{E_{5e}}{20}\Big) (x\cdot a) q^2\omega^2 a^4.
    \end{alignat}

\section{Conclusion}

Using the dynamical mass function worldline formalism, we derived formal expressions for the electromagnetic/gravitational Compton amplitudes of a generic spinning body to all orders in spin, with precise parameterized expressions in terms of Wilson coefficients for the amplitudes to order $S^3$ in electromagnetism and $S^5$ in gravity for bodies which match the \rk/Kerr three-point amplitude. In electromagnetism we found 1 Wilson coefficient and 1 independent structure in the Compton at $S^1$, 5 new Wilson coefficients and 5 independent structures at $S^2$, and 8 new Wilson coefficients but only 7 independent structures at $S^3$. In gravity we found 4 Wilson coefficients and 1 independent structure in the Compton at $S^2$, 2 new Wilson coefficients and 1 independent structure at $S^3$, 18 new Wilson coefficients and 7 independent structures at $S^4$, and 16 new Wilson coefficients and 11 independent structures at $S^5$. As well, one of the Wilson coefficients on an $S^4$ operator in gravity ($E_{4a}$) does not contribute to $S^4$ piece of the Compton but does contribute to the $S^5$ piece (whereas the other operator contributions which do not contribute at the order they are introduced actually do not contribute at all through $S^5$).

Dixon's multipole moment formalism provides additional physical constraints on the dynamical mass function beyond those required by naive multipole moments which are determined by the three-point amplitude. Many of the additional terms induced by Dixon's formalism (the operators with $E$ coefficients), especially at low orders in spin, happen to not affect the Compton amplitude. However, at sufficiently high orders in spin ($S^5$ and beyond in gravity) at least some of these additional terms do contribute to the Compton amplitude. Through $\OO(S^5)$ these combinations are linearly redundant in the Compton amplitude to contributions from Riemann squared operators. It would be interesting to see if these additional coefficients begin to contribute in linearly independent ways in the Compton at higher orders in spin or in higher-point processes, such as the five-point amplitude (with three graviton lines). 

In electromagnetism, using Dixon's multipole moments for \rk determines the dynamical mass function to be given by \eqref{eq:fulldmf} with no room for additional operators which are linear in the field strength. Because the stationary \rk solution only determines its multipole moments up to corrections which are linear in the field strength, they only determine the couplings in the action up to corrections which are quadratic in the field strength. Without some additional physical principle which specifies how the multipole moments of the \rk particle deform in the presence of a background field (which would determine its electromagnetic susceptibility tensors), it is not possible to determine the quadratic in field strength couplings using the multipole moment formalism. The couplings in \eqref{eq:fulldmf} contain all of the couplings of \eqref{eq:lsm} (in that the coefficients of operators which are present in both agree). The story is very similar in gravity. Dixon's multipole moment formalism applied to the Kerr solution determines the dynamical mass function to be given by \eqref{eq:grdmf} with no room for additional operators which are linear in the Riemann tensor. Because the stationary Kerr solution only determines its multipole moments up to corrections which are linear in the Riemann tensor, they only determine the couplings in the action up to corrections which are quaratic in the Riemann tensor. Without some further knowledge of the gravitational susceptibility tensors of a spinning black hole, it is not possible to determine the quadratic in Riemann couplings using the multipole moment formalism. The couplings in \eqref{eq:grdmf} contain all of the couplings of \eqref{eq:lsdmf}. 

It is uncertain what precise physical principles determine the correction couplings in the action for a spinning black hole in general. Through $\OO(S^4)$, spin-exponentiation, shift-symmetry, the results of the Teukolsky equation, and the results from the multipole moment formalism can all be maintained simultaneously by appropriately choosing the values of Wilson coefficients for quadratic in Riemann tensor operators. However, beginning at $\OO(S^5)$ these different principles cannot all be maintained. Spin-exponentiation is only possible to maintain for one of the two independent helicity combinations and is consistent with shift-symmetry at $\OO(S^5)$. However, shift-symmetry and the Teukolsky results are inconsistent with each other at $\OO(S^5)$. The couplings fixed by the multipole moment formalism are the unique values for the couplings so that the stress tensor they produce behaves correctly against test functions (meaning satisfies \eqref{eq:intstress} with the multipole moments of the Kerr solution used to form the generating function on the right hand side). In this way, only those couplings produce the Kerr multipole moments (up to corrections which are linear in the Riemann tensor) for a black hole which is in arbitrary nonuniform motion. Such couplings are consistent with maintaining a match to spin-exponentiation and the Teukolsky equation at $\OO(S^5)$.

\acknowledgments

We thank Zvi Bern, Juan Pablo Gatica, Lukas Lindwasser, and Richard Myers for enlightening discussions and Maor Ben-Shahar for sharing a preliminary draft of his work Ref.~\cite{Ben-Shahar:2023djm}. We also thank Henrik Johansson, Alexander Ochirov, and Yilber Fabian Bautista for alerting us of typos in an earlier version of this paper.
T.S. is supported by the U.S. Department of Energy (DOE) under award number DE-SC0009937. 
We are also grateful to the Mani L. Bhaumik Institute for Theoretical Physics for support.
%


\providecommand{\href}[2]{#2}\begingroup\raggedright\endgroup

\end{document}